%% file: root.tex
\documentstyle[11pt,fleqn,epsf,twoside,fancyhdr]{report}

\renewcommand{\baselinestretch}{1.50}

\newlength{\captsize}		\let\captsize=\normalsize
\newlength{\captwidth}		
\newlength{\beforetableskip} 	
\newcommand{\capt}[1]{
	\renewcommand{\baselinestretch}{0.9}		
        \begin{minipage}{\captwidth}
	\let\normalsize=\captsize
	\caption[#1]{\sf #1}
	\end{minipage}\\ \vspace{\beforetableskip}}

\makeatletter
	\long\def\@makecaption#1#2{\vskip 10\p@
	\setbox\@tempboxa\hbox{{\bf #1:} #2} 
	\ifdim \wd\@tempboxa >\hsize
           {\bf #1:} #2\par
        \else
           \hbox to\hsize{\box\@tempboxa\hfill}
        \fi}
\makeatother     

\begin{document}

\newcommand{\be}{\begin{equation}}
\newcommand{\ee}{\end{equation}}
\newcommand{\bea}{\begin{eqnarray}}
\newcommand{\eea}{\end{eqnarray}}
\newcommand{\sst}{\scriptscriptstyle}

\input{tesis9}           
\input{tesis10}        
\input{tesis12}            
\input{tesis1}           

\pagenumbering{roman}
\fancyhead[LE]{\thepage}
\fancyhead[RE]{\sl Indice}
\fancyhead[RO]{\bf \thepage}
\fancyfoot[CE,CO]{}

\tableofcontents
\clearpage
\thispagestyle{empty}
~

\newpage

\renewcommand{\chaptermark}[1]{\markboth{\sl #1}{}}
\renewcommand{\sectionmark}[1]{\markright{\sl \thesection . \hspace{0.05cm} 
#1}}

\fancyhead[LE,RO]{\bf \thepage}
\fancyhead[RE]{\leftmark}
\fancyhead[LO]{\rightmark}

\input{tesis3}               
\input{tesis4}               
\input{tesis5}               
\input{tesis6}               
\input{tesis7}               
\input{tesis8}               
\input{tesis11}                

\newpage
\input{tesis2}              

\end{document}

%% file: tesis9
\thispagestyle{empty}
\begin{center}
\vspace{-3.5cm}
{\Large  Universidad de Buenos Aires}\\
\vspace{0.3cm}
\hspace*{-0.5cm}{\Large Facultad de Ciencias Exactas y Naturales}\\
\vspace{0.3cm}
{\Large Departamento de F{\'\i}sica}\\
\vspace{3.0cm}
{\Huge Tesis Doctoral}\\
\vspace{1.5cm}
{\Huge\bf Transici\'on Cu\'antico-Cl\'asica en}\\
\vspace{0.5cm}
{\Huge\bf Teor\'\i a de Campos} \\
\vspace{3.cm}
{\Large Autor: Fernando C\'esar Lombardo}\\
\vspace{0.3cm}
{\Large Director: Dr. Francisco Diego Mazzitelli}\\
\vspace{5.0cm}
{\large Trabajo de Tesis para optar por el t\'{\i}tulo de Doctor en Ciencias
F\'{\i}sicas}\\
\vspace{0.3cm}
{\large Noviembre de 1998}

\end{center}

\newpage

\thispagestyle{empty}
~
\newpage

%% file: tesis10
\thispagestyle{empty}

\hfill \hfill {\it A mi familia} 

\newpage

\thispagestyle{empty}
~
\newpage

%% file: tesis12
\thispagestyle{empty}

{\Huge\bf Resumen}

\vspace{0.1cm}

En esta Tesis estudiamos el proceso de transici\'on cu\'antico-cl\'asica 
en teor\'\i a de campos. Extendiendo el formalismo de la funcional de 
influencia de Feynman y Vernon, estudiamos el proceso de p\'erdida de 
coherencia para campos autointeractuantes en espacios planos y para modelos 
de gravedad cu\'antica.

En primer lugar, despu\'es de destacar los principales resultados para 
una part\'\i cula Browniana cu\'antica acoplada a un entorno de osciladores 
arm\'onicos, consideramos una teor\'\i a de campos con autointeracci\'on 
en el espacio de Minkowski. Calculamos una acci\'on efectiva de granulado 
grueso integrando los modos del campos cuyas longitudes de onda sean menores 
que cierta longitud cr\'\i tica. A partir de esta acci\'on efectiva obtenemos 
la ecuaci\'on de evoluci\'on para la matriz densidad reducida. Calculamos los 
coeficientes de difusi\'on de esta ecuaci\'on y 
analizamos el proceso de p\'erdida de coherencia para las longitudes de 
onda mayores que la cr\'\i tica. Luego, con el 
objeto de analizar la formaci\'on de estructuras en modelos 
inflacionarios, generalizamos los resultados 
a campos acoplados conformemente a la m\'etrica de de Sitter. Mostramos 
que la p\'erdida de coherencia es efectiva si la longitud de onda cr\'\i tica
 es mayor que el radio de Hubble. 

Por otro lado, estudiamos el l\'\i mite cl\'asico en modelos de gravedad 
escalar-tensorial en dos dimensiones. Analizamos diferentes acoplamientos 
entre el campo dilat\'onico y el campo de materia. Discutimos el proceso de
radiaci\'on de Hawking en los distintos modelos. A partir de un c\'alculo 
exacto de la funcional de
 influencia, estudiamos las condiciones bajo las cuales es posible que 
la p\'erdida de coherencia sea efectiva en m\'etricas cosmol\'ogicas para 
asegurar la validez de la aproximaci\'on semicl\'asica.

Finalmente estudiamos modelos en cuatro dimensiones 
donde campos masivos se acoplan a la geometr\'\i a del espacio-tiempo de 
manera arbitraria. Calculamos las ecuaciones de Einstein-Langevin 
para estudiar el efecto de las fluctuaciones cu\'anticas inducidas por 
los campos de materia sobre la geometr\'\i a cl\'asica.


\noindent{\sl Palabras claves:} 

Transici\'on cu\'antico-cl\'asica  - Teor\'{\i}a cu\'antica de campos - 
Gravedad semicl\'asica - P\'erdida de coherencia - Agujeros negros - 
Cosmolog\'{\i}a 

\newpage

\thispagestyle{empty}
~
\newpage

%% file: tesis1
\thispagestyle{empty}

{\Huge\bf Abstract}

\vspace{0.5cm}

In this Thesis we study the quatum to classical transition 
process in the context of quantum field theory. Extending the influence 
functional 
formalism of Feynman and Vernon, we study the decoherence process for 
self-interacting quantum fields in flat space. We also use this formalism 
for arbitrary geometries to analyze the quantum to classical transition 
in quantum gravity.

After summarizing the main results known for the quantum Brownian motion, 
we consider a self-interacting field theory in Minkowski spacetime. 
We compute a coarse grained effective action by integrating out the field 
modes with wavelength shorter than a critical value. 
From this effective action we obtain the evolution equation
for the reduced density matrix (master equation).
We compute the diffusion coefficients for this equation and
analyze the decoherence induced on the long-
wavelength modes. 
We generalize the results to the case of a conformally
coupled scalar field in de Sitter spacetime.
We show that the decoherence is effective as long as the 
critical wavelength is taken to be not shorter than the Hubble 
radius. 

On the other hand, we study the classical limit for scalar-tensorial 
models in two dimensions. We consider different couplings between the 
dilaton and the scalar field. We discuss the Hawking radiation process and, 
from an exact evaluation of the influence functional, we study the conditions 
by which decoherence ensures the validity of the semiclassical approximation 
in cosmological metrics. 

Finally we consider four dimensional models with massive scalar fields,  
arbitrary coupled to the geometry. We compute the Einstein-Langevin equations 
in order to study the effect of the fluctuations induced by 
the quantum fields on the classical geometry.   

\vspace*{0.5cm}

\noindent{\sl Keywords:} 

Quantum to classical transition - Quantum field theory - Semiclassical 
gravity - Decoherence - Black holes  - Cosmology

\newpage

\thispagestyle{empty}
~
\newpage

%% file: tesis3
\newpage
\chapter{Introducci\'on}

\pagenumbering{arabic}
\thispagestyle{empty}

La motivaci\'on fundamental de la presente Tesis es avanzar en la 
comprensi\'on del origen y de los mecanismos por los cuales la transici\'on 
cu\'antico-cl\'asica tiene lugar en teor\'\i a de campos. En particular, 
utilizando una extensi\'on del formalismo de la funcional de influencia de 
Feynman y Vernon para teor\'\i a de campos,  
estudiamos este proceso para campos escalares en el espacio de Minkowski y 
para campos 
escalares acoplados a geometr\'\i as arbitrarias 
con el objeto de entender la transici\'on ``a lo cl\'asico'' de modelos 
de gravedad cu\'antica. 

La mec\'anica cu\'antica es una de las teor\'\i as m\'as exitosas de la
historia de la f\'\i sica: todas sus predicciones concuerdan con los
experimentos con gran precisi\'on y su aplicaci\'on ha transformado
el mundo tecnol\'ogico en diferentes
 \'areas. Por otra parte, su dominio de validez es notablemente amplio ya 
que se utiliza tanto para explicar la estructura de las estrellas de neutrones 
como tambi\'en para predecir las observaciones de experiencias que involucran 
part\'\i culas elementales que interact\'uan a muy altas energ\'\i as. 

Si bien la mec\'anica cu\'antica es imprescindible para 
una {\it descripci\'on microsc\'opica} de la na\-turaleza, con la mec\'anica 
cl\'asica bastar\'\i a, en principio,  para describir el comportamiento de 
sistemas a escalas macrosc\'opicas. Esta 
afirmaci\'on est\'a basada fundamentalmente en nuestro ``sentido com\'un'': a 
escala macrosc\'opica las cosas ``suceden'' o ``no suceden'' y los 
objetos materiales siempre tienen propiedades bien definidas. La f\'\i sica 
cl\'asica est\'a compuesta por un conjunto de axiomas compatibles con este 
tipo de afirmaciones. En cambio la mec\'anica cu\'antica es, a primera vista, 
incompatible con ellas: no es posible asignar propiedades bien definidas 
a los sistemas f\'\i sicos a menos que estas propiedades sean medidas. Pero 
la mec\'anica cu\'antica es absolutamente 
necesaria para explicar muchos fen\'omenos a\'un a escala macrosc\'opica, 
como los propios de las estrellas de neutrones antes mencionadas. 

En general, para que un sistema cu\'antico pueda considerarse como 
cl\'asico hay, al menos, dos condiciones que deben satisfacerse. Por un lado, 
la funci\'on de onda debe predecir que las variables can\'onicas est\'en 
fuertemente correlacionadas de acuerdo a las leyes cl\'asicas, o alguna 
distribuci\'on constru\'\i da a partir de ella (como ser la 
funcional de Wigner) debe presentar un ``pico'' alrededor de una o un 
conjunto de configuraciones cl\'asicas. Para ciertas funciones de onda que 
se describen frecuentemente como 
``semicl\'asicas'', se puede demostrar que predicen una fuerte 
correlaci\'on entre las coordenadas y los momentos. Entonces, nos referiremos 
a esta condici\'on simplemente como {\it correlaci\'on}. Por otro lado, la 
segunda 
condici\'on es que la interferencia entre las distintas configuraciones 
cl\'asicas debe ser despreciable, de modo tal que sea posible decir que el 
sistema est\'a en cierto estado definido, entre los muchos estados posibles. 
Esto involucra una {\it p\'erdida de coherencia}, es decir la destrucci\'on de
 los elementos no-diagonales de la matriz densidad, que representan los 
t\'erminos de interferencia.

El conflicto aparente entre la mec\'anica cu\'antica y nuestro sentido 
com\'un se basa 
en el hecho de que los efectos de interferencia cu\'antica entre estados 
macrosc\'opicamente distinguibles no son observados en la naturaleza. En todos
esos casos, la interferencia cu\'antica est\'a ausente y las probabilidades 
pueden sumarse, al igual que en la mec\'anica cl\'asica. Por otro lado, existe 
cierta ambig\"uedad cuando queremos establecer claramente la ``frontera'' 
entre lo que consideramos como cu\'antico y aquello que llamamos cl\'asico, 
por lo tanto el entendimiento de esta transici\'on es de gran importancia en 
muchas ramas de la f\'\i sica. 

En general la inexistencia de interferencia cu\'antica entre estados 
macrosc\'opicamente distinguibles puede ser explicada como consecuencia del 
proceso de {\it p\'erdida de coherencia}. Este proceso considera como 
aspecto fundamental que los objetos macrosc\'opicos siempre interact\'uan 
con un entorno formado por un gran n\'umero de variables irrelevantes. Esta
interacci\'on es la que produce que los efectos de interferencia cu\'antica 
desaparezcan muy r\'apidamente \cite{zurek81, zurek82, zurekpt} y emerja una 
descripci\'on en t\'erminos de variables cl\'asicas de lo que en principio
era un sistema cu\'antico. 

Por lo tanto, modelando en forma realista la interacci\'on entre sistemas
 macrosc\'opicos y sus entornos es posible tener una noci\'on clara de cu\'an 
eficiente es el mecanismo de p\'erdida de coherencia y cu\'al es la escala 
de tiempo en la que \'este act\'ua. En particular, se ha comprobado 
en varios ejemplos que la p\'erdida de coherencia puede tener lugar 
en tiempos mucho m\'as cortos que aquellos para los cuales el entorno 
comienza a producir efectos disipativos. En consecuencia no es dif\'\i cil 
entender el motivo por el cual no se observa interferencia cu\'antica de 
objetos macrosc\'opicos. 

Entre las motivaciones principales que llevaron a tratar de entender 
el proceso de p\'erdida de coherencia podemos mencionar, en primer lugar, 
al estudio de la din\'amica de la transici\'on del r\'egimen cu\'antico 
al cl\'asico: entender c\'omo y cu\'ando un sistema deja de comportarse 
cu\'anticamente (exhibiendo interferencias) para pasar a hacerlo 
cl\'asicamente. En este contexto, reviste gran inter\'es la 
comprensi\'on del proceso de transici\'on cu\'antico-cl\'asica en 
cosmolog\'\i a cu\'antica, el cual involucra p\'erdida de coherencia,
 procesos disipativos y correlaciones. Este estudio se encuadra 
principalmente en la fundamentaci\'on de la aproximaci\'on semicl\'asica 
de la gravedad cu\'antica, donde consideramos campos de naturaleza cu\'antica 
acoplados a una geometr\'\i a del espacio-tiempo que es cl\'asica. Esta 
aproximaci\'on contiene efectos muy interesantes como, por ejemplo, la   
creaci\'on de part\'\i culas, 
que a su vez tiene un rol preponderante en la transici\'on 
cu\'antico-cl\'asica.

Los problemas antes mencionados indican la necesidad de un mejor entendimiento 
de la naturaleza y estructura de los sistemas cu\'anticos abiertos, 
especialmente para teor\'\i a cu\'antica de campos. La relaci\'on entre los 
procesos estad\'\i sticos y cu\'anticos que involucran ruido y fluctuaciones, 
tales como disipaci\'on, p\'erdida de coherencia, correlaciones y  
creaci\'on de part\'\i culas debieron estar presentes en el 
Universo temprano, y su consecuente evoluci\'on debi\'o estar signada 
por la influencia de \'estos hasta arribar a nuestro Universo actual, que se
 comporta cl\'asicamente.    
     
Particularmente, distintas teor\'\i as pretenden explicar la formaci\'on de las
 estructuras en el Universo (galaxias, c\'umulos, etc) a partir de las 
inhomogeneidades primordiales, las cuales 
aparecieron debido a las fluctuaciones cu\'anticas de los campos de materia.  
 Es posible que las fluctuaciones cu\'anticas se hayan convertido en
perturbaciones cl\'asicas debido a la expansi\'on del Universo. Por lo tanto,  
la p\'erdida de coherencia de las inhomogeneidades primordiales ser\'\i a la
consecuencia de una combinaci\'on entre la expansi\'on del Universo y la 
existencia de interacciones no lineales que generan un acoplamiento entre 
aquellas inhomogeneidades que alcanzan la escala macrosc\'opica y aquellas 
que nunca crecen lo suficiente (las cuales conforman un entorno efectivo para 
las primeras). 

Por otro lado, y con el objeto de estudiar en detalle la frontera entre 
lo cu\'antico y lo cl\'asico, es posible concebir experiencias de ``p\'erdida 
de coherencia controlada'' en las que se controla con precisi\'on la 
intensidad de la interacci\'on entre el sistema cu\'antico y su entorno. 
Cuando podemos considerar el acoplamiento como d\'ebil, el sistema manifiesta 
interferencias, mientras que cuando el proceso de p\'erdida de coherencia es 
efectivo se comporta en forma cl\'asica. Recientemente se han realizado 
experiencias utilizando sistemas de iones atrapados en los que, mediante 
una combinaci\'on de campos electromagn\'eticos, se confina en el espacio 
a un conjunto de iones que permanece aislado de todo entorno (gracias a 
que los iones son enfriados lo suficiente). El entorno est\'a formado 
por los modos del campo electromagn\'etico debido a la aplicaci\'on de 
l\'aseres; y es la frecuencia de \'estos la que controla la transici\'on 
cu\'antico-cl\'asica \cite{iontrap}. 

El proceso inverso, aislar suficientemente a un sistema macrosc\'opico 
hasta que 
se comporte cu\'anticamente no es tan sencillo. Sin embargo existen propuestas
de utilizar superconductores para observar efectos cu\'anticos 
macrosc\'opicos, dado que estos materiales presentan importantes efectos 
colectivos que involucran la acci\'on coherente de un gran n\'umero de 
part\'\i culas \cite{leggett}. 

Como mencionamos al inicio de esta Introducci\'on, el objetivo principal en 
este trabajo consiste en estudiar el 
proceso de p\'erdida de coherencia en teor\'\i a de campos como primer 
paso hacia el entendimiento global de la transici\'on cu\'antico-cl\'asica en 
dicho contexto. 
En el Cap\'\i tulo 2 utilizamos una part\'\i cula Browniana cu\'antica 
acoplada 
a un entorno de osciladores arm\'onicos, como ejemplo para desarrollar un 
formalismo adecuado para estudiar sistemas cu\'anticos abiertos. Este 
formalismo, desarrollado por Feynman y Vernon \cite{feyver} es muy \'util para
describir los efectos disipativos y difusivos, caracter\'\i sticos de 
este tipo de modelos. Calculamos la ecuaci\'on maestra que da la evoluci\'on de
la matriz densidad reducida, la cual se obtiene integrando los 
grados de libertad correspondientes al entorno. 
A partir de la ecuaci\'on maestra podemos analizar la aparici\'on de 
efectos de ruido y disipaci\'on (debidos al acoplamiento con el 
entorno) que producen la transici\'on al r\'egimen cl\'asico de la 
part\'\i cula Browniana. Tambi\'en mostramos la deducci\'on de la ecuaci\'on 
de movimiento semicl\'asica y de la ecuaci\'on asociada de Langevin. Los 
resultados de este cap\'\i tulo est\'an basados, fundamentalmente en las 
referencias \cite{hpz1} y \cite{pazhabzu}.

En el Cap\'\i tulo 3, extendemos el formalismo de la funcional de influencia 
de Feynman y Vernon a teor\'\i a de campos. Consideramos un campo escalar 
con autointeracci\'on del tipo $\lambda \phi^4$. Introduciendo una 
escala artificial en el modelo, separamos los modos de dicho campo entre 
los modos de longitud de onda mayor y  menor que la escala introducida. De esta
manera podemos evaluar (mediante c\'alculos perturbativos) la ecuaci\'on 
maestra asociada a los modos de longitud de onda mayor que la longitud 
cr\'\i tica; luego de haber integrado los modos de longitud de onda 
corta, los que consideramos como entorno. Estudiando los coeficientes de 
difusi\'on de la ecuaci\'on maestra analizamos el proceso de p\'erdida de 
coherencia de los diferentes modos del ``campo-sistema''. 

Este modelo desarrollado en el espacio de Minskowski, lo extendemos a 
espacios curvos con el objeto de mostrar que la p\'erdida de coherencia 
es realmente efectiva para aquellos modos del campo escalar cuya 
longitud de onda sea mayor que el radio de Hubble en el espacio de de Sitter. 
Esto es importante para los modelos que intentan explicar la formaci\'on de 
estructuras en cosmolog\'\i a a partir de las fluctuaciones de los campos de 
materia cu\'anticos. Todo el Cap\'\i tulo 3 est\'a basado en el trabajo 
\cite{prd53}. 

Llegado a este punto comenzamos el estudio de campos cu\'anticos en espacios 
curvos, a los efectos de analizar el l\'\i mite semicl\'asico en gravedad 
cu\'antica. En el Cap\'\i tulo 4 estudiamos el l\'\i mite semicl\'asico en 
modelos 
de gravedad escalar-tensorial en dos dimensiones. Analizamos diferentes 
acoplamientos entre el 
campo dilat\'onico y el campo de materia. Estudiamos el proceso de 
radiaci\'on de Hawking en los distintos modelos. A partir de un c\'alculo 
exacto de la funcional de 
influencia discutimos las condiciones bajo las 
cuales es posible que la p\'erdida de coherencia sea efectiva en m\'etricas 
cosmol\'ogicas para asegurar la validez de la aproximaci\'on semicl\'asica. 
 Este cap\'\i tulo est\'a basado en \cite{tmunu} y \cite{if}.

Luego, en el Cap\'\i tulo 5 estudiamos modelos m\'as realistas en  
3+1 dimensiones. Con el objeto de entender la influencia de los 
campos cu\'anticos  de materia sobre la geometr\'\i a cl\'asica del 
espacio-tiempo, evaluamos las llamadas ecuaciones de 
Einstein-Langevin para un campo escalar masivo con acoplamiento arbitrario a 
la geometr\'\i a \cite{einstlang}. 

Finalmente en el Cap\'{\i}tulo 6 resumimos nuestras conclusiones.

%% file: tesis4
\newpage

\thispagestyle{empty}
~
\newpage

\chapter{El movimiento Browniano cu\'antico}

\thispagestyle{empty}

La relajaci\'on t\'ermica de sistemas en interacci\'on 
con entornos ha sido un tema de gran inter\'es en mec\'anica estad\'\i stica 
por mucho tiempo. En particular, es importante notar que 
la propia descripci\'on de la relajaci\'on de sistemas cu\'anticos acoplados a 
un entorno, tambi\'en debe ser ana\-lizada por medio de la mec\'anica 
cu\'antica 
\cite{lender22}. Las situaci\'on m\'as simple que podemos considerar 
en este contexto es aquella del movimiento Browniano de un oscilador 
arm\'onico cu\'antico en un entorno de la misma naturaleza. Los modelos de 
movimiento Browniano cu\'antico proveen un ejemplo t\'\i pico de 
sistemas cu\'anticos 
abiertos, y han sido muy utilizados para el entendimiento de la teor\'\i a 
de medici\'on en mec\'anica cu\'antica \cite{zurek81}, \'optica cu\'antica 
\cite{carmi} y p\'erdida de coherencia \cite{unruhzu}, por citar s\'olo 
algunos de los intereses que presentan estos modelos. El objeto central 
de estudio es la ecuaci\'on maestra para la matriz densidad reducida de la
part\'\i cula Browniana, que se obtiene luego de integrar los grados de 
libertad correspondientes al entorno.
 Una gran cantidad de trabajos en esta direcci\'on han sido hechos en el 
pasado \cite{calde,grabert, hakim, haake, linden}. La derivaci\'on m\'as 
general de esta ecuaci\'on maestra es la efectuada por B.L. Hu, J.P. Paz y 
Y. Zhang \cite{hpz1}, donde utilizaron la funcional de influencia de Feynman 
y Vernon \cite{feyver} provista por t\'ecnicas de integrales de camino, lo 
cual es de gran utilidad para extender el formalismo a teor\'\i as de campos, 
que es uno de nuestros objetivos fundamentales.


\section{El modelo: sistema y entorno}

El modelo est\'a compuesto por una part\'\i cula Browniana de masa $M$ y 
frecuencia natural $\Omega$, acoplada a un entorno representado por un 
conjunto de osciladores arm\'onicos de masa $m_n$ y 
frecuencia natural $\omega_n$. La part\'\i cula est\'a acoplada a cada uno
de los osciladores del entorno con una constante de acoplamiento $\lambda_n$. 
La acci\'on total del sistema-entorno est\'a dada por
\bea S[x,q] &=& S[x] + S_{\rm E}[q] + S_{\rm int}[x,q] \nonumber \\  
&=& \int_0^t ds \left[{1\over{2}} M ({\dot x}^2 - \Omega^2 x^2) + \sum_n 
[{1\over{2}}m_n ({{\dot q}_n}^2 - \omega_n^2 q_n^2)] - \sum_n \lambda_n x q_n 
\right], \label{accionqbm}\eea
donde $x$ y $q_n$ son las coordenadas de la part\'\i cula y de los 
osciladores respectivamente. Debemos se\~nalar que por simplicidad en este 
cap\'\i tulo, s\'olo nos referiremos a este tipo de acoplamiento lineal entre 
sistema y entorno. Nuestro objetivo fundamental es introducir el 
formalismo que luego emplearemos en teor\'\i a de campos, donde los 
acoplamientos ser\'an m\'as complicados. La ventaja de este modelo es que
la ecuaci\'on maestra puede obtenerse en forma exacta para todo tipo de 
entorno a considerarse \cite{hpz1,jpp}. Acoplamientos m\'as generales 
($x^k q_n^m$) han sido 
considerados en la literatura, donde la ecuaci\'on maestra ha sido obtenida 
perturbativamente \cite{hpz2}.  

En este modelo de movimiento Browniano, la evoluci\'on del sistema combinado 
(sistema-entorno), 
puede caracterizarse por cuatro escalas de tiempo diferentes: la primera 
est\'a asociada a la frecuencia natural de la part\'\i cula aislada $\Omega$; 
 la segunda est\'a representada por el tiempo de relajaci\'on (caracterizado 
por el acoplamiento entre la part\'\i cula y el entorno); la tercera 
corresponde al ``tiempo de memoria'' del entorno (en general asociado 
a la frecuencia m\'as alta presente en el entorno) y finalmente la escala 
de tiempo asociada con la temperatura del entorno, que mide la importancia 
relativa entre los efectos cu\'anticos y t\'ermicos. 

El efecto del entorno sobre la din\'amica del sistema est\'a caracterizado por
los fen\'omenos de fluctuaci\'on y disipaci\'on. Estos efectos pueden 
determinarse por una propiedad totalmente espec\'\i fica del entorno: la 
{\it densidad espectral} $I(\omega )$. Esta densidad da el n\'umeros de 
osciladores 
con una dada frecuencia y para una magnitud espec\'\i fica $\lambda_n$ de 
la constante de acoplamiento, presentes en el entorno, 

\be I(\omega )= \sum_n \delta (\omega - \omega_n) 
{\lambda_n^2\over{2 m_n \omega_n}}.\ee 
Por lo tanto, dando la densidad $I(\omega )$ y el estado inicial del entorno, 
tanto la disipaci\'on como las fluctuaciones quedan un\'\i vocamente 
determinadas, como podremos ver en la siguiente secci\'on. 

Diferentes $I(\omega )$ clasifican a los diferentes tipos de entornos; 
la constante de acoplamiento se ajusta a la frecuencia 
del entorno, como por ejemplo $\lambda_n = m_n \omega_n^\alpha$ 
para cada modelo de entorno \cite{grabert}. Por 
razones f\'\i sicas, uno no espera que un entorno real contenga un n\'umero 
infinito de frecuencias, y en general se introduce una escala 
arbitraria, que llamaremos {\it frecuencia de corte} $\Lambda$, que volver\'a 
cero a la densidad espectral para aquellas frecuencias mayores que 
esta frecuencia de corte; es decir $I(\omega ) \rightarrow 0$ cuando 
$\omega > \Lambda$. Por lo tanto, la escala de tiempo asociada a la 
memoria del entorno, mencionada en un p\'arrafo anterior, queda 
determinada por la inversa de esta frecuecia de corte.  El entorno se conoce 
usualmente como \'ohmico \cite{calde} si la densidad espectral es tal que 
$I(\omega ) \approx \omega$ ($\omega < \Lambda$); supra\'ohmico si $I(\omega ) 
\approx \omega^\alpha$ para $\alpha > 1$ o sub\'ohmico si $\alpha < 1$. Es 
simple ver que 
el caso \'ohmico (que es el m\'as estudiado en general) corresponde a la 
situaci\'on  
f\'\i sica en la que el entorno induce una fuerza lineal con la velocidad 
sobre el sistema. En este cap\'\i tulo se utilizar\'a la siguiente expresi\'on 
expl\'\i cita para la densidad espectral \cite{hpz1}      

\be I(\omega ) = {2\over{\pi}} M \gamma_0 \omega \left(
{\omega\over{\Lambda}}\right)^{\alpha - 1} e^{-{\omega^2\over{\Lambda^2}}},
\label{especden}\ee
donde con $\gamma_0$ representamos la constante de relajaci\'on del 
entorno que corresponde a la frecuencia asociada al 
acoplamiento entre sistema y entorno \cite{leggett}. 

 
\section{Funcional de influencia de Feynman y Vernon}

El problema central de estudio en esta secci\'on es desarrollar 
un formalismo general que nos permita encontrar todos los efectos cu\'anticos 
de un entorno sobre nuestro sistema de inter\'es. En trabajos previos, la 
ecuaci\'on de evoluci\'on para la matriz densidad reducida fue calculada
bajo ciertas aproximaciones, como por ejemplo para entornos a temperatura 
arbitraria pero s\'olo en el caso \'ohmico \cite{unruhzu}; o en el l\'\i mite 
de alta temperatura \cite{calde}. En esta secci\'on mostramos 
una representaci\'on funcional del operador de evoluci\'on de la matriz 
densidad reducida, que nos permitir\'a obtener la ecuaci\'on maestra 
exacta para un entorno general. 

El operador matriz densidad $\hat{\rho}$ evoluciona unitariamente 
bajo la acci\'on del operador de evoluci\'on $J(t, t_0)$, por lo tanto

\be \hat{\rho} (t) = J(t, t_0)\hat{\rho}(t_0),\ee
es la ecuaci\'on de evoluci\'on para la matriz densidad total $\hat{\rho}$. 
Utilizando la representaci\'on de la integral de caminos \cite{feyhibbs}, 
este propagador puede escribirse como
\bea &&J(x,q,x',q';t\vert x_i,q_i,x_i',q_i';t_0) = U(x,q;t\vert x_i,q_i;t_0) 
U^*(x',q';t\vert x_i',q_i';t_0)\nonumber \\
&& =\int_{x_i}^{x}{\cal D}x\int_{q_i}^{q}{\cal D}q \exp \left[{i\over
{\hbar}} S[x,q]\right] \int_{x_i'}^{x_f'}{\cal D}x'\int_{q_i'}^{q'}
{\cal D}q' \exp \left[-{i\over
{\hbar}} S[x',q']\right],\eea
donde $U$ es el operador de evoluci\'on de la funci\'on de onda. Las 
integrales funcionales del segundo t\'ermino de la ecuaci\'on est\'an 
realizadas sobre todas las posibles historias compatibles con las condiciones 
de contorno. Con $q$ estamos representando las coordenadas del conjunto 
completo de osciladores presentes en el entorno ($q_n$); y con $i$ notamos
 las variables en el instante inicial.

La matriz densidad reducida est\'a definida como

\be \rho_r(x,x') = \int_{-\infty}^{+\infty}dq \int_{-\infty}^{+\infty}dq'
\rho(x,q\vert x',q') \delta (q - q'),\ee
por lo tanto, su evoluci\'on temporal est\'a regida por un operador 
evoluci\'on reducido $J_r$, definido como: 

\be \rho_r(x,x';t) = \int_{-\infty}^{+\infty}dx_i\int_{-\infty}^{+\infty}dx_i'
~ J_r(x,x';t\vert x_i, x_i';t_0) ~ \rho_r(x_i,x_i';t_0).\ee

Considerando como condici\'on inicial que el sistema y el entormo 
no est\'an correlacionados \cite{hpz1}, $\hat{\rho}(t_0) = 
\hat{\rho}_{\rm sist}(t_0) 
\otimes \hat{\rho}_{\rm ent}(t_0)$, el operador de evoluci\'on reducido 
adopta la forma

\bea J_r(x_f,x_f';t\vert x_i, x_i';t_0) &=&  \int_{x_i}^{x_f}{\cal D}x 
 \int_{x_i'}^{x_f'}{\cal D}x' \exp\left[{i\over{\hbar}}(S[x] - S[x'])\right]
~ F[x,x']\nonumber \\
&=& \int_{x_i}^{x_f}{\cal D}x 
 \int_{x_i'}^{x_f'}{\cal D}x' \exp\left[{i\over{\hbar}}A[x,x']\right],\label
{evol0}\eea
donde con el sub-\'\i ndice $f$ denotamos a las variables en el momento final 
de la evoluci\'on. $A[x,x']$ es la llamada ``acci\'on efectiva'' para el 
sistema cu\'antico abierto y $F[x,x']$ es la {\it funcional de influencia} 
de Feynman y Vernon, definida expl\'\i citamente como \cite{feyver}
\bea F[x,x'] &=& \int_{-\infty}^{+\infty}dq_f \int_{-\infty}^{+\infty}dq_i  
\int_{-\infty}^{+\infty}dq_i' \int_{q_i}^{q_f}{\cal D}q 
\int_{q_i'}^{q_f'}{\cal D}q' \exp\left[{i\over{\hbar}} (S_{\rm ent}[q] + 
S_{\rm int}[x,q])\right]\nonumber \\
&& \times \exp\left[-{i\over{\hbar}} (S_{\rm ent}[q'] + 
S_{\rm int}[x',q'])\right] ~ \rho_{\rm ent}(q_i,q_i';t_0) \nonumber \\
&=& \exp\left[{i\over{\hbar}} \delta A[x,x']\right],\label{fvif}\eea
donde llamamos $\delta A[x,x']$ a la {\it acci\'on de influencia}. En 
consecuencia la acci\'on efectiva para el sistema cu\'antico abierto es 
$A[x,x'] = S[x] - S[x'] + \delta A[x,x']$. 

Debido a la estructura de esta funcional, podemos interpretar a las 
historias $x$ y $x'$ como movi\'endose hacia el futuro y pasado, 
respectivamente (gracias a la diferencia de signo que aparece en cada 
exponencial). Esta observaci\'on permite se\~nalar la similitud
entre el formalismo de la funcional de influencia y el de la 
funcional generatriz de camino temporal 
cerrado (CTC) \cite{ctp1} (similitud que utilizaremos en los cap\'\i tulo 4 
 y 5). Las
 reglas de Feynman que se deducen a partir del formalismo CTC son 
muy \'utiles para calcular la funcional de influencia en teor\'\i a de campos.

Gracias a la condici\'on inicial en la que no existen correlaciones, la 
funcional de 
influencia s\'olo depende del estado inicial del entorno. En este ejemplo
 mostramos la funcional $F[x,x']$ para un entorno que inicialmente 
est\'a en equilibrio termodin\'amico a temperatura $\beta^{-1}$ (condiciones
m\'as generales acerca de la condici\'on de equilibrio entre el 
sistema y el entorno pueden encontrarse en la Ref. \cite{hakim}; acerca de 
condiciones iniciales con correlaciones consultar las Refs. 
\cite{grabert,prom}). Para las presentes condiciones iniciales, la funcional 
de influencia puede calcularse exactamente \cite{feyver,calde,hpz1}. El 
resultado es:
\bea F[x,x'] &=& \exp\left[- {i\over{\hbar}}\int_0^{t}ds_1\int_0^{s_1}ds_2
[x(s_1)- x'(s_1)] \eta(s_1 - s_2) [x(s_2) + x'(s_2)]\right.\nonumber \\
&& \left. -{1\over{\hbar}} \int_0^{t}ds_1\int_0^{s_1}ds_2
[x(s_1)- x'(s_1)] \nu(s_1 - s_2) [x(s_2) - x'(s_2)]\right].\eea

Los n\'ucleos $\eta$ y $\nu$, son en general no-locales en el tiempo 
y est\'an definidos como

\be \nu (s) = \int_0^{+\infty}d\omega I(\omega ) coth{{\beta\hbar 
\omega\over{2}}} \cos{\omega s},\ee

\be \eta (s) = {d\over{ds}}\gamma (s),\ee
donde 

\be \gamma (s) = \int_0^{+\infty}d\omega {I(\omega )\over{\omega}} 
\cos{\omega s}.\ee 

Dada la expresi\'on de la funcional de influencia en t\'erminos de los 
n\'ucleos $\eta$ y $\nu$, podemos escribir la acci\'on de influencia como
\bea \delta A[x,x'] &=& - 2 \int_0^{t}ds_1\int_0^{s_1}ds_2 \Delta (s_1) \eta 
(s_1 - s_2) \Sigma (s_1)\nonumber \\
&+& i \int_0^{t}ds_1\int_0^{s_1}ds_2 \Delta (s_2) 
\nu (s_1 - s_2) \Delta (s_2),\eea
donde hemos introducido las variables: $\Delta = x - x'$ y 
$\Sigma = 1/2 (x + x')$.
 
Las partes real e imaginaria de la acci\'on efectiva $A[x,x']$ pueden 
interpretarse como respon\-sables de la disipaci\'on y el ruido, 
respectivamente. Los n\'ucleos de ruido y disipaci\'on est\'an siempre 
relacionados por una ecuaci\'on integral conocida como la relaci\'on de 
fluctuaci\'on-disipaci\'on \cite{linden}. Para el caso del movimiento 
Browniana cu\'antico, esta relaci\'on puede escribirse como

\be \nu (s) = \int_{-\infty}^{+\infty} ds' K(s - s') \gamma (s'),
\label{rfdqbm}\ee
donde el nucleo $K(s)$ est\'a definido por
\be K(s) = \int_0^{+\infty} d\omega {\omega\over{\pi}} 
coth{{\beta\hbar\omega\over{2}}} \cos{\omega s}.\ee

En el l\'\i mite cl\'asico de muy altas temperaturas, el nucleo $K$ es 
proporcional a una funci\'on delta de Dirac, $K(s)= 2 k_{\rm B}T\delta (s)$ y 
la relaci\'on de fluctuaci\'on-disipaci\'on no es m\'as que la relaci\'on de 
Einstein usual.

La ecuaci\'on de movimiento para la part\'\i cula Browniana puede 
deducirse a partir de la acci\'on efectiva $A[x,x']$ mediante 
$\left. {\delta A[x,x']\over{\delta x}}\right\vert_{x=x'} = 0$, 

\be \ddot{x}(t) + \Omega^2 x(t) + 2 \int_0^tds \eta(s - t) x(s) = 0,\ee
que en este caso corresponde a haber efectuado un promedio sobre todas 
las posibles realizaciones del ruido. Para poner de 
manifiesto este hecho podemos escribir la parte imaginaria de la acci\'on de 
influencia 
en t\'erminos de una fuerza estoc\'astica $\xi$, acoplada al 
sistema y definida por una distribuci\'on de probabilidad 
gaussiana

\be P[\xi (t)] = N_\xi \exp\left\{-{1\over{2}} \int_0^t ds_1\int_0^{s_1}ds_2 
\xi (s_1) \nu (s_1 - s_2)^{-1} \xi (s_2)\right\},\ee
donde $N_\xi$ es una constante de normalizaci\'on. Por lo tanto, la parte 
imaginaria de la acci\'on efectiva puede escribirse en funci\'on de la
 fuente de ruido estoc\'astica como

\be \int {\cal D}\xi (t) P[\xi ]\exp\left[-{i\over{\hbar}} \Delta (t) \xi (t)
\right] = \exp\left[{i\over{\hbar}}\int_0^tds_1\int_0^{s_1}ds_2 \Delta (s_1)
\nu (s_1 - s_2) \Delta (s_2)\right].\ee

Finalmente, habiendo escrito la parte imaginaria en funci\'on de $\xi$, 
podemos escribir una ecuaci\'on de movimiento explicitando la presencia 
de un t\'ermino de ruido. Esta es la ecuaci\'on asociada  
de Langevin, donde la din\'amica de la part\'\i cula cl\'asica est\'a afectada
 por la presencia de una fuente de ruido \cite{feyver,unruhzu,calde,grabert}:  

\be  \ddot{x}(t) + \Omega^2 x(t) + 2 \int_0^tds \eta(s - t) x(s) = \xi (t).
\label{lqbm}\ee

El ruido estoc\'astico $\xi$ est\'a caracterizado por su distribuci\'on de 
probabilidad $P[\xi]$ y por las funciones de correlaci\'on

\be \langle \xi (s) \rangle = 0  ~~~~~~~\mbox{y}~~~~~~
\langle \xi (s) \xi(s')\rangle = \nu (s - s').\ee


\section{La ecuaci\'on maestra}

En general, por consideraciones f\'\i sicas, el conocimiento de la 
ecuaci\'on maestra para la evoluci\'on de la matriz densidad reducida 
es m\'as \'util que la evoluci\'on exacta de la matriz densidad total misma. 
En realidad, a partir de la ecuaci\'on maestra podemos extraer muchos 
aspectos cualitativos acerca del comportamiento del sistema, los cuales 
son independientes de las condiciones iniciales.
 
El modelo presentado en este cap\'\i tulo tiene un acoplamiento lineal 
y puede resolverse exactamente. Las integrales de camino de la Ec. 
(\ref{evol0}) pueden evaluarse debido a que son integrales Gaussianas 
\cite{hpz1, jpp}. Por este motivo la deducci\'on de la ecuaci\'on maestra 
est\'a basada en la re\-presentaci\'on de integral funcional para el operador
de evoluci\'on de la matriz densidad reducida (Ec. (\ref{evol0})). La 
no-localidad de los n\'ucleos presentes en la funcional de influencia es 
la \'unica complicaci\'on; por lo tanto, obtener formalmente la 
ecuaci\'on maestra es conceptualmente equivalente a derivar la ecuaci\'on 
de Schr\"odinger a partir de la representaci\'on funcional del propagador 
en mec\'anica cu\'antica. 

Para hallar la ecuaci\'on maestra debemos evaluar la derivada temporal del 
operador de evoluci\'on reducido. El hecho antes mencionado que la funcional 
de influencia sea no-local, implica que no podamos calcular esta derivada 
simplemente expandiendo en $dt$ al propagador $J_{\rm r}(t + dt, t)$ y 
rest\'andole 
$J_{\rm r}(t,t)$; dado que el propagador $J_{\rm r}(t + dt, t)$ depende del 
estado 
del sistema a tiempo $t$. Esto s\'olo puede hacerse en el caso de alta 
temperatura porque en este l\'\i mite la funcional de influencia se torna 
local en el tiempo \cite{calde}. 

Los resultados basados en el procedimiento funcional seguido por B.L. Hu, 
J.P. Paz y Y. Zhang 
en la Ref. \cite{hpz1} pueden re-obtenerse de una manera extremadamente 
sencilla mediante un m\'etodo perturbativo (pero que en el caso de 
acoplamiento lineal provee la ecuaci\'on maestra exacta) \cite{jpp2}. En 
esta secci\'on mostramos dicho procedimiento debido 
a que estamos interesados, fundamentalmente, en discutir las 
caracter\'\i sticas m\'as importantes de la ecuaci\'on maestra, y no 
de su deducci\'on. El m\'etodo funcional ser\'a empleado en el cap\'\i tulo 
siguiente donde evaluaremos la ecuaci\'on maestra en teor\'\i a de campos.

En la representaci\'on de interacci\'on, la matriz densidad total 
(sistema-entorno) evoluciona de acuerdo a la ecuaci\'on
\vfill\eject
\be i\hbar\dot\rho =[V(t),\rho ],\label{eq1jpp}\ee
donde el potencial de interacci\'on $V(t)$, en la representaci\'on de 
interacci\'on es, simplemente

\be V(t)=\exp\left[{i\over{\hbar}}(H_{\rm sist}+ H_{\rm ent})t\right]\ V 
\ \exp\left[-{i\over{\hbar}}(H_{\rm sist}+H_{\rm ent})t\right],\ee
y la matriz densidad $\rho$ es

\be\rho (t)=\exp\left[{i\over{\hbar}}(H_{\rm sist}+H_{\rm ent})t\right]
\ \rho \ \exp\left[-{i\over{\hbar}}(H_{\rm sist}+H_{\rm ent})t\right].\ee

La soluci\'on de la ecuaci\'on (\ref{eq1jpp}) puede obtenerse 
perturbativamente mediante la serie de Dyson:

\be \rho (t)=\sum_{n\ge 0}\int_0^tdt_1\int_0^{t_1}dt_2\ldots\int_0^{t_n}
({1\over{i\hbar}})^n[V(t_1),[V(t_2),[\ldots,[V(t_n),\rho (0)]\ldots]].
\label{eq2jpp}\ee

A partir de esta serie, en sencillo deducir la ecuaci\'on de evoluci\'on para 
la matriz densidad reducida. La matriz densidad reducida 
$\rho_{\rm r} = {\mbox Tr}_{\rm ent}\rho$, en la 
representaci\'on de interacci\'on, puede escribirse (conservando 
t\'erminos hasta segundo orden en la constante de interacci\'on) como:
\bea \rho_{\rm r}(t)&\approx &
\rho_{\rm r}(0)+{1\over{i\hbar}}\int_0^t dt_1 {\mbox Tr}_{\rm ent}\bigl(
[V(t_1),\rho (0)]
\bigr)\nonumber \\
&-&{1\over{\hbar^2}}\int_0^tdt_1 \int_0^{t_1}dt_2 {\mbox Tr}
_{\rm ent}\bigl(
[V(t_1),[V(t_2), 
\rho (0)]]
\bigr).\label{eq3jpp}\eea

Para hallar la ecuaci\'on maestra debemos tomar la derivada temporal 
de la Ec. (\ref{eq3jpp}), obteniendo

\be
\dot{\rho}_{\rm r}={1\over{i\hbar}} {\mbox Tr}_{\rm ent}[V(t),\rho (0)] -
{1\over{\hbar^2}} 
\int_0^t dt_1 {\mbox Tr}_{\rm ent}[V(t),[V(t_1),\rho (0)]].\label{eq4jpp}\ee
Ahora bien, como estamos trabajando con la condici\'on inicial de 
no-correlaci\'on  $\rho (0)=\rho_{\rm sist}(0)\otimes
\rho_{\rm ent}(0)$, la ecuaci\'on para matriz densidad reducida puede 
re-escribirse como 

\be \dot{\rho}_{\rm r}={1\over{i\hbar}} {\mbox Tr}_{\rm ent}
[V(t),\rho_{\rm sist}(0)\otimes
\rho_{\rm ent}(0)] 
-{1\over{\hbar^2}} \int_0^t
dt_1 {\mbox Tr}_{\rm ent}[V(t),[V(t_1),\rho_{\rm sist}(0)\otimes\rho_
{\rm ent}(0)]].
\label{eq5jpp}\ee

El estado inicial $\rho_{\rm sist}(t_0= 0)$ que aparece en el lado 
derecho de la 
Ec. (\ref{eq5jpp}) puede expresarse nuevamente en t\'erminos de 
$\rho_{\rm r}(t)$ 
gracias a la expansi\'on (\ref{eq3jpp}). De esta manera el lado derecho de
la ecuaci\'on (\ref{eq5jpp}) puede escribirse completamente en t\'erminos de 
la matriz densidad reducida a tiempo $t$. De esta manera podemos 
escribir la ecuaci\'on maestra (v\'alida a segundo orden en un desarrollo
 perturbativo y para la condici\'on inicial donde no existe correlaci\'on) 
como:
\bea   \dot{\rho}_{\rm r} &=&{1\over{i\hbar}} {\mbox Tr}_{\rm ent}[V(t),
\rho_{\rm r}\otimes
\rho_{\rm ent}(0)] 
-{1\over{\hbar^2}} \int_0^t
dt_1 {\mbox Tr}_{\rm ent}[V(t),[V(t_1),\rho_{\rm r}\otimes\rho_{\rm ent}(0)]]
\nonumber \\
&+&{1\over{\hbar^2}} \int_0^t
dt_1 {\mbox Tr}_{\rm ent}\left\{[V(t),{\mbox Tr}_{\rm ent}\bigl([V(t_1),
\rho_{\rm r}\otimes\rho_{\rm ent}(0)]\bigr)
\otimes\rho_{\rm ent}]\right\}.\label{eq6jpp}\eea

En el caso de nuestro particular inter\'es, $V = \sum_n \lambda_n q_n x$.
 Si el entorno est\'a inicialmente en equilibrio a temperatura 
$T=1/k_{\rm B}\beta$, la ecuaci\'on maestra se reduce a

\be \dot{\rho}_{\rm r}=-{1\over{\hbar^2}} \int_0^t
dt_1 {\mbox Tr}_{\rm ent}[V(t),[V(t_1),\rho_{\rm r}\otimes\rho_{\rm ent}(0)]],
\label{eq7jpp}\ee
y el t\'ermino dentro de la integral temporal puede escribirse como
\bea 
{\mbox Tr}_{\rm ent}[V(t),[V(t_1),\rho_{\rm r}\otimes\rho_{\rm ent}(0)]]
&=&{1\over 2} 
\sum_{n}\lambda_n^2
\Bigl(\langle \{q_n(t),q_n(t_1)\}\rangle [x(t),[x(t_1),\rho_{\rm r}]]
\nonumber \\
&+&\langle [q_n(t),q_n(t_1)]\rangle [x(t),\{x(t_1),\rho_{\rm r}\}]\Bigr),\eea
donde $\langle ..... \rangle$ denota el promedio sobre 
el estado inicial del entorno. Retornando a la representaci\'on de 
Schr\"odinger, la ecuaci\'on maestra puede escribirse

\be
\dot{\rho}_{\rm r} ={1\over{i\hbar}} [H_{\rm sist},\rho_{\rm r}] 
-\int_0^t dt_1 \Bigl(\nu(t-t_1) [x,[x(t_1-t),\rho]]
-i \eta(t-t_1) [x,\{x(t_1-t),\rho\}]\Bigr).
\label{eq8jpp}\ee
en funci\'on de los n\'ucleos de ruido y disipaci\'on, que est\'an dados por:
\bea
\nu(t)&=&{1\over{2\hbar^2}}\sum_n\lambda_n^2\langle \{q_n(t),q_n(0)\}
\rangle =\sum_n
{\lambda_n^2\over{2m_n\hbar\omega_n}} \cos(\omega_n t) (1+2N_n)\nonumber \\
\eta(t)&=& {i\over{2\hbar^2}}\sum_n\lambda_n^2\langle [q_n(t),q_n(0)]\rangle 
=\sum_n{\lambda^2\over{2m_n\hbar\omega_n}}\sin(\omega_n t)
\eea
Usando que $1 + 2 N_n = coth{\beta\hbar \omega\over{2}}$, la expresi\'on 
para la densidad espectral y la relaci\'on 

$$\sum_{n=1}^N {\lambda_n^2\over{2 m_n \omega_n}}f(\omega_n) = \int_0^\infty 
d\omega I(\omega )f(\omega ),$$ obtenemos

$$
\nu(t)=\int_0^\infty d\omega I(\omega)\cos(\omega t) (1+2N(\omega)), 
$$
$$
\eta(t)=\int_0^\infty d\omega I(\omega)\sin(\omega t).
$$
Resolviendo las ecuaciones de Heisenberg para el sistema y determinando 
el operador $x(t)$

$$
x(t)=x \cos(\Omega t) + {1\over{M\Omega}} p\sin(\Omega t),
$$
podemos escribir la expresi\'on final de la ecuaci\'on maestra:

\be \dot{\rho}_{\rm r} =-{i\over\hbar}\bigl[H_{\rm sist}+
{1\over2}M\tilde\Omega^2(t)
x^2,\rho_{\rm r}\bigr]
+2i\gamma(t)\bigl[x,\bigl\{ p,\rho_{\rm r}\bigr\}\bigr]
-D(t)\bigl[x,\bigl[ x,\rho_{\rm r}\bigr]\bigr] 
 -f(t)\bigl[x,\bigl[ p,\rho_{\rm r}\bigr]\bigr],
\label{master}\ee
donde los coeficientes temporales est\'an dados por:
\bea
\tilde\Omega^2(t) &=& -{2\hbar\over M}\int_0^t dt'\cos(\Omega t') 
\eta(t')\nonumber \\
\gamma(t) &=& -{1\over 2M\Omega}\int_0^t dt'\sin(\Omega t') \eta(t')\nonumber 
\\
D(t) &=& \int_0^t dt'\cos(\Omega t') \nu(t')\label{coef} \\
f(t) &=& -{1\over M\Omega}\int_0^t dt'\sin(\Omega t') \eta(t'),\nonumber\eea
aqu\'\i \ ${\tilde \Omega}$ es la frecuencia natural renormalizada por el 
entorno; $\gamma (t)$ es la tasa de relajaci\'on; $D(t)$ y $f(t)$ son los 
coeficientes de difusi\'on (responsables de los efectos de p\'erdida de 
coherencia).

Esta ecuaci\'on, derivada perturbativamente, es la misma que se deduce en 
el c\'alculo exacto para el movimiento Browniano con acoplamiento 
lineal \cite{hpz1, jpp, jpp2}. 

\section{P\'erdida de coherencia}

En esta secci\'on haremos una breve discusi\'on acerca de los principales 
resultados obtenidos en relaci\'on al proceso de p\'erdida de 
coherencia en el movimiento Browniano cu\'antico. 
La relevancia del proceso de p\'erdida de coherencia en el contexto de la
 transici\'on cu\'antico-cl\'asica ha sido reconocida en los \'ultimos 
a\~nos \cite{zurek81,zurek82,zurekpt,joos}. La idea b\'asica de este 
proceso es que los aspectos cl\'asicos son una propiedad 
emergente en los sistemas cu\'anticos abiertos inducida por los entornos. 
Debido a la interacci\'on con el entorno, la gran mayor\'\i a de los estados 
del espacio de Hilbert de un sistema cu\'antico se vuelven altamente 
inestables frente a la interacci\'on. Despu\'es del tiempo de p\'erdida de 
coherencia, que para objetos macrosc\'opicos es m\'as corto que cualquier 
escala din\'amica del sistema, un estado cu\'antico gen\'erico decae en una 
mezcla de ``estados indicadores'' \cite{zurek82}. De esta forma el entorno 
induce una regla de selecci\'on efectiva sobre el sistema advirtiendo de la 
existencia de una superposici\'on de estados indicadores que es estable. Estos 
estados indicadores se distinguen por su ``habilidad'' de permanecer 
despu\'es del monitoreo efectuado por el entorno, y por eso son los 
estados en los cuales el sistema cu\'antico abierto ser\'a observado (o 
medido).

Para estudiar el proceso de p\'erdida de coherencia a partir de la ecuaci\'on 
maestra (\ref{master}), consideremos la siguiente superposici\'on 
inicial de estados coherentes:

\be \psi(x,t=0) = N \exp\left(-{(x -L_0)^2\over{2\delta^2}}+iP_0x\right) + 
N  \exp\left(-{(x + L_0)^2\over{2\delta^2}}-iP_0x\right),\ee
donde $N$ es una constante. En este caso es posible resolver la ecuaci\'on 
maestra y mostrar que la funci\'on de Wigner constru\'\i da a partir de la 
matriz densidad como \cite{pazhabzu}

\be W(x,p) = {1\over{2 \pi \hbar}}\int_{-\infty}^{+\infty}dy~ 
e^{{ipy\over{\hbar}}}~ \psi^{\star}(x + {y\over{2}})\psi (x - {y\over{2}}),\ee
est\'a dada, en general por:

\be W(x,p;t) = W_1(x,p,;t) + W_2(x,p;t) + W_{\rm int}(x,p;t),\ee
donde 
\bea 
W_1 &=& {\tilde N}^2 {\delta_2\over{\delta_1}} \exp\left(-
{(x - x_c)^2\over{\delta_1^2}} - \delta_2^2[p-p_c-\beta(x-x_c)]^2\right)
\nonumber \\
W_2 &=&   {\tilde N}^2 {\delta_2\over{\delta_1}} \exp\left(-
{(x + x_c)^2\over{\delta_1^2}} - \delta_2^2[p + p_c-\beta(x + x_c)]^2\right)
\nonumber \\
W_{\rm int} &=& 2 {\tilde N}^2{\delta_2\over{\delta_1}} \exp\left(-{x^2\over{
\delta_1^2}}-\delta_2^2 (p - \beta x)^2\right) \cos{[\phi_p+(\phi_x
-\beta\phi_p)x]} \exp{(- A_{\rm int})}.\eea 
Las funciones $x_c(t)$, $p_c(t)$, $\delta_1(t)$, $\delta_2(t)$, $\beta (t)$, 
$\phi_x(t)$, $\phi_p(t)$ y $A_{\rm int}(t)$ dependen del entorno (y de las 
constantes $L_0$, $P_0$ y $\delta$) de una manera complicada \cite{pazhabzu}. 
Nos concentraremos en $A_{\rm int}$ dado que es relevante para el 
estudio de la p\'erdida de coherencia. 

Para un an\'alisis cuantitativo del proceso de p\'erdida de coherencia es 
necesario definir una funci\'on que permita evaluar la importancia de 
la interferencia a un dado tiempo. Esta herramienta est\'a dada por el 
cociente entre el t\'ermino de interferencia 
y los t\'erminos directos de la funci\'on de Wigner, evaluados en sus 
respectivos picos, la cual es una 
cantidad determinada por $A_{\rm int}$, 

\be \exp{(-A_{\rm int})}={1\over{2}}{W_{\rm int}\vert_{\rm pico}\over{
\left[W_1(x,p)\vert_{\rm pico} W_2(x,p)\vert_{\rm  pico}\right]^
{1\over{2}}}}.\label{aint}\ee
 
Asumiendo como condiciones iniciales que 
\bea
\delta_1^2 &=& \delta_2^2 = \delta^2,\nonumber \\
\phi_x &=& P_0 = p_c ~~~,~~~ \phi_p = L_0 = x_c \\
A_{\rm int} &=& 0,\nonumber \eea
la funci\'on $A_{\rm int}$ est\'a dada por la siguiente expresi\'on

\be A_{\rm int} = {L_0^2\over{\delta^2}}+ \delta^2 P_0^2 - {\phi_p^2\over{
\delta_2^2}} - \delta_1^2 \phi_x^2,\ee
por lo tanto, se anula inicialmente y est\'a siempre acotada:

\be A_{\rm int} \leq {L_0^2\over{\delta^2}} + \delta^2 P_0^2 = \left. 
A_{\rm int}\right\vert_{\rm max}.\ee
El proceso de p\'erdida de coherencia destruye la interferencia entre los
 estados indicadores, la cual est\'a representada  por los t\'erminos 
fuera de la 
diagonal de la matriz densidad reducida. Como $A_{\rm int}$ crece con 
el tiempo, los efectos de interferencia se tornan menos importantes, as\'\i \
que el estado del sistema se acerca a una mezcla entre los estados 
indicadores.    

Para analizar c\'omo la evoluci\'on de $A_{\rm int}$ se ve afectada por el 
entorno es conveniente utilizar la ecuaci\'on maestra (\ref{master}). 
En la Ref. \cite{hpz1} fue demostrado que para un entorno acoplado 
linealmente a una temperatura arbitraria y para cualquier densidad 
espectral, $A_{\rm int}$ satisface la siguiente identidad:

\be \dot{A}_{\rm int} = D(t) \phi_p^2 - 2 f(t) \phi_p (\phi_x - \beta\phi_p).
\label{aintdot}\ee
El primer t\'ermino contiene el efecto de la difusi\'on e implica que 
$A_{\rm int}$ siempre aumente. Por el contrario, el signo del segundo 
t\'ermino 
puede variar con el tiempo debido a la relaci\'on entre $\phi_x$ y $\phi_p$. 
La ecuaci\'on (\ref{aintdot}) puede resolverse aproximadamente despreciando 
el coeficiente de difusi\'on anomala $f(t)$, y considerando a $D(t)$ como 
una constante (esta aproximaci\'on se corresponde al caso de entorno \'ohmico 
y de alta temperatura). Si en el estado inicial ponemos $P_0= 0$, 

\be A_{\rm int} \approx {4 L_0^2 D t\over{(1 + 4 D \delta^2 t)}},\ee
entonces, la tasa de p\'erdida de coherencia 
$\Gamma_{\rm dec} = 1/t_{\rm dec}$ con $t_{\rm dec}$ tal que 
$A_{\rm int}(t_{\rm dec})=1$, est\'a dada por \cite{pazhabzu}

\be \Gamma_{\rm dec} = 4 L_0^2 D \approx 8 L_0^2 m \gamma_0 k_{\rm B}T,\ee
donde se us\'o que $D= 2 \gamma_0k_{\rm B} T$. 
Esto muestra que el coeficiente de difusi\'on $D(t)$ de la ecuaci\'on 
maestra es el que regula el proceso de p\'erdida de coherencia y provee 
tambi\'en de la escala de tiempo en la cual \'esta se torna efectiva para 
destruir las interferencias cu\'anticas. Por lo tanto los t\'erminos fuera 
de la diagonal decaen como $\Gamma_{\rm dec}$, y las coherencias cu\'anticas 
desaparecen exponencialmente en una escala que podemos re-escribir como 

\be t_{\rm dec} = \gamma_0^{-1} \left({\lambda_{\rm T}\over{L_0}}\right)^2,\ee
donde $\lambda_{\rm T} = \hbar/\sqrt{2 m k_{\rm B} T}$ es la longitud de 
onda t\'ermica de de Broglie. Entonces, para un objeto macrosc\'opico 
el tiempo de p\'erdida de coherencia $t_{\rm dec}$ es, t\'\i picamente, 
varios \'ordenes de magnitud menor que el tiempo de relajaci\'on 
$t_{\rm r} = \gamma_0^{-1}$. Por ejemplo, para un sistema en una 
habitaci\'on a una temperatura $T = 300 K$, con una masa de $m = 1 g$ y 
con una separaci\'on $L_0 = 1 cm$, el cociente 
$t_{\rm dec}/t_{\rm r}= 10^{-40}$. En consecuencia, aunque el tiempo de 
relajaci\'on sea del orden de la edad del Universo, 
$t_{\rm r} \sim 10^{17} {\mbox segs}$, la coherencia cu\'antica ser\'\i a 
destru\'\i da en $t_{\rm dec} \sim 10^{-23} {\mbox segs}$. Una diferencia 
tan grande puede obtenerse s\'olo para objetos macrosc\'opicos, y puede 
aceptarse s\'olo cuando las condiciones por las cuales se deriv\'o la 
ecuaci\'on maestra se satisfacen. No obstante, es simple entender,  
en este contexto, por qu\'e la p\'erdida de coherencia entre posiciones 
macrosc\'opicamente distinguibles es casi instant\'anea; a\'un para sistemas 
casi aislados.

%% file: tesis5
\newpage

\thispagestyle{empty}
~
\newpage

\chapter{Granulado grueso y p\'erdida de coherencia en teor\'\i a de campos}

\thispagestyle{empty}

En el cap\'\i tulo precedente iniciamos el estudio del proceso de 
p\'erdida de coherencia en el caso del movimiento Browniano cu\'antico 
como un ejemplo sencillo de sistema cu\'antico abierto en el que la 
presencia de muchos de los efectos interesantes de modelos m\'as complicados 
est\'an puestos de manifiesto. Como mencionamos en la Introducci\'on, este 
estudio es tambi\'en de mucho inter\'es en el 
contexto de la cosmolog\'\i a cu\'antica \cite{5prd53}. Se han realizado 
muchos esfuerzos para explicar el surgimiento de una m\'etrica 
cl\'asica del 
espacio-tiempo a partir de una teor\'\i a completamente cu\'antica, 
y para derivar, a partir de \'esta, las ecuaciones de Einstein (o la 
versi\'on cu\'anticamente corregida de ellas). Son los efectos  
de ruido, disipaci\'on, creaci\'on de part\'\i culas y 
su acci\'on sobre la geometr\'\i a (en modelos de cosmolog\'\i a cu\'antica), 
los aspectos relevantes de 
estudio para lograr un mejor entendimiento de la transici\'on 
cu\'antico-cl\'asica, tanto 
para modelos de movimiento Browniano cu\'antico como para teor\'\i a 
de campos \cite{7-9prd53}. 

De acuerdo al modelo inflacionario de evoluci\'on del Universo, las 
fluctuaciones cu\'anticas de los campos de materia son el origen de las 
inhomogeneidades que dan lugar (gracias a la expansi\'on del Universo) a la 
formaci\'on de estructuras \cite{10prd53}. Una suposici\'on 
vital en estos modelos es que las fluctuaciones cu\'anticas se vuelven 
cl\'asicas cuando sus longitudes de onda asociadas se vuelven m\'as grandes 
que el radio de Hubble $H^{-1}$. Por lo tanto, es necesario 
un an\'alisis detallado del proceso de transici\'on cu\'antico-cl\'asico 
para lograr un entendimiento completo del mecanismo de formaci\'on de 
estructuras. 

Hubieron numerosos intentos en la direcci\'on de probar 
la suposici\'on anteriormente mencionada \cite{111213prd53} utilizando 
modelos simplificados en los cuales la interacci\'on entre dos campos escalares
 es tal que desaparece despu\'es de una redefinici\'on de los campos; o 
usando un acoplamiento bi-cuadr\'atico entre dos campos escalares 
independientes.

En este cap\'\i tulo mostramos una manera consistente de iniciar el 
estudio de la transici\'on cu\'antico-cl\'asica de las inhomogeneidades 
primordiales a partir de las fluctuaciones cu\'anticas. Para comenzar 
consideramos una teor\'\i a con autointeracci\'on $\lambda \phi^4$ en 
el espacio-tiempo de Minkowski. Siguiendo la idea original de A. Starobinsky 
\cite{14prd53} acerca del modelo de inflaci\'on estoc\'astica, se\-paramos 
el campo escalar en dos partes: el {\it sistema} $\phi_<$ que contiene 
los modos del campo escalar cuyas longitudes de onda son m\'as ``largas'' que 
cierta longitud cr\'\i tica $\Lambda^{-1}$ y el {\it entorno} $\phi_>$ que 
contiene los modos con longitud de onda menor que la cr\'\i tica \cite{mataz}.
 Calculamos 
la funcional de influencia y la ecuaci\'on maestra asociada al sistema. 
Tambi\'en 
extendemos los c\'alculos a espacio-tiempos curvos, donde para un campo 
escalar acoplado conformemente a la m\'etrica, podemos demostrar que 
el valor m\'\i nimo de la longitud de onda cr\'\i tica debe ser del orden
 del radio de Hubble para que el sistema pierda coherencia. Para longitudes 
cr\'\i ticas m\'as chicas existen modos del sistema que no pierden 
coherencia cu\'antica \cite{prd53}. 


\section{La funcional de influencia para teor\'\i a de campos}

La t\'ecnica m\'as conveniente para la formulaci\'on del problema 
y para la derivaci\'on de un ``modelo efectivo'' de evoluci\'on de lo
que llamaremos sistema, es la acci\'on efectiva de granulado grueso (AEGG). 
Esta acci\'on efectiva se obtiene de manera totalmente an\'aloga al caso del 
movimiento Browniano cu\'antico, al integrar los grados de libertad 
asociados al entorno. Por lo tanto, esta acci\'on efectiva puede 
hallarse a partir de la funcional de influencia de Feynman y Vernon 
\cite{feyver}. Es interesante notar que la AEGG coincide con la acci\'on 
efectica de camino temporal cerrado (CTC) de Schwinger y Keldysh 
($S_{\rm ef}^{\rm CTC}$) cuando la aproximaci\'on 
cl\'asica es 
v\'alida \cite{ctp1,calzethu1,jordan}. 
En este cap\'\i tulo 
mostramos la deducci\'on de la acci\'on efectiva de granulado grueso 
y su utilidad para el estudio del proceso de p\'erdida de coherencia de 
un sistema acoplado a un entorno.  

Consideramos un campo escalar cu\'antico en un espacio-tiempo plano 
(signatura $(+,-,-,-)$) con una autointeracci\'on cu\'artica. Su acc\'on 
cl\'asica est\'a dada por 
\begin{equation}S[\phi] = \int d^4x \{{1\over{2}} \partial_\nu \phi(x) 
\partial^\nu \phi(x) - {1\over{2}} m^2 \phi^2(x) - {1\over{4!}} \lambda 
\phi^4(x)\},\label{action}\end{equation}
donde $m$ es la masa {\it desnuda} del campo y $\lambda$ es la 
constante de acoplamiento, tambi\'en desnuda. Se\-parando el campo entre 
sistema y entorno, como 
mencionamos en la introducci\'on de este cap\'\i tulo,

\begin{equation}\phi(x) = \phi_<(x) + 
\phi_>(x),\label{splitting}\end{equation}
el campo sistema queda definido como 

\begin{equation}\phi_<(\vec x, t) = \int_{\vert \vec k\vert < \Lambda} 
{d^3\vec k\over{(2 \pi)^3}} \phi(\vec k, t) \exp{i \vec k . \vec 
x},\label{sys}\end{equation} y el entorno es 
\begin{equation}\phi_>(\vec x, t) = \int_{\vert \vec k\vert > \Lambda} 
{d^3\vec k\over{(2 \pi)^3}} \phi(\vec k, t) \exp{i \vec k . \vec 
x}.\label{env}\end{equation} 
El campo sistema contiene aquellos modos menores que $\Lambda$, mientras que 
el entorno contiene aquellos mayores a $\Lambda$. Despu\'es de haber 
separado al campo 
en modos, la acci\'on cl\'asica puede expresarse como

 \begin{equation}S[\phi] = S_0[\phi_<] + S_0[\phi_>] + S_{\rm int}[\phi_<, 
\phi_>],\label{actions}\end{equation} donde con $S_0$ estamos notando a la 
acci\'on del campo libre y el t\'ermino de interacci\'on es 
\begin{eqnarray}S_{\rm int}[\phi_<, \phi_>] &=& - \int d^4x 
\{{\lambda\over{4!}}\phi_<^4(x) 
+ {\lambda\over{4!}}\phi_>^4(x) + {\lambda\over{4}} \phi_<^2(x) \phi_>
^2(x) \nonumber \\ &&+ {\lambda\over{6}} \phi_<^3(x) 
\phi_>(x) + {\lambda\over{6}} \phi_<(x) 
\phi_>^3(x)\}.\label{inter}\end{eqnarray}

La matriz densidad total (para el sistema-entorno) est\'a definida 
como: 
\begin{equation} \rho[\phi_<,\phi_>,\phi_<',\phi_>',t]=\langle\phi_< 
\phi_>\vert {\hat\rho} \vert \phi_<' 
\phi_>'\rangle,\label{matrix}
\end{equation} donde $\vert \phi_<\rangle$ y $\vert \phi_>\rangle$ son los 
autoestados de los operadores de campo ${\hat\phi}_<$ y ${\hat\phi}_>$, 
respectivamente. Vamos a considerar, s\'olo por simplicidad, la misma 
condici\'on inicial que en el caso del movimiento Browniano cu\'antico del 
Cap\'\i tulo 2: el sistema y entorno se acoplan a tiempo inicial $t_0$. Por 
lo tanto, la matriz densidad 
puede escribirse como el producto de la matriz densidad correspondiente al 
sistema y al entorno a tiempo $t_0$: 

\begin{equation}{\hat\rho}[t_0] = {\hat\rho}_{<}[t_0] 
{\hat\rho}_{>}[t_0].\label{sincorr}\end{equation} 
Posteriormente asumiremos que el campo entorno est\'a, inicialmente, en su 
estado de vac\'\i o.

Como nos interesa la evoluci\'on del sistema teniendo en cuenta la 
influencia del entorno, ser\'a la matriz densidad reducida nuestro 
objeto de real importancia. La matriz densidad reducida est\'a definida
integrando funcionalmente todas las configuraciones posibles del 
campo entorno, sobre la matriz densidad total 
\begin{equation}\rho_{\rm red}[\phi_<,\phi'_<,t] = \int {\cal D}\phi_> 
\rho[\phi_<,\phi_>,\phi_<',\phi_>,t].\label{red}
\end{equation}

La evoluci\'on temporal de la matriz $\rho_{\rm r}$ est\'a dada por 

\begin{equation}\rho_{\rm r}[\phi_{<f},\phi_{<f}',t] = \int d\phi_{<i}\int 
d\phi_{<i}' ~J_{\rm r}[\phi_{<f},\phi_{<f}',t\vert \phi_{<i},\phi_{<i}',t_0] 
~\rho_{\rm r}[\phi_{<i}\phi_{<i}',t_0],\label{evol}
\end{equation} donde $J_{\rm r}[t,t_0]$ es el operador de evoluci\'on reducido 
definido, a su vez por

\begin{equation}J_{\rm r}[\phi_{<f},\phi_{<f}',t\vert \phi_{<i},\phi_{<i}',
t_0] = 
\int_{\phi_{<i}}^{\phi_{<f}}{\cal D}\phi_< \int_{\phi'_{<i}}^{\phi_{<f}'}{\cal 
D}\phi_<' \exp\left\{i[S[\phi_<] - S[\phi_<']]\right\} 
F[\phi_<,\phi_<'].\label{evolred}
\end{equation}

En esta expresi\'on encontramos, al igual que en el caso del movimiento 
Browniano (ecuaci\'on (\ref{evol0})), la funcional de influencia de Feynman y
 Vernon $F[\phi_<,\phi_<']$, 
que tiene en cuenta el efecto del entorno sobre el sistema; ahora definida 
como:
\begin{eqnarray}
F[\phi_<,\phi_<'] &=& \int d\phi_{>i} \int 
d\phi_{>i}' \rho_{>}[\phi_{>i},\phi_{>i}',t_0] \int d\phi_{>f}
\int_{\phi_{>i}}^{\phi_{>f}}{\cal D}\phi_>\int_{\phi_{>i}'}^{\phi_{>f}}{\cal 
D}\phi_>'\nonumber \\
&&\times \exp\left\{i[S[\phi_>] + S_{\rm int}[\phi_<,\phi_>] - 
S[\phi'_>] - S_{\rm int}[\phi'_< , \phi'_>]]\right\}.
\end{eqnarray}
 
Podemos definir tambi\'en la {\it acci\'on de influencia} $\delta 
A[\phi_<,\phi_<']$ y la {\it acci\'on efectiva de granulado grueso} 
 $A[\phi_<,\phi_<']$ como

\begin{equation}F[\phi_<,\phi_<'] = \exp\left\{i 
\delta A[\phi_<,\phi_<']\right\},\label{IA}\end{equation}
\begin{equation}A[\phi_<,\phi_<'] = S[\phi_<] - S[\phi_<'] + \delta 
A[\phi_<,\phi_<'].\label{CTPEA}\end{equation} 

En el presente cap\'\i tulo calcularemos la acci\'on de influencia 
perturbativamente en la constante de acoplamiento, hasta orden cuadr\'atico. 
 Desarrollando en potencias de $\lambda$ obtenemos, 
\begin{eqnarray}
\delta A[\phi_<,\phi_<'] &=&\{\langle 
S_{\rm int}[\phi_<,\phi_>]\rangle_0 - \langle 
S_{\rm int}[\phi_<',\phi_>']\rangle_0\}\nonumber \\
&&+ {i\over{2}}\{\langle S_{\rm int}^2[\phi_<,\phi_>]\rangle_0 - \big[\langle 
S_{\rm int}[\phi_<,\phi_>]\rangle_0\big]^2\}\nonumber \\
&&- i\{\langle S_{\rm int}[\phi_<,\phi_>] S_{\rm int}[\phi_<',\phi_>']
\rangle_0 - 
\langle S_{\rm int}[\phi_<,\phi_>]\rangle_0\langle 
S_{\rm int}[\phi_<',\phi_>']\rangle_0\} \label{inflac} \\
&&+ {i\over{2}}\{S^2_{\rm int}[\phi_<',\phi_>']\rangle_0 - 
\big[\langle 
S_{\rm int}[\phi_<',\phi_>']\rangle_0\big]^2\},\nonumber
\end{eqnarray} 
donde el valor medio cu\'antico $\langle ..... \rangle$ de una 
funcional de los campos se define como 
\begin{eqnarray}\langle B[\phi_>,\phi_>'] \rangle_0 &=& \int 
d\phi_{>i} \int d\phi_{>i}' 
~~ \rho_{>}[\phi_{>i},\phi'_{>i},t_0] \nonumber \\
&& \times \int d\phi_{>f} 
\int_{\phi_{>i}}^ {\phi_{>f}}{\cal D}\phi_>\int_{\phi_{>i}'}^{\phi_{>f}}{\cal 
D}\phi_>' \exp\left\{i[S_0[\phi_>] - S_0[\phi_>']]\right\} 
B.\label{averag}\end{eqnarray} 
Los propagadores para el campo entorno est\'an definidos por:

\begin{equation}\langle \phi_>(x),\phi_>(y)\rangle_0 = i 
G_{++}^\Lambda(x-y),\label{feyn}\end{equation}
\begin{equation}\langle \phi_>(x),\phi_>'(y)\rangle_0 = - i G_{+-}^\Lambda 
(x-y),\label{frecmas}\end{equation} 
\begin{equation}\langle 
\phi_>'(x),\phi_>'(y)\rangle_0 = - i G_{--
}^\Lambda(x-y).\label{dyson}\end{equation} 
Estos propagadores no son los propagadores usuales de Feynman, Dyson y 
Wightman, debido a que en este caso, la integraci\'on en los momentos 
est\'a restringida por la presencia de la frecuencia de corte (infraroja) 
$\Lambda$. Las expresiones expl\'\i citas para cada uno son las siguientes:

 \begin{equation}G_{++}^\Lambda (x-y)= \int_{\vert \vec p\vert > \Lambda} 
{d^4p\over{(2 \pi)^4}} e^{i p (x - y)} {1\over{p^2 - m^2 + i 
\epsilon}},\label{feypro}\end{equation} 
\begin{equation}G_{+-}^\Lambda (x-y) = -\int_{\vert \vec p\vert > \Lambda} 
{d^4p\over{(2 \pi)^4}} e^{i p (x - y)} 2 \pi i 
\delta (p^2 - m^2) \Theta(p^0),\label{frecmaspro}\end{equation}
\begin{equation}G_{--}^\Lambda (x-y) = \int_{\vert \vec p\vert > \Lambda} 
{d^4p\over{(2 \pi)^4}} e^{i p (x - y)} {1\over{p^2 - m^2 - i 
\epsilon}}.\label{dysonprop}\end{equation} 

Como ejemplo ilustrativo mostramos la expresi\'on  de 
$G_{++}^{\Lambda}$, en el caso no masivo. El propagador de Feynman usual 
est\'a dado por 

\begin{equation}G_{++}(x)={i\over{8 \pi^2}}{1\over{\sigma}} - 
{1\over{8 \pi}} 
\delta (\sigma),\nonumber \end{equation} mientras que 
\begin{eqnarray}
G^\Lambda_{++}(x) &=& {i\over{8 \pi^2}}\Big[{cos[\Lambda (r - t)]\over{r (r - 
t)}} + {cos[\Lambda (r + t)]\over{r(r+t)}}\Big] 
-{1\over{8 \pi^2}}\Big[{sin[\Lambda (r - t)]\over{r (r - t)}} - {sin[\Lambda 
(r + t)]\over{r (r + t)}}\Big]\nonumber \\ &\equiv & G_{++}(x) - G_{++}^{\vert 
\vec p\vert <\Lambda}(x), \end{eqnarray} donde $\sigma = {1\over{2}}x^2$ es 
un medio de la distancia geod\'esica. Por supuesto, el propagador 
$G_{++}^{\Lambda}$ coincide con el propagador $G_{++}$ cuando $\Lambda 
\rightarrow 0$.

La acci\'on de influencia puede calcularse a partir de las ecuaciones 
(\ref{inflac})-(\ref{dyson}) usando t\'ecnicas usuales, quedando determinada 
por:
\begin{eqnarray} \delta A[\phi_<, \phi'_<] &=& -\lambda \int d^4x 
\left\{{1\over{24}} [\phi^4_<(x) - \phi_<^{'4}(x)] + {1\over{4}} i 
G_{++}^\Lambda(0) [\phi_<^2(x) - \phi_<^{'2}(x)]\right\}\nonumber \\ &&+ 
\lambda^2\int d^4x \int d^4y \left\{-{1\over{72}} \phi_<^3(x) G_{++}^ 
\Lambda(x-y) \phi_<^3(y)\right.\nonumber \\
&& - \left. {1\over{36}} \phi^3_<(x) G_{+-}^\Lambda(x-y) 
\phi_<^{'3}(y) + {1\over{72}} \phi_<^{'3}(x) G_{--
}^\Lambda(x-y) \phi_<^{'3}(y)\right.\nonumber \\
&& - \left.{1\over{16}} \phi_<^2(x) i G_{++}^{\Lambda 
2}(x-y) \phi_<^2(y) + {1\over{8}}\phi_<^2(x)i 
G_{+-}^{\Lambda 
2}(x-y) \phi_<^{'2}(y)\right.\nonumber \\
&&- \left.{1\over{16}}\phi_<^{'2}(x) i G_{--}^{\Lambda 2} 
(x-y)\phi_<^{'2}(y) + 
{1\over{18}} \phi_<(x) G_{++}^{\Lambda 
3}(x-y)\phi_<(y)\right. \nonumber \\
&& + \left. {1\over{9}} \phi_<(x) G_{+-}^{\Lambda 
3}(x-y)\phi_<'(y)- 
{1\over{18}}\phi_<'(x) G_{--}^{\Lambda 
3}(x-y) \phi'_<(y) \right\}.\end{eqnarray} 


Definiendo nuevas variables:

$$\Sigma P ={1\over{2}}({\phi_<}^4 
+ \phi_<^{'4}) ~~~;~~~ \Delta P = {1\over{2}}({\phi_<}^4 
- \phi_<^{'4})$$
$$\Sigma R ={1\over{2}}(\phi_<^3 + \phi_<^{'3}) ~~~;~~~ 
\Delta R ={1\over{2}}({\phi_<}^3 - \phi_<^{'3})$$ 
$$\Sigma Q
={1\over{2}}({\phi_<}^2 + \phi_<^{'2}) ~~~;~~~  \Delta Q
={1\over{2}}({\phi_<}^2 - \phi_<^{'2})$$
$$\Sigma \Phi ={1\over{2}}(\phi_< +  
\phi_<^{'}) ~~~;~~~  \Delta \Phi = {1\over{2}}(\phi_< -  
\phi_<^{'}),$$
y usando las identidades que satisfacen los propagadores, podemos 
separar las partes real e ima\-ginaria de la acci\'on de influencia: 
\begin{eqnarray} {\mbox Re} \delta 
A&=& - \lambda \int 
d^4x\left\{{1\over{12}}~\Delta P(x) + {i\over{2}}~G_{++}^\Lambda(0)~ 
\Delta Q(x)\right\}
\nonumber \\ 
&&+ 2 \lambda^2 \int d^4x\int d^4y ~\theta (x^0 - y^0)\left\{-{1\over{18}} 
~\Sigma R(x)~ {\mbox Re} 
G^\Lambda_{++}(x-y) 
~\Delta R(y)\right.\nonumber \\ &&+ \left.{1\over{4}} ~\Sigma Q(x)~ {\mbox Im} 
G^{\Lambda 2}_{++}(x-y) 
~\Delta Q(y) + 
{1\over{3}} ~\Sigma \Phi(x)~ {\mbox Re} G^{\Lambda 3}_{++}(x-y)~ \Delta \Phi 
(y)\right\},\label{inff}\\
{\mbox Im}\delta A &=&
\lambda^2 \int 
d^4x \int d^4y \left\{-{1\over{18}}~ \Delta R(x)~ {\mbox Im} 
G^\Lambda_{++}(x-y) 
~\Delta R(y)\right.\nonumber 
\\ &&-\left.{1\over{4}}~ \Delta Q(x)~ {\mbox Re} G^{\Lambda 2}_{++}(x-y)~ 
\Delta Q(y) 
+ {1\over{3}} 
~\Delta \Phi (x)~ {\mbox Im} G^{\Lambda 3}_{++}(x-y)~ \Delta \Phi (y)\right\}. 
\end{eqnarray} 
Esta expresi\'on ha sido calculada a orden cuadr\'atico en la constante 
$\lambda$. En lo que resta del presente cap\'\i tulo especializaremos 
nuestros c\'alculos en el caso no-masivo por simplicidad \cite{prd53}.

La parte real de la acci\'on de influencia tiene divergencias, las que deben 
renormalizarse si\-guiendo alguno de los procedimientos usuales en teor\'\i a 
de campos. Como los propagadores definidos en las ecuaciones 
(\ref{feyn})-(\ref{dyson}) difieren de los usuales por la presencia de la 
longitud cr\'\i tica $\Lambda$, las divergencias ultravioletas coinciden 
con las usuales de la teor\'\i a $\lambda \phi^4$. En consecuencia, la 
acci\'on de influencia se renormaliza con los mismos contrat\'erminos que 
se utilizan normalmente. El t\'ermino $G_{++}^{\Lambda}(0)\Delta Q(x)$ es 
divergente. Como 
$G_{++}^{\Lambda}= G_{++} - G_{++}^{\vert \vec p\vert <\Lambda}$ y  
$G_{++}^{\vert \vec p\vert <\Lambda}(0)$ es una cantidad finita, las 
divergencias de  $G_{++}^{\Lambda}$ y $G_{++}$ son coincidentes. Este 
t\'ermino representa la renormalizaci\'on de la masa.
 
Por medio de la regularizaci\'on dimensional, el cuadrado del 
propagador de Feynman puede escribirse como 

\begin{equation}G^{\Lambda 2}_{++}(x) = G_{++}^2(x) + G_{++}^{(\vert \vec 
p\vert <\Lambda) 2}(x) - 2 G_{++}(x) G_{++}^{(\vert \vec p\vert 
<\Lambda)}(x),\end{equation}
donde
\begin{equation}G_{++}^2(x)={1\over{16 \pi^2}}\left\{\left[{i\over{n-4}}+
i\psi(1) - 4 
\pi i\right]\delta^4(x)+i\Sigma(x) - \eta(x) - Log[4 \pi 
\mu^2]\delta^4(x)\right\},\nonumber\end{equation}
\begin{equation}\Sigma (x)={1\over{(2 \pi)^4}}\int d^4p e^{i p x} Log \vert 
p^2\vert,\nonumber\end{equation}
\begin{equation}\eta (x)={\pi\over{(2 \pi)^4}}\int d^4p e^{i p x} \theta 
(p^2).\nonumber\end{equation}
El t\'ermino proporcional a ${1\over n-4}\delta^4(x-y)$ representa 
otra de las divergencias usuales, y es independiente de $\Lambda$. As\'\i , el 
t\'ermino ${\mbox Im} G_{++}^{\Lambda 2}(x-y)\Sigma Q(x)\Delta Q(y)$ presente 
en la ecuaci\'on (\ref{inff}) tambi\'en diverge y renormaliza a la constante 
$\lambda$. Las dem\'as divergencias se deben tratar de la misma manera. Una 
caracter\'\i stica importante es que la parte imaginaria ${\mbox Im} \delta A$ 
es finita. 


\section{Ecuaciones de movimiento semicl\'asicas}

El m\'etodo que estamos desarrollando en el presente cap\'\i tulo nos 
permitir\'a describir la din\'amica efectiva de los modos relevantes 
del campo cu\'antico, en t\'erminos de una acci\'on disipativa y 
estoc\'astica. En la ecuaci\'on de movimiento aparecer\'an t\'erminos 
asociados a fuerzas estoc\'asticas que describen el intercambio de 
energ\'\i a (y otros posibles n\'umeros cu\'anticos) con los modos 
del entorno, tratados perturbativamente \cite{gleiram,grainermu}. Los 
t\'erminos 
disipativos describir\'an una eventual tendencia al equilibrio 
de los modos del sistema. La existencia de una relaci\'on de 
fluctuaci\'on-disipaci\'on asegura que los modos del sistema 
obtengan la misma temperatura que aquellos del entorno con los 
que interact\'uan. 

La derivaci\'on de una ecuaci\'on de movimiento efectiva para los modos 
relevantes corresponde a hacer un granulado grueso del campo cu\'antico, 
lo que qued\'o puesto de manifiesto en el c\'alculo de la acci\'on efectiva
 de granulado grueso. En el esp\'\i ritu general de la mec\'anica 
estad\'\i stica, asumir equilibrio t\'ermico para los modos que han sido 
eliminados puede interpretarse como asumir la m\'\i nima 
cantidad de informaci\'on acerca del estado cu\'antico 
microsc\'opico del entorno. El granulado grueso aporta los siguientes 
caracter\'\i sticas a la din\'amica efectiva del sistema:

1. par\'ametros como la masa y la constante de acoplamiento son 
renormalizadas y aparecen nuevas interacciones efectivas, en general 
no-locales. Lo mismo sucede en el caso del movimiento Browniano cu\'antico, 
donde aparece un corrimiento de la frecuencia natural de la 
part\'\i cula debido al entorno \cite{hpz1}.

2. la din\'amica es disipativa y contiene t\'erminos de ruido que satisfacen 
la relaci\'on de fluctuaci\'on-disipaci\'on. La evoluci\'on din\'amica 
semicl\'asica es inherentemente estoc\'astica, reflejando el acoplamiento
 entre los modos relevantes y un entorno t\'ermico. La ecuaci\'on 
(\ref{rfdqbm}) muestra tal relaci\'on en el caso del movimiento Browniano.

3. los t\'erminos disipativos son, en general, no-locales en el tiempo debido
a que el entorno requiere de cierto tiempo para ``responder''. Un cambio en 
el sistema produce una influencia sobre el entorno y esta se 
comporta {\it causalmente}.

Las partes real e imaginaria de $\delta A[\phi_<^+,\phi_<^-]$ pueden asociarse 
con la disipaci\'on y el ruido respectivamente \cite{feyver}. En particular, 
la parte imaginaria puede re-escribirse en t\'erminos de tres fuentes 
estoc\'asticas de ruido $\nu(x)$, $\xi(x)$ y $\eta(x)$ con una distribuci\'on 
de probabilidad gaussiana dada por
\begin{eqnarray}
P[\nu(x), \xi(x), \eta(x)] &=& N_\nu N_\xi N_\eta \exp\left\{
-{1\over{2}}\int d^4x\int d^4y ~\nu(x)~\left[{\lambda^2\over{9}} 
{\mbox Im} G_{++}^\Lambda\right]^{-1}~\nu(y)~\right\}\nonumber \\
&&\times \exp\left\{-{1\over{2}}\int d^4x\int d^4y ~\xi(x)~
\left[{\lambda^2\over{2}} {\mbox Re} G_{++}^{\Lambda 2}\right]^{-1}
~\xi(y)\right\}\nonumber \\
&&\times \exp\left\{-{1\over{2}}\int d^4x\int d^4y ~\eta(x)~\left[{-2
\lambda^2\over{3}} 
{\mbox Im} G_{++}^{\Lambda^3}\right]^{-1}~\eta(y)\right\},\end{eqnarray}
donde $N_\nu$, $N_\xi$ y $N_\eta$ son constantes de normalizaci\'on. De 
esta manera, la parte imaginaria puede escribirse como tres integrales 
funcionales sobre las fuentes de ruido gaussianas
\begin{eqnarray}
&&\int {\cal D}\nu(x)\int {\cal D}\xi (x) \int {\cal D}\eta(x) 
~P[\nu,\xi,\eta] \exp\left\{ -i \left[\Delta R(x)\nu(x) + 
\Delta Q(x) \xi(x) 
+ \Delta \Phi(x) \eta(x)\right]\right\}\nonumber \\
&&= \exp\left\{-i\int d^4x\int d^4y 
\left[{\lambda^2\over{18}}\Delta R(x) {\mbox Im} G_{++}^\Lambda(x,y)\Delta 
R(y)\right.\right.\nonumber \\
&&+ \left.\left. {\lambda^2\over{4}}\Delta Q(x) 
{\mbox Re} G_{++}^{\Lambda 2}(x,y)
\Delta Q(y) - 
{\lambda^2\over{3}}\Delta \Phi(x) {\mbox Im} G_{++}^{\Lambda 3}(x,y) 
\Delta \Phi(y)\right]\right\}.\end{eqnarray}

Por lo tanto, la acci\'on efectiva de granulado grueso $A$, puede escribirse 
\begin{equation}A[\phi_<^+,\phi_<^-]=-{1\over{i}} \ln \int {\cal D}
\nu ~P[\nu]~\int {\cal D}\xi ~P[\xi]~\int {\cal D}\eta ~P[\eta] 
~\exp\left\{i S_{\rm eff}[\phi_<^+,\phi_<^-, \nu, \xi, \eta]\right\},
\end{equation}
donde hemos definido 

\begin{equation}S_{\rm ef}[\phi_<^+,\phi_<^-, \nu, \xi, \eta]= 
{\mbox Re} A[\phi_<^+,\phi_<^-]- \int d^4x~\left[\Delta R(x)~ \nu (x) + 
\Delta Q(x)~ \xi(x) + \Delta \Phi(x) ~\eta(x)\right].\end{equation}

La ecuaci\'on de movimiento se obtiene tomando la variaci\'on funcional:

\be \left.{\delta S_{\rm ef}[\phi_<^+,\phi_<^-, \nu, \xi, \eta]\over{\delta 
\phi_<^+}}\right\vert_{\phi_<^+=\phi_<^-}=0,\ee obteni\'endose
\bea
&&\Box \phi_<(x) + {1\over{6}}\phi_<^3(x) +{1\over{12}}
 \lambda^2_{\rm ren}\phi_<^2(x)
\int d^4y ~{\mbox Re}G_{++}^\Lambda (x,y)~\phi_<^3(y)
\nonumber \\
&&+ {1\over{4}} \lambda^2_{\rm ren}\phi_<(x)\int d^4y ~{\mbox Im} 
G_{++}^{\Lambda 2}~ \phi_<^2(y) 
+ {1\over{6}} \lambda^2_{\rm ren}\int d^4y {\mbox Re} G_{++}^{\Lambda 3} 
~\phi_<(y)\nonumber \\
&&= {3\over{2}} \phi_<^2(x) \nu (x) + \phi_<(x) \xi (x) 
+ {1\over{2}} \eta (x).\eea
Esta es la llamada ecuaci\'on de Langevin \cite{gleiram} donde 
la fuerza estoc\'astica aparece determinada por ruido aditivo y 
multiplicativo. Esta, es la generalizaci\'on a teor\'\i a de campos de la 
ecuaci\'on (\ref{lqbm}) para el movimiento Browniano cu\'antico.

A partir de la acci\'on efectiva de granulado grueso (o de la acci\'on 
efectiva de camino temporal cerrado) y de esta ecuaci\'on semicl\'asica 
de movimiento, es posible concluir que el efecto total del 
entorno puede resumirse en tres aspectos fundamentales: renormaliza 
las constantes desnudas de la teor\'\i a, agrega disipaci\'on y 
produce fluctuaciones (ruido) que hacen ``termalizar'' al sistema. En la
 pr\'oxima secci\'on vamos a estudiar c\'omo los n\'ucleos de ruido 
destruyen las coherencias cu\'anticas del sistema originando la 
llamada transici\'on cu\'antico-cl\'asica.  


\section{La ecuaci\'on maestra}

Al igual que en el caso del movimiento Browniano cu\'antico, en esta 
secci\'on obtendremos la ecuaci\'on de evoluci\'on (ecuaci\'on maestra) para 
la matriz densidad reducida; por lo tanto gene\-ralizaremos los resultados del 
Cap\'\i tulo 2 a teor\'\i a de campos. 

Para el caso del movimiento Browniano con acoplamiento lineal ($x q_i$) 
sabemos que una soluci\'on aproximada de la ecuaci\'on maestra 
est\'a representada por \cite{hpz1,jpp}

\begin{equation}
\rho_{\rm r}[x,x';t] \approx \rho_{\rm r}[x,x',0] \exp{\Big[-(x - 
x')^2\int_0^t D(s) ds}\Big],\label{qbmdecay}\end{equation} de donde es simple
 ver que los t\'erminos fuera de la diagonal de la matriz densidad reducida 
van disminuyendo a medida que $\int_0^t D(s) ds$ se hace suficientemente 
grande. Para acoplamientos no lineales, tales como $x^n q_i^m$, es de esperar 
que la ecuaci\'on maestra desarrolle t\'erminos difusivos de la forma 
 $iD^{(n,m)}(t)(x^n-x'^n)^2\rho_{\rm r}$. En esta observaci\'on basaremos 
nuestro 
estudio de los coeficientes de difusi\'on que se obtienen en nuestro 
ejemplo, como primer paso hacia el entendimiento del proceso de p\'erdida 
de coherencia.

Para ello debemos calcular el operador de evoluci\'on reducido de la 
ecuaci\'on (\ref{evolred}). Como este operador est\'a definido por medio de
integrales de camino, podemos obtener una estimaci\'on utilizando 
la aproximaci\'on de punto de ensilladura

\begin{equation}
J_{\rm r}[\phi^+_{<f},\phi^-_{<f},t\vert\phi^+_{<i},\phi^-_{<i},t_0] 
\approx \exp\left\{i A[\phi_<^{+{\rm cl}},\phi_<^{-{\rm cl}}]
\right\},\label{prosadle}\end{equation} donde 
 $\phi_<^{+{\rm cl}} (\phi_<^{-{\rm cl}})$ es la soluci\'on de la ecuaci\'on de
movimiento ${\delta {\mbox Re} A\over\delta\phi_<^+}\vert_{\phi_<^+=\phi^-_<
}=0$ con condiciones de contorno $\phi_<^{+{\rm cl}}(t_0)=\phi_{<i}
(\phi^-_{<i})$ y $\phi_<^{+{\rm cl}}(t)=\phi^+_{<f} (\phi^-_{<f})$.

Los efectos de p\'erdida de coherencia est\'an contenidos en la parte 
imaginaria de la acci\'on efectiva de granulado grueso, la que hemos 
calculado perturbativamente hasta orden $\lambda^2$. Por lo tanto, al evaluar 
${\mbox Im} A$ podemos aproximar $\phi_<^{\rm cl}$ por la soluci\'on de la 
ecuaci\'on de campo libre, con las condiciones de contorno adecuadas. De esta 
manera, la soluci\'on cl\'asica que utilizamos en la aproximaci\'on de 
punto de ensilladura es:

\begin{equation}\phi_<^{cl}(\vec x, s) =  \left[\phi_{<f} {\sin (k_0 s)
\over {\sin (k_0 
t)}} + \phi_{<i} {\sin [k_0 (t - s)]\over{\sin (k_0 t)}}\right] \cos
 (\vec k_0 . 
\vec 
x)\equiv \phi_<^{cl}(s) \cos (\vec k_0 . \vec 
x)
,\label{classicphi}\end{equation} donde hemos asumido que el campo sistema 
contiene s\'olo un modo Fourier (aunque arbitrario) con $\vec k = \vec k_0$.
 
Siguiendo el procedimiento empleado en el Cap\'\i tulo 2, debemos calcular la 
derivada temporal del operador de evoluci\'on reducido $J_{\rm r}$, y eliminar 
la dependencia en las configuraciones iniciales de los campos 
 $\phi^+_{<i}$ y $\phi^-_{<i}$ que aparecen debido a la soluci\'on cl\'asica. 
El operador libre, definido por 

\begin{equation}J_0[\phi^+_{<f}, \phi^-_{<f}, t\vert \phi^+_{<i}, \phi^-_{<i}
, 0] = \int_{\phi^+_{<i}}^{\phi^+_{<f}}{\cal D}\phi^+_< \int_{\phi^-_{<i}}
^{\phi^-_{<f}} 
{\cal D}\phi^-_< \exp\left\{i \left[ S_0(\phi^+_<) - 
S_0(\phi^-_<)\right]\right\},\label{propdeJ0}\end{equation} satisface 
las siguientes identidades

\begin{equation}\phi_<^{+{\rm cl}}(\vec x, s) J_0 = \left[ \phi^+_{<f} 
\cos [k_0(t - s)] + {\sin [k_0(t - s)]\over{k_0}} i 
\partial_{\phi^+_{<f}}\right]J_0,\label{rel1}\end{equation}
y
\begin{equation}\phi_<^{-{\rm cl}}(\vec x, s) J_0 = \left[ \phi^-_{<f} 
\cos [k_0(t - s)] - {\sin [k_0(t - s)]\over{k_0}} i 
\partial_{\phi^-_{<f}}\right]J_0.\label{rel2}\end{equation}
 
La derivada temporal est\'a dada por 
\begin{eqnarray}&&i \partial_t J_r[\phi^+_{<f},\phi^-_{<f},t\vert 
\phi^+_{<i},\phi^-_{<i},0] = \left\{h_{ren}[\phi^+_<] - h_{ren}[\phi^-_<]
\right.\nonumber \\ 
&&- \left. i 
\lambda^2 \left[{(\phi_{<f}^{+3} - \phi_{<f}^{-3})V\over{1152}}
 \int_0^t ds \Delta R^{\rm cl}(s) {\mbox Im} G_{++}^{\Lambda}
(3k_0;t-s)\right.\right.\nonumber \\ && 
+ \left.\left.{(\phi_{<f}^{+2} - \phi_{<f}^{-2})V\over{32}} 
\int_0^t ds \Delta Q^{\rm cl}
(s)({\mbox Re} G_{++}^{\Lambda 2}(2k_0;t-s)+ 2 {\mbox Re} G_{++}^{\Lambda 2}
(0;t-s))\right.\right.
\nonumber \\ && - \left.\left.{(\phi^+_{<f} 
- \phi^-_{<f})V\over{6}} \int_0^t ds \Delta \Phi^{\rm cl}(s) 
{\mbox Im} G_{++}^{\Lambda 3}(k_0;t-s)]\right.\right.\nonumber \\
&&- \left.\left.\lambda^2 [-{(\phi_{<f}^{+3} + \phi_{<f}^{-3})V\over{1152}}
\int_0^t ds \Delta R^{\rm cl}(s) 
{\mbox Re} G_{++}^{\Lambda}(3k_0;t-s)\right.\right. \nonumber \\ && 
+ \left.\left.{(\phi_{<f}^{+2} + \phi_{<f}^{-2})V\over{32}} 
\int_0^t ds \Delta Q^{\rm cl}
(s)({\mbox Im} G_{++}^{\Lambda 2}(2k_0;t-s)+ 2 {\mbox Im} G_{++}^{\Lambda 2}
(0;t-s))\right.\right.\nonumber \\ && + \left.\left.{(\phi^+_{<f} 
- \phi^-_{<f})V\over{6}} \int_0^t ds \Delta \Phi^{\rm cl}(s) 
{\mbox Re} G_{++}^{\Lambda 3}(k_0;t-s)\right]+ ... \right\}J_r[\phi^+_{<f},
\phi^-_{<f},t\vert 
\phi^+_{<i},\phi^-_{<i},0]\label{timeder} \end{eqnarray}
donde $G_{++}^{\Lambda n}(k;t-s)$ es la transformada de Fourier de la
 n-\'esima 
potencia del propagador (para $n = 1, 2 ~{\mbox o}~3$). Los puntos suspensivos 
est\'an denotando otros t\'erminos que provienen de la derivada temporal que 
no contribuyen a los coeficientes de difusi\'on (es decir, vamos a ignorar 
todos aquellos t\'erminos que no sean proporcionales a $(\phi_{<f}^{+n} 
- \phi_{<f}^{-n})^2$ en la ecuaci\'on final). El factor $V$ que est\'a 
presente en la ecuaci\'on (\ref{timeder}) proviene de haber asumido que 
el campo s\'olo tiene un modo Fourier. Este desaparece cuando uno considera 
paquetes de ondas. 

Utilizando las identidades (\ref{rel1}) y (\ref{rel2}) podemos extraer la 
dependencia con las configuraciones iniciales de campo y obtener la 
ecuaci\'on maestra. Como esta ecuaci\'on es muy complicada, s\'olo nos 
concentramos en las correcciones que a la evoluci\'on unitaria usual 
aportan los n\'ucleos de ruido:
\begin{eqnarray}&& i \partial_t \rho_{\rm r} 
[\phi^+_{<f},\phi^-_{<f},t] = 
\langle \phi^+_{<f}\vert\Big[{\hat H}_{\rm ren}, 
{\hat\rho}_{\rm r}\Big]\vert\phi^-_{<f}\rangle 
- i \lambda^2 \left[{(\phi_{<f}^{+3} 
- \phi_{<f}^{-3})^2 V\over{1152}}D_1(k_0;t)\right.\nonumber \\
&& + \left.{(\phi_{<f}^{+2} - 
\phi_{<f}^{-2})^2 V\over{32}}D_2(k_0;t) - {(\phi^+_{<f} - 
\phi^-_{<f})^2 V\over{6}}D_3(k_0;t)\right] \rho_{\rm r}
[\phi^+_{<f},\phi^-_{<f},t]
 + ...~ .\label{masterf}\end{eqnarray} 
La ecuaci\'on (\ref{masterf}) es la generalizaci\'on de la ecuaci\'on 
maestra para el movimiento Browniano cu\'antico (\ref{master}) presentada 
en el Cap\'\i tulo 2. En este caso, el acoplamiento del sistema 
es no-lineal. Debido a la presencia de tres t\'erminos de interacci\'on
  ($\phi_<^3 \phi_>$, 
$\phi_<^2 \phi_>^2$, and $\phi_< \phi_>^3$) hay tres coeficientes de 
difusi\'on en la ecuaci\'on maestra. La forma de cada uno est\'a 
determinada por estos acoplamientos y por la elecci\'on particular del 
estado cu\'antico del entorno. 

De los tres coeficientes de difusi\'on dependientes del 
tiempo $D_i(t)$ contenidos en la ecuaci\'on (\ref{masterf}), si consideramos 
s\'olo los coeficientes presentes en el 
desarrollo a un lazo para el campo cu\'antico irrelevante, s\'olo sobreviven 
los coeficientes $D_1$ y $D_2$, los cuales est\'an dados por
\begin{eqnarray}D_1(k_0;t) &=& \int_0^t ds ~~ \cos^3(k_0 s){\mbox Im} 
G_{++}^{\Lambda}(3k_0;t-s)\nonumber \\ &&= {1\over{6 k_0}}\int_0^t ds ~~ 
\cos^3(k_0 s) ~~ \cos (3 k_0 s) ~~ \theta(3 k_0 - \Lambda)\nonumber \\ &&= {2 
k_0 t + 3 \sin (2 k_0 t) + {3\over{2}} \sin (4 k_0 t) + {1\over{3}} \sin 
(6 k_0 
t)\over{576 k_0^2}},~~~~ {\Lambda\over{3}}<k_0<\Lambda\label{d1}\end{eqnarray} 

\begin{equation}D_2(k_0;t) = \int_0^t ds ~~ \cos^2(k_0 s)({\mbox Re} G_{++}^
{\Lambda 2}(2k_0;t-s)+ 
2{\mbox Re} G_{++}^{\Lambda 2}(0;t-s)).\label{d2prev}\end{equation} 
Como los cuadrados de los propagadores son:
\begin{eqnarray}{\mbox Re} G_{++}^{\Lambda 2}(2k_0;t-s)&=& 
{\pi\over{k_0}}\left\{\int_\Lambda^{2k_0+\Lambda}dp\int_
\Lambda^{2k_0 + p}dz 
\cos [(p + z)s]\right.\nonumber \\
&&+ \left. \int_{2k_0+\Lambda}^{\infty}dp \int_{p-2k_0}^{p+2k_0}dz 
\cos [(p+z)s]\right\},\end{eqnarray} 
\begin{equation}{\mbox Re} G_{++}^{\Lambda 2}(0;t-s)= 
\pi\left\{2\pi \delta(s) - 2 {\sin (2 \Lambda 
s)\over{s}}\right\},\end{equation} 
el coeficiente de difusi\'on $D_2$ es
\begin{eqnarray}D_2(k_0;t) &=& {\pi\over{4}}\left\{3 \pi - ({3\over{2}} - 
{\Lambda\over{2k_0}}){\mbox Si}[2 t(\Lambda - k_0)]- (2 - 
{\Lambda\over{2 k_0}}) {\mbox Si}[2 \Lambda t] \right.\nonumber \\ && - 
\left.({3\over{2}} + 
{\Lambda\over{2k_0}}){\mbox Si}[2 t(\Lambda + k_0)] -(1 +
{\Lambda\over{2k_0}}){\mbox Si}[2t(2k_0 + \Lambda)]\right.\nonumber \\ 
&&+ \left.{\cos [2\Lambda 
t]\over{4k_0t}} - {\cos [2 t(\Lambda + k_0)]\over{4k_0t}}+{\cos [2 t(\Lambda - 
k_0)]\over{4k_0t}}-{\cos [2 
t(2k_0+\Lambda)]\over{4k_0t}}\right\},\label{d2}\end{eqnarray}
donde las funciones ${\mbox Si}[z]$ son las funciones 
seno-integral \cite{abra}. 


\section{P\'erdida de coherencia y perturbaciones cosmol\'ogicas}

En esta secci\'on analizamos el comportamiento de los coeficientes 
de difusi\'on $D_1$ y $D_2$. Luego, extendemos los resultados al 
espacio-tiempo de de Sitter para discutir la transici\'on cu\'antico-cl\'asica 
de las fluctuaciones cu\'anticas primordiales. 

Desarrollar un an\'alisis pormenorizado del proceso de 
p\'erdida de coherencia en el modelo que estamos considerando en 
este cap\'\i tulo es una tarea muy complicada. Realmente, uno deber\'\i a 
estudiar los elementos fuera de la diagonal de la matiz densidad 
reducida y mostrar si ellos desaparecen o no. Tambi\'en podr\'\i amos 
analizar la funcional de Wigner y determinar la aparici\'on o no de 
correlaciones cl\'asicas. Teniendo en mente los resultados obtenidos para 
el movimiento Browniano 
cu\'antico, nos concentraremos s\'olo en los t\'erminos de 
difusi\'on de la ecuaci\'on maestra. Tomaremos, en esta aproximaci\'on,  al 
valor de dichos coeficientes como se\~nal de la p\'erdida de coherencia 
sufrida por la matriz densidad reducida.

\subsection{Espacio-tiempo plano}

El coeficiente $D_1$ est\'a asociado al t\'ermino de interacci\'on 
$\phi_<^3\phi_>$. Como el campo entorno ($\phi_>$) aparece linealmente 
en esta interacci\'on, y s\'olo contiene modos con $k>\Lambda$ 
(por definici\'on), el campo sistema ($\phi_<$) s\'olo est\'a acoplado 
con el modo  $\vec k =3 \vec k_0$ del entorno. Esto implica que $D_1$ 
es diferente de cero s\'olo si ${\Lambda\over 3}<k_0<\Lambda$. En la Figura 3.1
 hemos graficado la evoluci\'on temporal de $D_1(k_0,t)$. Como puede 
verse en esta figura, los efectos difusivos inducidos por $D_1$ aumentan con 
el tiempo proporcionalmente a ${t\over k_0}$ (como estamos haciendo 
un c\'alculo perturbativo, ninguna estimaci\'on ser\'a v\'alida para 
tiempos muy largos). En la Figura 3.2 graficamos $D_1(k_0,t)$ como una 
funci\'on de $k_0$ a tiempo fijo. Aqu\'\i , la curva depende de $k_0$ y 
tiene su valor m\'aximo cuando $k_0={\Lambda\over 3}$. 

\begin{figure}[ht]
\centering \leavevmode
\epsfxsize=12cm
\epsfbox{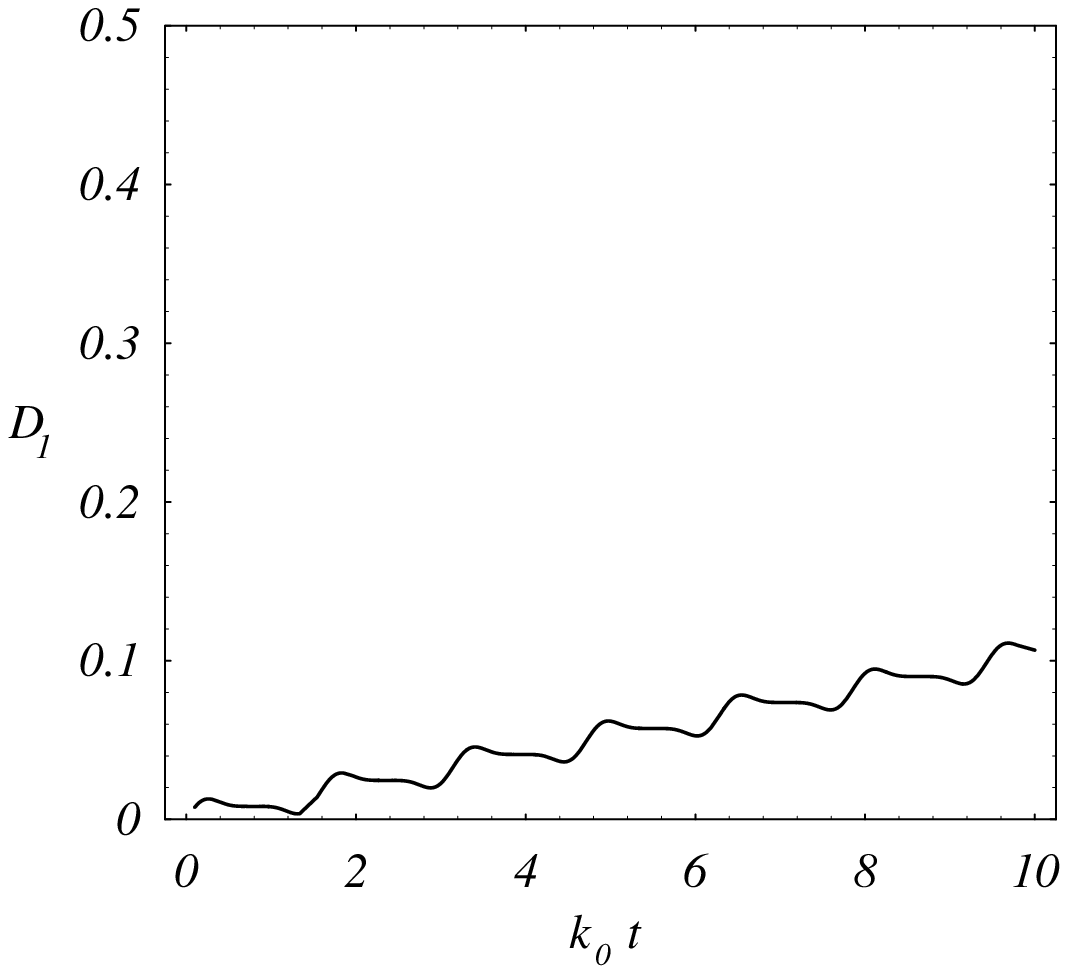}
\setlength{\captwidth}{12cm}
\capt{Evoluci\'on temporal del coeficiente de difusi\'on $D_1$ para 
${\Lambda\over{3}}<k_0<\Lambda$ fijo.}
\end{figure}

\begin{figure}[h]
\centering \leavevmode
\epsfxsize=12cm
\epsfbox{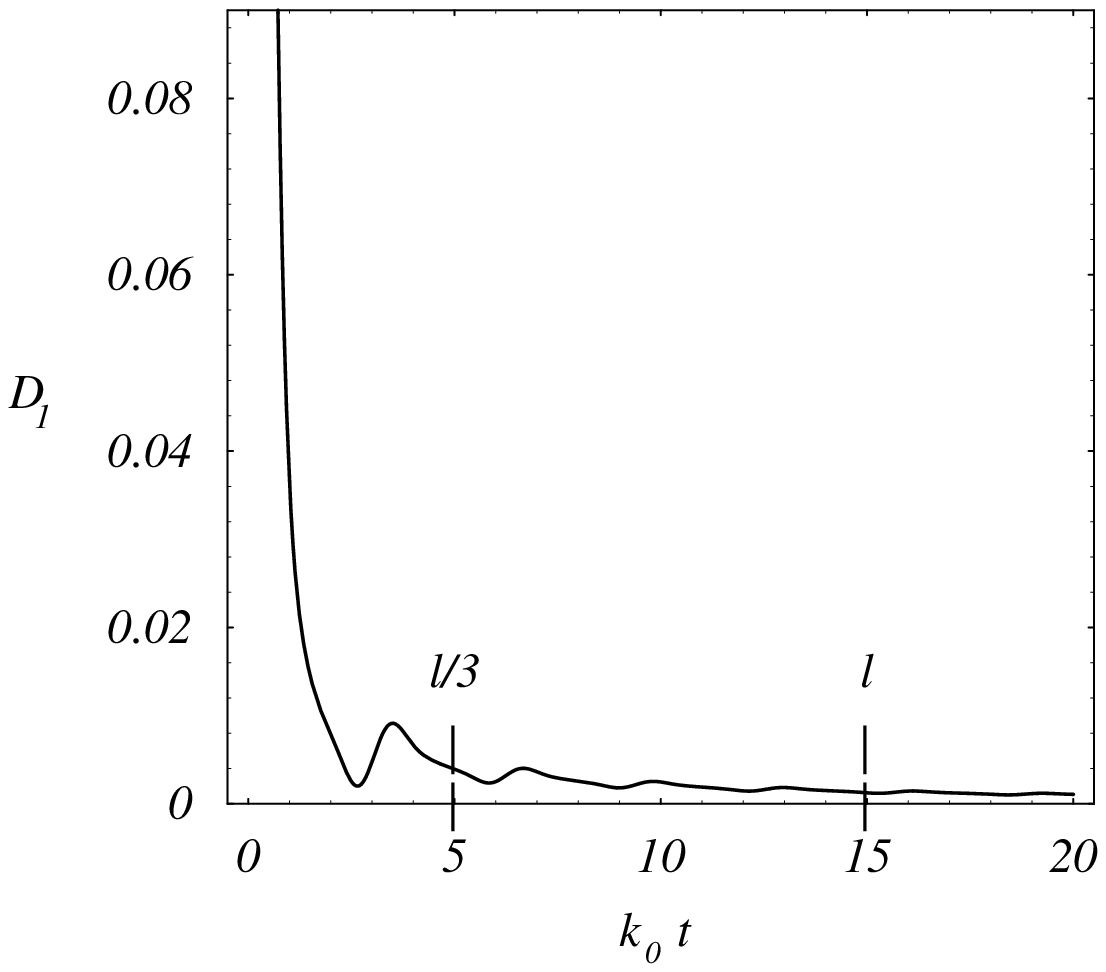}
\setlength{\captwidth}{12cm}
\capt{Coeficiente $D_1$ en funci\'on de $k_0$ para un valor fijo del tiempo. 
$D_1$ es diferente de cero s\'olo dentro de la ventana 
${\Lambda\over{3}}<k_0<\Lambda$. En la figura indicamos dicha ventana para 
un valor particular $l = \Lambda t = 15$.}
\end{figure}

Como analog\'\i a, es interesante notar que en el caso del movimiento 
Browniano cu\'antico con interacci\'on $x^3 q$ y densidad espectral 
$I(\omega )\sim \delta (\omega - 3k_0) \theta (\omega -\Lambda)$, se 
obtiene un coeficiente de difusi\'on similar \cite{hpz2}.

El coeficiente de difusi\'on $D_2$ proviene del t\'ermino de interacci\'on  
$\phi_<^2\phi_>^2$. Como la interacci\'on es cuadr\'atica en el campo 
entorno, no existe restricci\'on acerca de lo posibles valores de $k_0$ para 
que $D_2\neq 0$. En la Figura 3.3 graficamos la evoluci\'on temporal de 
este coeficiente para varios valores de $y={k_0\over\Lambda}$ (ahora 
con $y$ identificamos cada una de las curvas). Formalmente, $D_2$ deber\'\i a 
anularse cuando $t \rightarrow 0$ (ver ecuaci\'on (\ref{d2prev})), sin 
embargo el gr\'afico muestra un salto inicial para todo valor de la variable 
$y$. Una posible explicaci\'on para la aparici\'on de este salto reside en la 
condici\'on inicial de 
no-correlaci\'on. A $t_0 = 0$ estamos asumiendo que la interacci\'on se 
enciende instant\'aneamente, es decir, m\'as r\'apido que cualquier otra 
escala temporal presente en el sistema y en el entorno.  Esta condici\'on 
inicial presupone, impl\'\i citamente, la existencia de una 
frecuencia de corte ultravioleta $\Lambda_{\rm uv}$ para el entorno. Si 
uno considera tal $\Lambda_{\rm uv}$, $D_2$ se anula a $t_0 =0$ y 
desarrolla un pico durante una escala temporal del orden de     
$t\sim{1\over\Lambda_{\rm uv}}$. Sin embargo, el origen de este salto 
inicial es a\'un un tema de debate \cite{prom}, dado que se han encontrado 
estos saltos a\'un para condiciones iniciales que contienen correlaci\'on 
entre sistema y entorno.

\begin{figure}[h]
\centering \leavevmode
\epsfxsize=12cm
\epsfbox{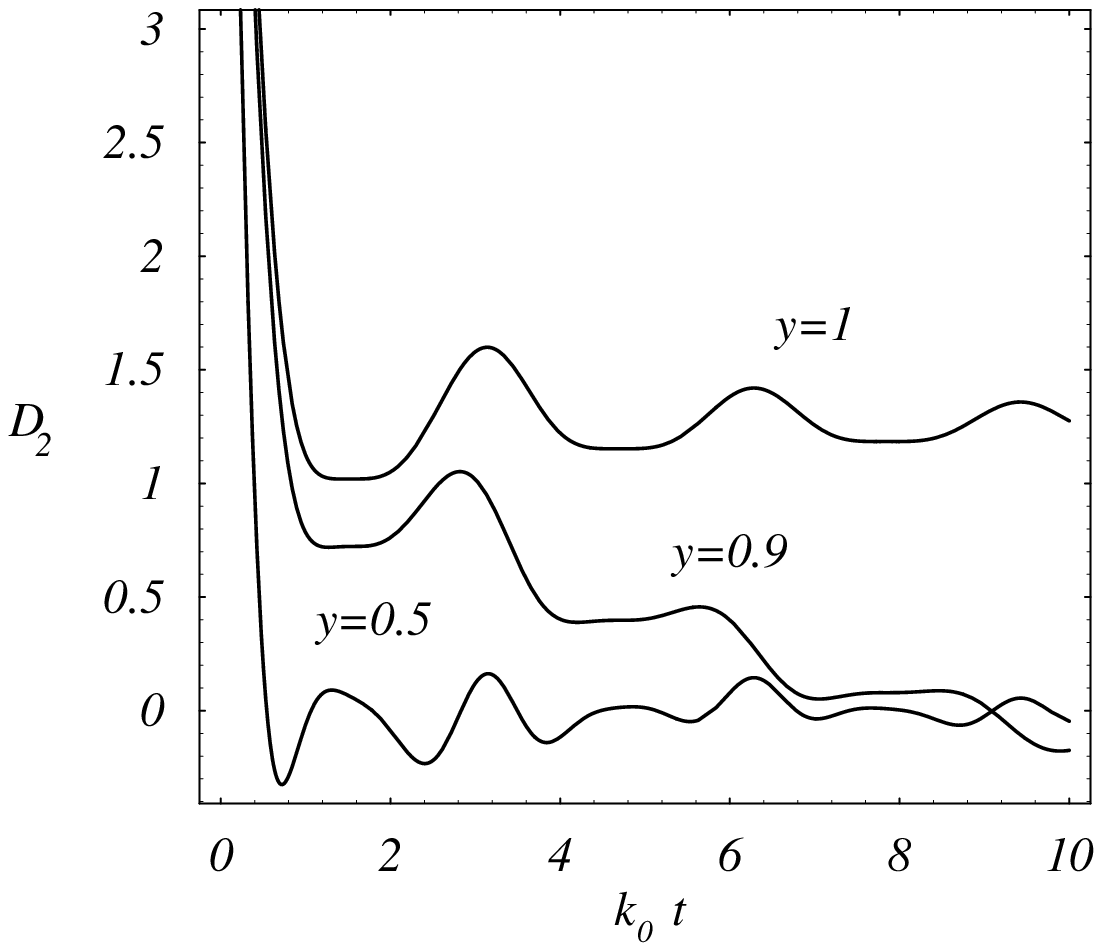}
\setlength{\captwidth}{12cm}
\capt{Evoluci\'on temporal del coeficiente de difusi\'on $D_2$ para 
diferentes valores de $y = k_0/\Lambda$.}
\end{figure}

Entonces, despu\'es del salto inicial, la Figura 3.3 muestra que 
el efecto difusivo crece con $y$, y es m\'aximo cuando $k_0\sim \Lambda$. 
La interpretaci\'on f\'\i sica de este hecho es clara. Podemos pensar que 
los efectos difusivos son debido a las part\'\i culas creadas por el 
entorno gracias a su interacci\'on con el sistema \cite{calzmazzi}. Cuando 
en el entorno existe un umbral de frecuencias, s\'olo aquellos modos del 
sistema con frecuencias cercanas a dicho umbral tienen la posibilidad 
de excitar al entorno; producir creaci\'on de part\'\i culas, y por lo 
tanto, de perder coherencia. Es por \'esto que el coeficiente de difusi\'on es 
m\'as importante cuando $k_0\sim\Lambda$.

Las Figuras 3.4 y 3.5 muestran la dependencia de $D_2$ con $k_0$ a tiempo 
fijo. 
Para valores grandes de $ l = \Lambda t$
$(\Lambda\rightarrow\infty)$, el entorno contiene s\'olo frecuencias muy 
grandes y el coeficiente de difusi\'on es muy chico, a menos que 
$k_0 \sim \Lambda$ (Figura 3.4). Cuando $l \leq 1$, $D_2$ es apreciablemente 
distinto de cero para todos los modos que pertenecen al sistema, dado que
 el entorno s\'olo puede producir p\'erdida de coherencia para modos 
$0<k_0<\Lambda$ (Figura 3.5). Finalmente, para  $l \ll 1$, $D_2(k_0,t)$ 
tiende a un valor no nulo en el l\'\i mite  $k_0<\Lambda\rightarrow 0$.

\begin{figure}[h]
\centering \leavevmode
\epsfxsize=12cm
\epsfbox{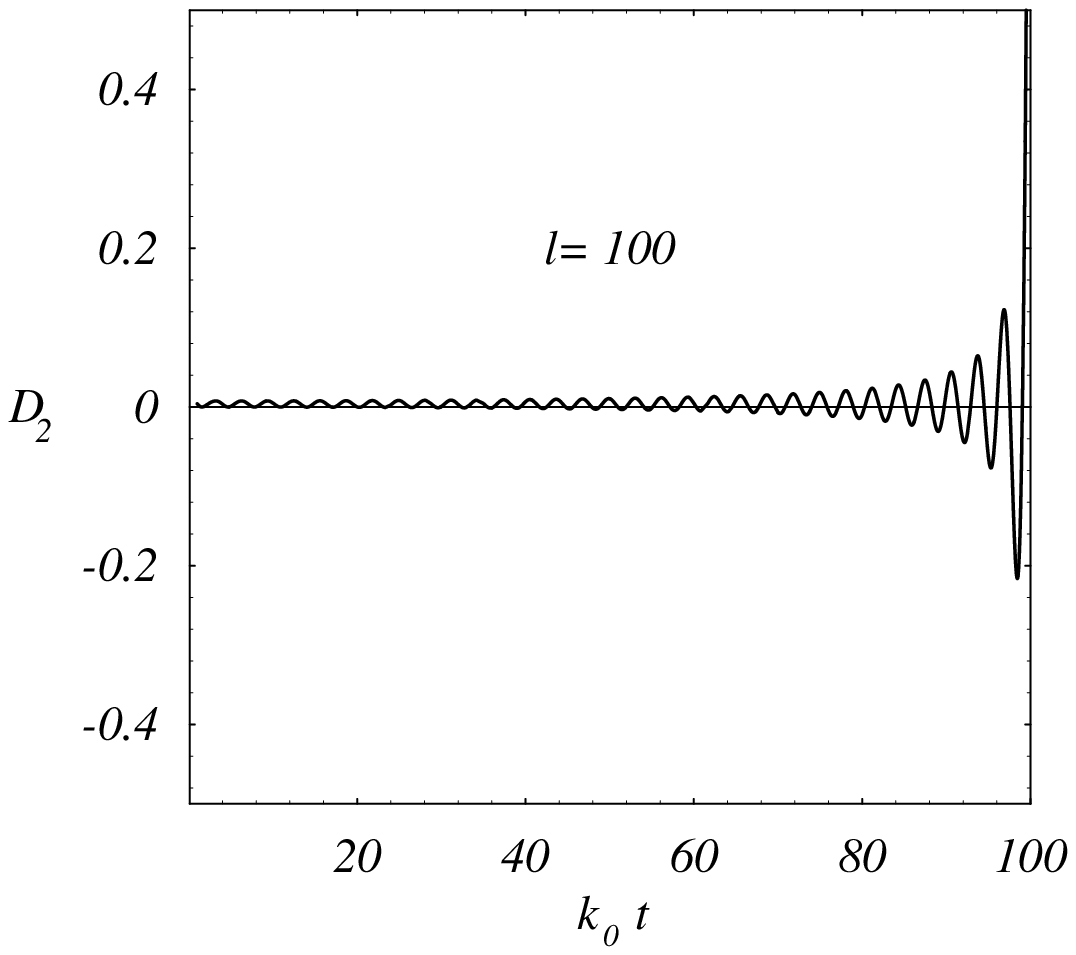}
\setlength{\captwidth}{12cm}
\capt{Coeficiente de difusi\'on $D_2$ en funci\'on de $k_0$ para un valor 
fijo del tiempo y $l = \Lambda t = 100$.}
\end{figure}

\begin{figure}[h]
\centering \leavevmode
\epsfxsize=12cm
\epsfbox{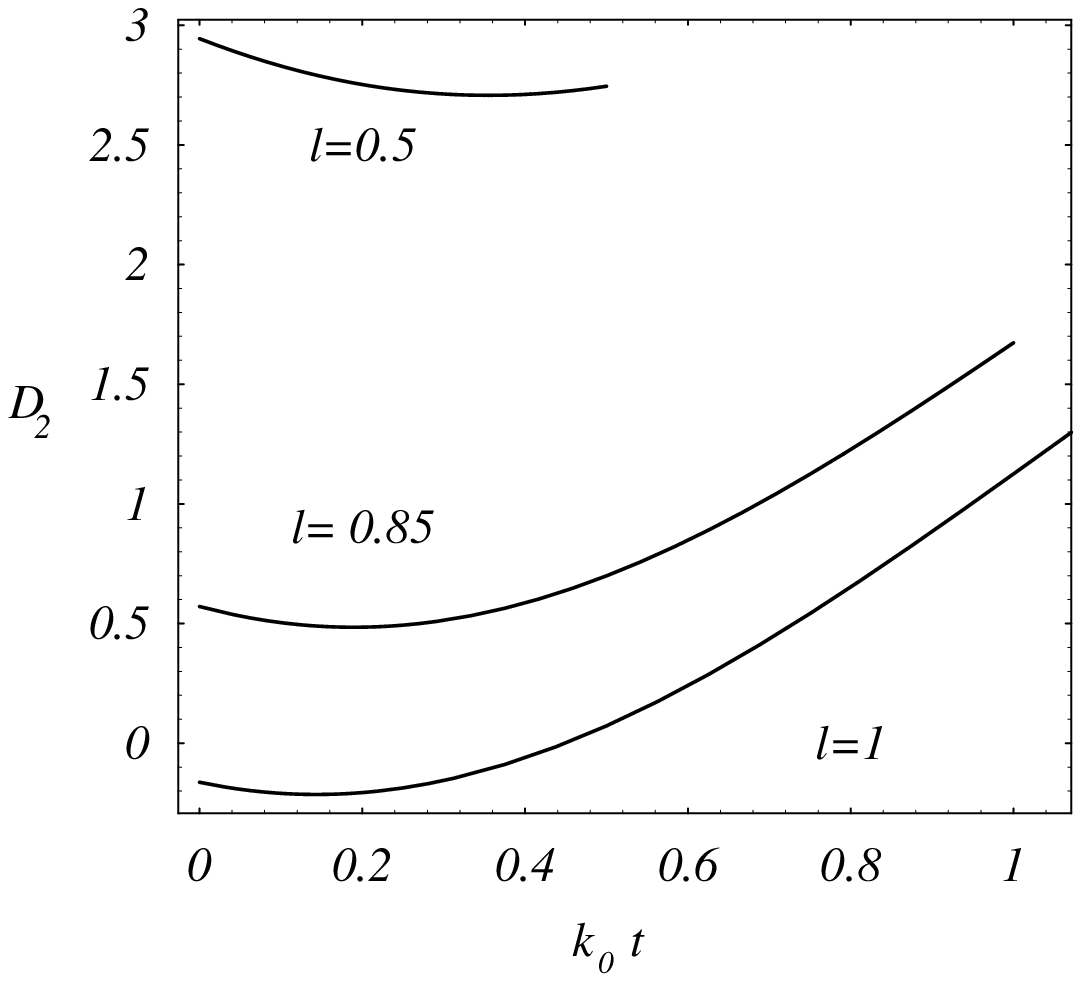}
\setlength{\captwidth}{12cm}
\capt{Misma figura que 3.4 pero para valores m\'as peque\~nos de $l$.}
\end{figure}

En el caso estudiado en la Ref. \cite{13prd53}, donde se consideraron 
dos campos escalares acoplados bi-cuadraticamente, se demostr\'o que los
 efectos difusivos eran m\'as importantes en el sector infrarojo. En 
nuestro ejemplo, el an\'alisis de $D_1$ y de $D_2$ muestra 
que la p\'erdida de coherencia es m\'as importante para $k_0\sim\Lambda$, 
en contraposici\'on con la conclusi\'on de \cite{13prd53}.


\subsection{Espacio-tiempo de de Sitter}

Como primer paso en hacia la comprensi\'on del proceso de p\'erdida de 
coherencia 
de las perturbaciones cosmol\'ogicas haremos una simple 
extensi\'on de los c\'alculos efectuados en las \'ultimas secciones. La 
acci\'on para un campo escalar con auto-interacci\'on es un espacio-tiempo 
curvo es 

\begin{equation}
S=\int d^4x \sqrt{-g}[{1\over 2}g^{\mu\nu}\partial_{\mu}\phi
\partial_{\nu}\phi-{1\over 2}m^2\phi^2-{\xi\over 2}R\phi^2
-{\lambda\over 4!}\phi^4]
\end{equation} 
donde $\xi$ representa el acoplamiento a la curvatura del espacio-tiempo. Para 
una m\'etrica de Robertson-Walker  $ds^2=dt^2-a^2(t)d\vec x^2=a^2(\eta)
[d\eta^2- d\vec x^2]$, esta acci\'on puede escribirse como

\begin{equation}
S=\int d^4x [{1\over 2}\eta^{\mu\nu}\partial_{\mu}\chi
\partial_{\nu}\chi-{1\over 2}m^2 a^2\chi^2-{1\over 2}(\xi-{1\over 6})R a^2 
\chi^2-{\lambda\over 4!}\chi^4]
\end{equation} 
donde $\chi =a\phi$.

Dada la acci\'on precedente, es directo notar que los resultados obtenidos 
en el espacio Minkowski pueden gene\-ralizarse a este caso cuando el campo 
escalar no tiene masa y est\'a acoplado de manera conforme ($\xi =1/6$). Por 
lo tanto, las expresiones de los coeficientes de difusi\'on $D_1$ y $D_2$ 
ser\'an las mismas que en el espacio plano, ecuaciones  
(\ref{d1}) y (\ref{d2}), reemplazando 
 $t\rightarrow \eta=\int_0^t {dt'\over a(t')}$. En el caso particular de la 
m\'etrica de de Sitter, $a(t)=\exp (Ht)$ y 
$\eta={1\over H }(1-{1\over a})$. De esta ma\-nera, nuestras variables 
adimensionales $l = \Lambda t$ y $k_0 t$ en las ecuaciones (\ref{d1}) y 
(\ref{d2})
pasan a ser $l = {\Lambda\over H }(1 - {1\over{a}})$
y
${k_0\over H}(1 - {1\over{a}})$ respectivamente. Si  $a(t=0)=1$, $\Lambda$ 
y $k_0$ son los valores f\'\i sicos iniciales de la frecuencia de corte y 
del vector de onda. 

Un modo del sistema pierde coherencia si los coeficientes de difusi\'on son 
diferentes de cero durante un tiempo apreciable. Por lo tanto la pregunta 
en el caso cosmol\'ogico pasa a ser: ?`cu\'al es el valor m\'aximo de 
$\Lambda$
 tal que, unos pocos instantes despu\'es del tiempo inicial, {\bf todos} los
 modos con $k_0\leq\Lambda$ sufren a\'un efectos difusivos? El valor de los 
coeficientes de difusi\'on a un dado tiempo depende de la variable $l$, la 
cual est\'a dada aproximadamente por $l=\Lambda/H$ (despu\'es del salto 
inicial). Existen dos posibilidades. Cuando $l\sim {\cal O}(1)$ (Figura 3.6), 
el an\'alisis hecho en el espacio plano muestra que ambos coeficientes, 
$D_1$ y $D_2$ son apreciablemente diferentes de cero para todos los posibles 
valores de $k_0$. Por otro lado, cuando  $l\gg 1$ la situaci\'on se vuelve 
totalmente diferente: $D_1$ es siempre muy peque\~no mientras que $D_2$ 
es chico en el sector infrarojo. Los efectos difusivos s\'olo importan cuando 
 $k_0\sim\Lambda$ (Figura 3.7). Como conclusi\'on podemos decir que para 
obtener p\'erdida de coherencia en todos los modos con $k_0<\Lambda$, s\'olo 
podemos incluir en nuestro sistema aquellos modos con longitud de onda 
mayor que $H^{-1}$. Esta es la propuesta original de A. Starobinsky 
\cite{14prd53}. Si inclu\'\i mos modos de longitud de onda menor que $H^{-1}$ 
en nuestro sistema, el umbral del entorno aumenta, y el sector infrarojo 
del sistema nunca puede exitar al entorno lo suficiente y, por lo tanto, 
no pierde coherencia.

\begin{figure}[h]
\centering \leavevmode
\epsfxsize=12cm
\epsfbox{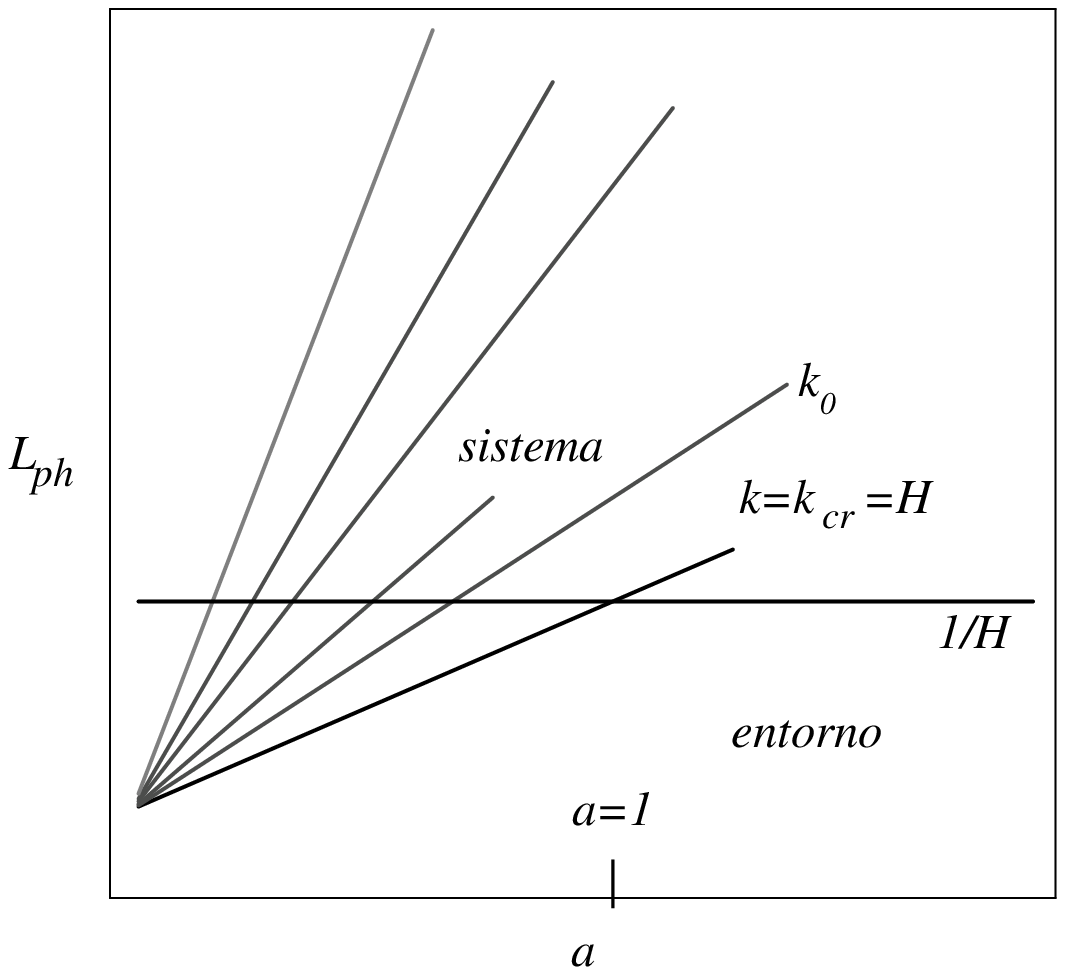}
\setlength{\captwidth}{12cm}
\capt{Longitudes de onda f\'\i sicas en funci\'on del factor de escala
 durante el per\'\i odo inflacionario. La l\'\i nea horizontal $L_{\rm ph} = 
H^{-1}$ corresponde al radio de Hubble. La l\'\i nea $k = 
k_{\rm cr} \equiv \Lambda$ divide sistema de entorno. A tiempo inicial 
($a = 1$), la longitud de onda cr\'\i tica es igual al radio de Hubble 
($k_{\rm cr} = H$). Todos los modos con $k_0 < k_{\rm cr}$ (l\'\i nea 
s\'olida) pierden coherencia.}
\end{figure}

\begin{figure}[h]
\centering \leavevmode
\epsfxsize=12cm
\epsfbox{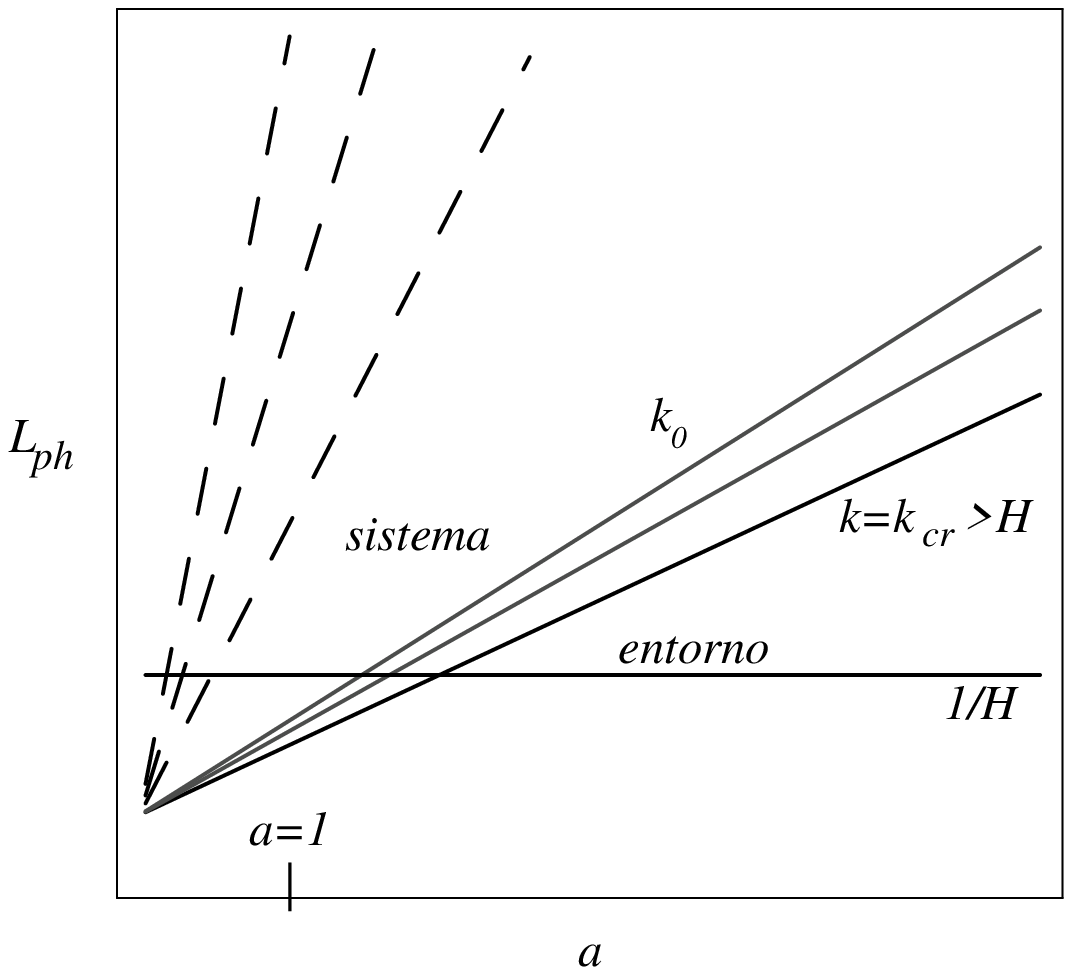}
\setlength{\captwidth}{12cm}
\capt{Lo mismo que en la figura 3.6, pero para el caso en que $k_{\rm cr}>H$. 
Despu\'es del tiempo inicial, los efectos difusivos son importantes
 s\'olo para los modos con $k_0 \sim k_{\rm cr}$ (l\'\i nea 
s\'olida). Los modos con $k_0 \ll k_{\rm cr}$ no pierden coherencia.}
\end{figure}
Para establecer una relaci\'on estrecha con los modelos inflacionarios 
estudiados en la literatura, es necesario extender los
c\'alculos al caso m\'\i nimamente acoplado ($\xi =0$). 
 Un c\'alculo m\'as realista, en ese caso, deber\'a incluir una frecuencia 
de corte 
$\Lambda$ que dependa del tiempo \cite{mataz}, dado que el sistema 
puede contener modos con $k_{ph}={k_0\over a}<H$ a cada tiempo. Sin 
embargo, el ejemplo discutido en esta secci\'on representa un modelo sencillo 
que ilustra los principales aspectos del problema. 
 

\section{Discusi\'on}

Como s\'\i ntesis de este cap\'\i tulo debemos se\~nalar que hemos
 obtenido la AEGG para los modos de frecuencia baja integrando los modos de 
alta frecuencia de un campo escalar con autointeracci\'on. A partir de la 
parte imaginaria de AEGG obtuvimos los coeficientes de difusi\'on de la 
ecuaci\'on maestra. Como sistema y entorno son dos sectores de un mismo campo 
escalar, los resultados dependen fuertemente del ``tama\~no'' de estos 
sectores, el cual est\'a determinado por la longitud de onda cr\'\i tica 
$\Lambda^{-1}$. 

Posteriormente hemos analizado la dependencia con $\Lambda$ de cada 
coeficiente. La p\'erdida de coherencia es efectiva 
para aquellos modos del sistema cuyas longitudes de onda asociadas sean 
del orden de la cr\'\i tica. Tambi\'en generalizamos los resultados a la 
m\'etrica de de Sitter mostrando que, cuando la longitud de onda cr\'\i tica 
es igual al radio de Hubble, todos los modos del sistema evidencian 
p\'erdida de coherencia y por lo tanto la transici\'on a lo cl\'asico 
se torna posible. 

Por otro lado, existe una importante 
conexi\'on entre nuestro trabajo, el grupo de renorma\-lizaci\'on y las 
transformaciones de bloques de esp\'\i n. La AEGG es similar a la 
acci\'on efectiva promediada (AEP) propuesta en Refs. 
\cite{21prd53,22prd53,23prd53}. La AEP es una acci\'on efectiva para 
promedios del campo sobre vol\'umenes finitos del espacio-tiempo Eucl\'\i deo. 
 Est\'a definida a partir de una longitud de onda cr\'\i tica, similar 
a la introducida en este cap\'\i tulo, dada por la relaci\'on Eucl\'\i dea
    $p_{\mu}p_{\mu}=\Lambda^{2}$, donde $\Lambda^{-1}$ es el tama\~no 
t\'\i pico del vol\'umen Eucl\'\i deo. Es posible escribir una ecuaci\'on de 
evoluci\'on exacta para la dependencia de
AEP con $\Lambda$. Esta ecuaci\'on fue discutida originalmente por Wegner y 
Houghton \cite{25prd53} y por  Polchinski \cite{26prd53} cuando prob\'o la 
renormalizaci\'on de la teor\'\i a $\lambda \phi^4$. La AEGG es la versi\'on 
de camino temporal cerrado de la AEP, donde el 
promedio sobre el campo se hace sobre vol\'umenes en tres dimensiones con una 
superficie de $t = {\mbox cte.}$ Evaluamos la AEGG en forma 
perturbativa, no obstante es posible obtener una ecuaci\'on exacta para 
la dependencia con $\Lambda$ de AEGG; y de esta forma poder evaluar de 
manera no perturbativa la acci\'on de influencia. Un primer intento 
en esta direcci\'on ha sido efectuado en la Ref. \cite{dalvitctp}.  
 
Para finalizar el cap\'\i tulo, deseamos mencionar una posible aplicaci\'on 
de la AEGG para el an\'alisis de transiciones de fase en teor\'\i a de 
campos. Cuando un campo sufre una transici\'on de fase con ruptura de la 
simetr\'\i a, desarrolla una estructura espacial de dominios que 
crece con el tiempo. El crecimiento de los dominios est\'a asociado a
 la din\'amica de los defectos topol\'ogicos que aparecen. Ahora bien, el 
objeto que contiene gran parte de la informaci\'on es el par\'ametro de 
orden de la transici\'on y est\'a representado por el valor promedio 
del campo sobre el tama\~no de cada dominio. Por lo tanto, es la AEGG 
la herramienta indicada para estudiar las transiciones de fase, incluyendo 
 el proceso de transici\'on 
cu\'antico-cl\'asico del par\'ametro de orden y la formaci\'on de 
defectos topol\'ogicos \cite{rivers,vorti}. 

%% file: tesis6
\newpage

\thispagestyle{empty}
~
\newpage

\chapter{La acci\'on de influencia en gravedad semicl\'asica}

\thispagestyle{empty}

Pese a todos los intentos por lograr una descripci\'on unificada 
de las interacciones fundamentales de la Naturaleza, la gravitatoria 
contin\'ua siendo la interacci\'on m\'as problem\'atica de vincular con las 
dem\'as en una manera consistente. Por lo tanto, ante la ausencia de 
una teor\'\i a cu\'antica de la gravitaci\'on, una pregunta importante es: 
?`podemos decir algo acerca de la influencia de los campos cu\'anticos de 
materia sobre la geometr\'\i a cl\'asica del espacio-tiempo? Con el 
advenimiento de la mec\'anica cu\'antica, 
muchos c\'alculos fueron efectuados asumiendo al campo electromagn\'etico 
como un campo cl\'asico de fondo, interactuando con materia cu\'antica. Bajo 
ciertas circunstancias esta 
aproximaci\'on semicl\'asica produjo resultados 
en total acuerdo con la posterior electrodin\'amica cu\'antica. En 
consecuencia, uno puede esperar que 
un r\'egimen similar exista para los aspectos cu\'anticos de la gravedad, en 
el cual el campo gravitatorio es considerado como un campo cl\'asico de fondo,
 mientras que los campos de materia est\'an cuantizados de la manera usual. 

A partir de la propuesta de J. Hartle y S. Hawking para la 
funci\'on de onda del Universo \cite{hh}, la validez de la aproximaci\'on 
semicl\'asica ha sido ampliamente estudiada, principalmente en modelos 
cosmol\'ogicos de {\it minisuperespacio} en cuatro dimensiones. Como 
mencionamos en la Introducci\'on, el l\'\i mite semicl\'asico est\'a basado 
en dos ingredientes principales: correlaciones y p\'erdida de coherencia
 \cite{pazsinha}. Al igual que en el caso del movimiento Browniano cu\'antico, 
las correlaciones entre las diferentes variables fue analizada por medio de la
funcional de Wigner \cite{hallicorr}, mientras que la p\'erdida de coherencia 
fue estudiada a partir de la matriz densidad reducida \cite{hallideco}. Un 
exceso en la p\'erdida de coherencia puede producir ausencia en las 
correlaciones cl\'asicas \cite{8if}, por lo tanto ambos aspectos no son 
manifestaciones independientes. Posteriormente, la transici\'on 
cu\'antico-cl\'asica fue analizada utilizando la funcional de p\'erdida de 
coherencia de J. Hartle y M. Gell-Mann \cite{9if}, la cual es una 
funcional de dos historias cl\'asicas de la geometr\'\i a 
${\cal D}[g_{\mu\nu}^+,g_{\mu\nu}^-]$, despu\'es de haber integrado las 
variables cu\'anticas. La m\'etrica $g_{\mu\nu}$ puede 
considerarse como cl\'asica si la funcional de p\'erdida de coherencia es 
diagonal (o aproximadamente diagonal). En ese caso, las probabilidades 
para diferentes historias satisfacen la regla cl\'asica de suma de 
probabilidades, y entonces los t\'erminos de interferencia cu\'antica son 
despreciables.

Si bien no contamos con una teor\'\i a cu\'antica de la gravitaci\'on, podemos 
 obtener informaci\'on acerca de la transici\'on cu\'antico-cl\'asica. De la 
misma manera que en el caso de la part\'\i cula Browniana cu\'antica, se 
pueden conocer los efectos de las fluctuaciones cu\'anticas del entorno sobre 
la part\'\i cula, siendo \'esta de naturaleza cl\'asica. 
Siguiendo nuestra formulaci\'on de la funcional de influencia para sistemas 
cu\'anticos abiertos, la m\'etrica del espacio-tiempo $g_{\mu\nu}$ juega 
el rol del ``sistema'', y los campos cu\'anticos del ``entorno''. De esta 
manera, integrando s\'olo los grados de libertad del entorno podemos 
obtener informaci\'on relevante acerca de la disipaci\'on, el ruido y la 
transici\'on cu\'antico-cl\'asica que sufre el sistema. Como veremos en este 
cap\'\i tulo, la funcional de influencia est\'a relacionada con la funcional 
de p\'erdida de coherencia de Hartle y Gell-Mann.      

El efecto de los campos cu\'anticos sobre la geometr\'\i a del 
espacio-tiempo constituye un problema fundamental en el marco de 
las teor\'\i as semicl\'asicas. La manera de 
estudiar estos efectos cu\'anticos de los campos sobre la geometr\'\i a 
(que suponemos como cl\'asica) est\'a basada en el estudio de las ecuaciones 
de Einstein semicl\'asicas \cite{bd}

\be
{1\over{8\pi G}} G_{\mu\nu}= T_{\mu\nu}^{\rm clas}+ 
\langle T_{\mu\nu}\rangle ,\label{see}\ee
donde $G_{\mu\nu}$ es el tensor de Einstein; $T_{\mu\nu}^{\rm clas}$ es el 
tensor de energ\'\i a-impulso cl\'asico y con el valor medio 
$\langle T_{\mu\nu}\rangle$ incorporamos los efectos de los campos 
cu\'anticos. 

El estudio de estas ecuaciones es fundamental para entender procesos tales 
como el estado final de la evaporaci\'on de agujeros negros y  problemas de 
p\'erdida de informaci\'on; o para entender el efecto de los 
campos cu\'anticos sobre las singularidades cl\'asicas de la relatividad 
general. 

El an\'alisis de las ecuaciones de Einstein semicl\'asicas en modelos 
realistas es una tarea muy complicada. Por esta raz\'on es importante 
formular modelos simplificados donde tales dificultades no se manifiesten, 
pero 
que provean de resultados que permitan inducir los efectos de los modelos 
generales. Los modelos de gravedad escalar-tensorial en dos dimensiones son 
muy 
\'utiles en este sentido. A partir de estos modelos es posible entender 
algunos de los principales problemas referidos a las propiedades cu\'anticas 
de los agujeros negros y la influencia de los efectos cu\'anticos en 
situaciones cosmol\'ogicas \cite{todo2d}. 

Estas ecuaciones semicl\'asicas no proveen una 
descripci\'on completa del 
problema, por ejemplo cuando el estado de los campos cu\'anticos es tal 
que el tensor de energ\'\i a-impulso presenta importantes fluctuaciones 
alrededor de su valor medio \cite{ford}, motivo por el cual en el 
Cap\'\i tulo 5 estudiaremos las correcciones estoc\'asticas a (\ref{see}) que 
deben incluirse en los modelos m\'as realistas.

En el presente cap\'\i tulo estudiaremos fundamentalmente cu\'al en el 
dominio de aplicaci\'on de la aproximaci\'on semicl\'asica y en qu\'e 
circunstancias dicha aproximaci\'on puede justificarse. En este contexto, 
los modelos en dos dimensiones proveen importantes simplificaciones para 
el c\'alculo de la funcional de influencia en gravedad semicl\'asica.
La organizaci\'on de este cap\'\i tulo puede resumirse de la siguiente 
manera: en primer lugar presentamos el modelo bi-dimensional de 
Callan-Giddings-Harvey-Strominger; derivamos las ecuaciones de movimiento 
cl\'asicas y calculamos la acci\'on efectiva exacta. Seguidamente
 mostramos c\'omo obtener la acci\'on efectiva de camino temporal cerrado 
 para este modelo 
a partir de la expresi\'on de la acci\'on efectiva Eucl\'\i dea. Deducimos 
las ecuaciones de movimiento de manera covariante, evaluando la contribuci\'on 
 de las correcciones cu\'anticas aportadas por los campos de materia. 
Analizamos la 
radiaci\'on de Hawking para m\'etricas de agujeros negros. Posteriormente 
estudiamos modelos donde el dilat\'on est\'a 
acoplado a los campos de materia, re-analizando los problemas 
antes mencionados, destacando resultados f\'\i sicos importantes. Finalmente 
estudiamos la transici\'on cu\'antico-cl\'asica 
a partir del c\'alculo de la funcional de influencia.  

\section{El modelo de Callan-Giddings-Harvey-Strominger}

Los trabajos de S. Hawking acerca de la evaporaci\'on de 
agujeros negros \cite{haw1,haw2} sugieren que el proceso de formaci\'on y 
subsiguiente evaporaci\'on de un agujero negro no est\'a gobernado por las 
leyes de la mec\'anica cu\'antica ordinaria: estados puros podr\'\i an 
evolucionar en estados mixtos. Esta conjetura es dif\'\i cil de probar 
debido a la 
gran cantidad de grados de libertad y a la complejidad del espacio-tiempo en
 cuatro dimensiones. Un modelo donde se pueda controlar anal\'\i ticamente 
todos los grados de libertad es de mucha utilidad. C. Callan, S. Giddings, 
J. Harvey y A. Strominger (CGHS) \cite{cghs} propusieron 
un modelo bi-dimensional inspirado en teor\'\i as efectivas de cuerdas 
a bajas energ\'\i as \cite{witten} que despert\'o gran inter\'es. El modelo 
(CGHS) es  
un modelo bi-dimensional de la gravedad, acoplado a un campo escalar llamado 
el {\it dilat\'on} y a N campos conformes $f_i$, que contiene soluciones de 
agujero negro y radiaci\'on de Hawking. En este modelo puede calcularse 
el efecto completo de los campos cu\'anticos 
$f_i$ sobre la geometr\'\i a del espacio-tiempo, y en consecuencia el problema 
semicl\'asico puede resolverse exactamente \cite{rst,bpp1}. CGHS propusieron 
la siguiente acci\'on cl\'asica:
\be S_{\rm CGHS}= {1\over{2 \pi}} \int d^2x \sqrt{-g(x)} 
\left\{e^{-2\phi}\left[R + 4 (\partial \phi)^2 + 4 \lambda^2\right] - 
{1\over{2}} \sum_{i = 1}^{N} (\partial f_i)^2\right\},\label{cghs}\ee
donde $\phi$ es el dilat\'on, $R$ es el escalar de Ricci en dos dimensiones, 
$\lambda$ es una constante positiva y los $f_i$ son N 
campos escalares no-masivos, acoplados conformemente a la geometr\'\i a 
bi-dimensional. Esta acci\'on representa una acci\'on efectiva que describe 
a los modos radiales de agujeros negros dilat\'onicos extremales en 
cuatro o m\'as dimensiones, y tambi\'en est\'a relacionada a la acci\'on  de 
cuerdas no cr\'\i ticas a bajas energ\'\i as. Independientemente de su 
motivaci\'on, es un modelo 
simplificado interesante para estudiar el problema en cuesti\'on.

\subsection{Ecuaciones cl\'asicas de movimiento}

Las ecuaciones de movimiento para cada uno de los campos de la teor\'\i a: 
$g^{\mu\nu}$, $\phi$ y $f$ se obtienen tomando la variaci\'on funcional de 
la acci\'on respecto de cada campo. Definiendo nuevas coordenadas,   

\be x^+ = t + x ~~, ~~~~~~ x^- = t - x,\ee
las ecuaciones de movimiento son

\be T_{+-} = {\delta S_{\rm CGHS}\over{\delta g_{+-}}}=0,\ee
\be T_{++}= {\delta S_{\rm CGHS}\over{\delta g_{++}}}=0,\ee
\be T_{--}= {\delta S_{\rm CGHS}\over{\delta g_{--}}}=0.\ee  

En dos dimensiones (1+1) siempre es posible elegir la medida conforme
\be g_{\mu\nu} = e^{2\rho}\eta_{\mu\nu},\ee
donde $\eta_{\mu\nu}$ es la m\'etrica del espacio plano de Minkowski 
(con signatura ($-$,+)) y el factor $e^{2\rho}$ se conoce como el {\it factor 
conforme} o factor de Weyl; el intervalo puede escribirse como

\be ds^2 = e^{2\rho}[-dt^2 + dx^2] = - e^{2\rho}dx^+dx^-.\ee
Expl\'\i citamente, las ecuaciones de 
movimiento, en esta medida son:

\be T_{\pm\pm} = e^{-2\phi}(4\partial_{\pm}\rho\partial_{\pm}\phi - 2 
\partial^2_{\pm}\phi )+ {1\over{2}}\partial_{\pm}\partial_{\pm}f=0,
\label{217}\ee
\be T_{+-} = e^{-2\phi}(2\partial_{+}\rho\partial_{-}\phi - 4 
\partial_{+}\phi\partial_-\phi - \lambda^2 e^{2\rho})= 0,\label{218}\ee
\be -4 \partial_+\partial_-\phi + 4\partial_+\phi\partial_-\phi + 2\partial_+
\partial_-\rho + \lambda^2 e^{2\rho}=0,\label{219}\ee
\be \partial_+\partial_-f = 0.\label{2110}\ee

Sumando miembro a miembro las ecuaciones ({\ref{218}) y (\ref{219}) se obtiene

\be \partial_+\partial_-(\rho - \phi) = 0,\label{2111}\ee
por lo tanto $\rho - \phi$ es un campo libre y, en 
consecuencia, 

\be \rho - \phi = {1\over{2}}W_+(x^+) + {1\over{2}}W_-(x^-)\equiv {1\over{2}}W.
\label{2112}\ee
 
Teniendo en cuenta (\ref{2111}) podemos escribir:

\be \partial_+\partial_-e^{-2\phi}+ \lambda^2e^{2(\rho - \phi)}=0,\ee
de donde se deduce que
\be \partial_+\partial_-e^{-2\phi}= - \lambda^2e^{2(\rho - \phi)} = 
- \lambda^2 e^{(W_+ + W_-)}.\ee

Integrando esta ecuaci\'on obtenemos, 

\be e^{-2\phi}= - \lambda^2\int dx^- e^{W_-(x^-)}\int dx^+ e^{W_+(x^+)} + 
u_+(x^+) + u_-(x^-),\label{2113}\ee
donde $u_+$ y $u_-$ son funciones a determinar a partir de las ecuaciones 
(\ref{217}) y (\ref{2110}). Definiendo:

\be u = u_+(x^+)+ u_-(x^-),\ee
\be h_\pm(x^{\pm})= \lambda \int^{x^{\pm}} dx^{\pm} e^{W_{\pm}},\ee
se puede obtener formalmente que 

\be e^{-2\phi} = u - h_+h_-,\label{2116}\ee
y en consecuencia

\be e^{-2\rho} = e^{-W}( u - h_+h_-).\label{2117}\ee

Para completar la soluci\'on, es necesario conocer las funciones $u_+$ y 
$u_-$. La ecuaci\'on (\ref{2110}) para los campos de materia implica que

\be f = f_+(x^+) + f_-(x^-),\ee
por lo que utilizando las ecuaciones de v\'\i nculo (\ref{217}), y a partir 
de (\ref{2116}) y (\ref{2117}) encontramos que:

\be u_+= {M\over{2\lambda}}- {1\over{2}}\int dx^+ e^{W_+}\int dx^- e^{-W_+}
\partial_+f\partial_+f,\label{2121}\ee
\be u_-= {M\over{2\lambda}}- {1\over{2}}\int dx^+ e^{W_-}\int dx^- e^{-W_-}
\partial_-f\partial_-f.\label{2122}\ee
En ambas ecuaciones, el t\'ermino ${M\over{2\lambda}}$ es una constante de 
integraci\'on cuyo significado quedar\'a claro en lo que sigue de este cap\'\i 
tulo. 

Consideremos primero las soluciones de vac\'\i o, es decir, con $f = 0$. En 
este caso es trivial notar que 

\be u = u_+ + u_- = {M\over{2\lambda}},\ee
y que, adem\'as
\be h_\pm = \lambda x_\pm,\ee
(a menos de traslaciones constantes en $x^+$ y $x^-$). Fijando completamente 
la medida, es decir eligiendo $W=0$, el campo de Liouville $\rho$ (que 
frecuentemente llamamos m\'etrica) y el dilat\'on pueden escribirse como

\be e^{-2\rho} = e^{-2\phi}= {M\over{\lambda}}- \lambda^2 x^+x^-,\label{2123}
\ee
y el elemento de l\'\i nea es

\be ds^2 = {dx^+ dx^-\over{(\lambda^2 x^+x^- - {M\over{\lambda}})}}.
\label{2124}\ee

La soluci\'on (\ref{2123}) con $M\not= 0$ corresponde a la soluci\'on de 
agujero negro bi-dimensional encontrada por Witten. $M$ resulta ser la masa 
del agujero negro \cite{witten}. La m\'etrica (\ref{2123}) puede llevarse a 
una forma manifiestamente est\'atica mediante el siguiente cambio de 
coordenadas:

\be \lambda x^+ = e^{\lambda \sigma^+}= e^{\lambda(\tau + \sigma)},\ee
\be \lambda x^- =- e^{-\lambda \sigma^-}= -e^{-\lambda(\tau - \sigma)},\ee   
de donde es simple obtener:

\be ds^2 = (-d\tau^2 + d\sigma^2)\left[{1\over{1+{M\over{\lambda}}
e^{-2\lambda\sigma}}}\right].\label{2126}\ee

En el caso particular en que $M=0$, la m\'etrica es plana y el dilat\'on 
resulta lineal en la coordenada espacial,

\be \phi = - {1\over{2}}\ln [-\lambda^2x^+x^-]= -{1\over{2}}\lambda(\sigma^+
-\sigma^-)=-\lambda\sigma,\ee
a esta soluci\'on con $M=0$ se la conoce como ``vac\'\i o de dilat\'on 
lineal''. Cuando $M\not= 0$ la m\'etrica (\ref{2126}) representa a la de un 
agujero negro de masa $M$ \cite{cghs}. El horizonte est\'a en el l\'\i mite 
$\sigma \rightarrow -\infty$.

Para poder estudiar la formaci\'on y eventual evaporaci\'on de un agujero 
negro es necesario considerar aquellas soluciones con $f\not= 0$. Para ello 
necesitamos considerar el tensor de energ\'\i a-impulso de los campos de 
materia, definido por:

\be T^f_{\mu\nu}={2\over{\sqrt{g}}}{\delta S^f\over{\delta g_{\mu\nu}}},\ee
donde $S^f$ es la parte de la acci\'on CGHS que corresponde a los campos de 
materia $f$. En la medida conforme, 

\be T^f_{\pm\pm}= \partial_\pm f \partial_\pm f,\ee
\be T^f_{+-} = T^f_{-+}=0.\ee

Del resultado anterior para la soluci\'on de vac\'\i o es de esperar que 
una perturbaci\'on debida a la presencia de campos de materia resulte 
en la formaci\'on de un agujero negro. Por lo tanto es instructivo considerar 
una onda de choque (o pulso de materia) de amplitud $a$ viajando en 
la direcci\'on $x^-$, descripta por el tensor de 
energ\'\i a-impulso:

\be {1\over{2}} \partial_+f\partial_+f = a\delta (x^+ - x_0^+).\ee

Nuevamente, eligiendo la medida a partir de $W=0$, y utilizando las 
ecuaciones anteriores, es f\'acil probar que 

\be u_+''= - a \delta (x^+ - x_0^+),\ee
donde las primas denotan derivadas respecto de $x^+$. Integrando 
dos veces obtenemos

\be u_+= -a (x^+ - x_0^+)\theta (x^+ - x_0^+).\ee
La ecuaci\'on an\'aloga para $u_-$ implica que $u_-= \alpha x_- + \beta$. 
As\'\i , la m\'etrica y el dilat\'on quedan dados por la siguiente expresi\'on:

\be e^{-2\phi}=e^{-2\rho}=-a(x^+ - x_0^+)\theta (x^+ - x_0^+)-\lambda^2x^+x^-.
\label{2133}\ee

Es importante notar que para $x^+ < x_0^+$ se re-obtiene el vac\'\i o de 
dilat\'on lineal, mientras que para $x^+ > x_0^+$ la soluci\'on es la 
correspondiente a la de un agujero negro de masa $ax_0^+\lambda$ (luego de
redefinir a $x^-$ por $x^- - {a\over{\lambda^2}}$). En conclusi\'on, 
cualquier onda de materia incidente en gravedad dilat\'onica cl\'asica 
produce un agujero negro con horizonte y singularidad.

Para demostrar esta \'ultima afirmaci\'on, calculamos expl\'\i citamente el 
escalar de curvatura, que en el modelo bi-dimensional siempre est\'a 
dado en t\'erminos de derivadas del factor de Liouville (en la medida
 conforme).

\be R = 8 e^{-2\rho}\partial_+\partial_-\rho.\label{R}\ee
Luego, utilizando la ecuaci\'on (\ref{2133}) el escalar $R$ puede escribirse 

\be R= 4\left[{u_+'u_-' + \lambda^2(u - x^+u_+' - x^- u_-')
\over{u - \lambda^2 x^+x^-}}\right],\label{2134}\ee
donde podemos notar que el escalar de curvatura diverge a lo largo de
 la curva $u - \lambda^2 x^+x^-= 0$. 

La eventual existencia de horizontes puede ponerse de manifiesto por 
medio de un sistema de coordenadas en el cual la m\'etrica es 
asint\'oticamente plana. Haciendo el siguiente cambio de coordenadas:

\be e^{\lambda \sigma^+}= \lambda x^+,\ee
\be e^{-\lambda \sigma^-}=-\lambda x^--{a\over{\lambda}},\ee
la m\'etrica toma la forma
\bea
e^{2\rho} = \left\{\begin{array}{ll} 
\left[1+{a\over{\lambda}}e^{\lambda \sigma^-}\right]^{-1} & ~~ 
\mbox{si} ~ \sigma^+ < \sigma_0^+ \\ 
\left[1+{a\over{\lambda}}e^{\lambda (\sigma^-- \sigma^+ + \sigma_0^+)}
\right]^{-1} & ~~ 
\mbox{si} ~ \sigma^+ > \sigma_0^+
\end{array}\right. \label{2136}\eea 
de donde es f\'acil ver que en coordenadas $(\sigma^+,\sigma^-)$ el factor 
conforme $e^{2\rho}$ se anula cuando $\sigma^- \rightarrow \infty$, que al 
pasar a las coordenadas $(x^+,x^-)$ implica la existencia de un horizonte 
sobre la curva $x^- = - {a\over{\lambda^2}}$ \cite{cghs,witten}.

\subsection{Efectos cu\'anticos: la acci\'on efectiva Eucl\'\i dea}

Para poder incluir los efectos cu\'anticos debido a los campos de 
materia $f$, debemos calcular las correcciones que aparecen en el 
tensor de energ\'\i a-impulso. Es decir, podemos descomponer al tensor 
en una contribuci\'on cl\'asica y una cu\'antica:

\be T_{\mu\nu} = T^{\rm clas}_{\mu\nu} + \langle T_{\mu\nu}\rangle .\ee
Donde $\langle T_{\mu\nu}\rangle$ proviene de la variaci\'on funcional de 
la correspondiente acci\'on efectiva definida a partir de 

\be e^{-S_{\rm ef}^{\rm E}}= {\mbox det}[-\Box ]^{-{1\over{2}}} = {\cal N}
\int {\cal D}f e^{-{1\over{2}}\int d^2x f (-\Box ) f},\ee
debido a que estamos integrando funcionalmente sobre campos escalares 
no-masivos m\'\i nimamente acoplados en dos dimensiones.

Debido a la simetr\'\i a conforme, la traza del tensor de energ\'\i a-impulso 
de los campos escalares se anula cl\'asicamente, pero existe a nivel 
cu\'antico una anomal\'\i a. 
En dos dimensiones, debido a que 
todas las m\'etricas son conformemente planas, una 
teor\'\i a de campos no-masivos libre ser\'\i a trivial sin anomal\'\i as 
conformes. Por otro lado, como la m\'etrica bi-dimensional tiene s\'olo una 
componente desconocida, la acci\'on efectiva est\'a univocamente determinada 
por la anomal\'\i a de traza de su variaci\'on funcional (tensor de 
energ\'\i a-impulso). Como la traza del tensor de energ\'\i a-impulso es 
proporcional a $R$ (que determina completamente al tensor de Riemann en 
dos dimensiones), la contribuci\'on a la acci\'on efectiva 
a partir de un campo no-masivo en dos dimensiones puede 
conocerse de manera {\it exacta}. En consecuencia, como 

\be T_\mu^\mu = g^{\mu\nu}T_{\mu\nu} = {2\over{\sqrt{g}}}g^{\mu\nu}
{\delta S_{\rm ef}\over{\delta g_{\mu\nu}}}={R\over{24 \pi}},
\label{2dtraceano}\ee
en la medida conforme podemos 
obtener la acci\'on efectiva debida a los campos de materia cu\'anticos 
integrando la ecuaci\'on (\ref{2dtraceano}):

\be S_{\rm ef}^{\rm E}=  S_{\rm CGHS}^{\rm E} - {N\over{96 \pi}}\int 
d^2x \sqrt{g(x)}\int d^2x' \sqrt{g(x')} R(x) {1\over{\Box}} 
R(x'). \label{effact}\ee Esta es la conocida acci\'on de Polyakov. 

En las secciones siguientes del presente cap\'\i tulo haremos un an\'alisis 
detallado del tensor de energ\'\i a-impulso asociado a esta acci\'on efectiva 
y de las ecuaciones de movimiento obtenidas a partir de \'el. Pero ahora 
centraremos nuestra atenci\'on en el c\'alculo del tensor de 
energ\'\i a-impulso de los campos de materia.

Calculando la variaci\'on funcional de (\ref{effact}) podemos 
obtener la contribuci\'on cu\'antica al tensor de energ\'\i a-impulso:
\begin{eqnarray}
\langle T_{\mu\nu}\rangle &=& -{1\over{24\pi}}  
\int d^2y \sqrt{g} \left[ \nabla_\mu \nabla_\nu - 
g_{\mu\nu} \Box \right]_{(x)} R(y) {1\over{\Box}} \nonumber \\
&+& {1\over{96\pi}} \int d^2x \sqrt{g}\int d^2y \sqrt{g}~\left\{
- g_{\mu\nu}\partial^\alpha {R(x)\over{\Box}}\partial_\alpha {R(y)\over{\Box}}
+ 2 \partial_\mu {R(x)\over{\Box}} \partial_\nu {R(y)\over{\Box}}\right\},
\label{f-v}\end{eqnarray}
donde el primer t\'ermino contiene la informaci\'on acerca de la anomal\'\i a 
de traza y el segundo, si bien tiene traza nula, contiene efectos no triviales 
como por ejemplo, es el t\'ermino que aporta la radiaci\'on de Hawking para 
m\'etricas de colapso de agujero negro. Volveremos a este punto m\'as 
adelante.

\subsection{La acci\'on efectiva de camino temporal cerrado}

La manera usual de obtener las ecuaciones semicl\'asicas de movimiento 
se basa en tomar la variaci\'on funcional de (\ref{effact}). Estrictamente, 
\'esta 
no es la manera correcta. Si pasamos del espacio Eucl\'\i deo al de Minkowski 
reemplazando el propagador Eucl\'\i deo 
por el de Feynman obtenemos la acci\'on efectiva in-out usual (como 
veremos m\'as adelante al evaluar la creaci\'on de part\'\i culas en 
m\'etricas 
cosmol\'ogicas). Las ecuaciones de movimiento derivadas a partir 
de esta acci\'on no son reales ni causales porque 
ellas son ecuaciones para los elementos de matriz in-out y no para 
valores medios de los campos. La soluci\'on de este problema radica en el 
formalismo de camino temporal cerrado (CTC) \cite{ctp1,calzethu1,jordan}. Con 
este formalismo 
podemos construir una acci\'on efectiva in-in para los valores de 
expectaci\'on. Asumiendo el punto de vista semicl\'asico, donde no integramos 
sobre las diferentes configuraciones de la m\'etrica, la acci\'on efectiva 
CTC se define como \cite{calzethu1,jordan}:

\begin{equation} e^{i S_{\rm ef}^{\rm CTC}[g^{\pm},f^{\pm}]} = {\cal N} 
e^{i(S_{\rm CGHS}[g^+
, f^+]
-S_{\rm CGHS}[g^-,f^-])} 
\int {\cal D} {\hat f}^+{\cal D} {\hat f}^- e^{i(S_{\rm mat}[g^+,{\hat f
}^+]
- S_{\rm mat}[g^-,{\hat f}^-])}. 
\label{ctpeff}\end{equation} 
Las ecuaciones de movimiento se obtienen tomando la variaci\'on 
de esta acci\'on con respecto a la m\'etrica $g_{\mu\nu}^+$, e imponiendo 
que  $g_{\mu\nu}^+=g_{\mu\nu}^-$. Las integrales de la ecuaci\'on 
(\ref{ctpeff}) deben realizarse sobre las fluctuaciones cu\'anticas de los 
campos. Por lo tanto ${\hat f}^+$ y ${\hat f}^-$ deben tener modos de 
frecuencia negativa y positiva, respectivamente, en el pasado remoto (estas 
son las llamadas condiciones de contorno in) y deben coincidir sobre 
una hipersuperficie espacial en el futuro (a un tiempo finito). Esta 
hipersuperficie, la cual debe ser una hipersuperficie de Cauchy, la llamaremos 
$\Sigma$. En consecuencia, la integral de camino puede pensarse como 
la integral de camino de dos campos independientes que evolucionan en dos 
ramas 
temporales diferentes; una en la direcci\'on positiva del tiempo en presencia 
de la fuente  $g_{\mu\nu}^+$ desde el vac\'\i o in hasta $\Sigma$, y la otra
en la direcci\'on temporal contraria, desde $\Sigma$ hasta el vac\'\i o in 
en presencia de la fuente $g_{\mu\nu}^-$. El v\'\i nculo que debe imponerse 
a la evoluci\'on de los campos es ${\hat f}^+\vert_{\Sigma}={\hat f}^
-\vert_{\Sigma}$. Es importante notar que ${\hat f}$ y $g_{\mu\nu}$ son 
independientes sobre las ramas + y $-$. 

En la aproximaci\'on semicl\'asica, la AECTC coincide con la acci\'on 
efectiva de granulado grueso definida en el Cap\'\i tulo 3, y puede 
escribirse, sin p\'erdida de generalidad como \cite{calzethu}

\begin{equation}S_{\rm ef}^{\rm CTC}= S_{\rm CGHS}[g_{\mu\nu}^+,f^+]-
S_{\rm CGHS}
[g_{\mu\nu}^-,f^-]
+\Gamma_{\rm FI}[g_{\mu\nu}^{\pm}],\label{if}\end{equation}
donde $\Gamma_{\rm FI}[g_{\mu\nu}^{\pm}]$ es la acci\'on de influencia, 
presentada en los cap\'\i tulos 2 y 3. Como veremos, esta acci\'on depende 
de la hipersuperficie $\Sigma$, donde los campos deben coincidir. Como 
en todo sistema cu\'antico abierto donde realizamos la integraci\'on sobre 
las variables correspondientes al entorno (en este caso las fluctuaciones 
cu\'anticas de los campos de materia  ${\hat f}_i$), encontramos 
una descripci\'on ``efectiva'' para el sistema (la m\'etrica $g_{\mu\nu}$, 
el dilat\'on, y la configuraci\'on cl\'asica de fondo de los campos de materia 
$f_i$). El objeto $e^{iS_{\rm ef}^{\rm CTC}}$ es b\'asicamente la funcional de
 influencia y coincide con la {\it funcional de p\'erdida de coherencia} de 
Hartle y Gell-Mann \cite{9if}. 

En nuestro caso presente, elegimos como condici\'on inicial para el 
estado cu\'antico de los campos escalares el estado de vac\'\i o in en el 
pasado remoto $\vert 0,{\mbox in}\rangle$. Para 
elecciones m\'as generales, la acci\'on de influencia es un objeto m\'as 
complicado que depende fuertemente de las mismas \cite{prd53}. 

La funcional de influencia puede escribirse alternativamente como:

 \begin{equation} e^{iS_{\rm ef}^{\rm CTC}}=\sum_{\alpha}\langle 0, 
in\vert\alpha,T
\rangle_{g^-}\langle \alpha ,T\vert 0, in\rangle _{g^+},\label{fieldprod}
\end{equation} y en consecuencia puede interpretarse como el producto 
escalar, sobre 
$\Sigma$, entre los estados evolucionados temporalmente (sobre cada una de las 
m\'etricas $g_{\mu\nu}^{\pm}$) desde el estado inicial com\'un hasta la 
hipersuperficie futura com\'un $\Sigma$. La AECTC tambi\'en puede escribirse 
alternativamente en t\'ermino de los coeficientes de Bogolubov que 
``conectan'' las bases in y out de cada rama temporal. Esto implica 
que existe p\'erdida de coherencia si y s\'olo si existe creaci\'on de 
part\'\i culas durante la evoluci\'on del campo \cite{calzmazzi}. 

La ecuaci\'on (\ref{ctpeff}) puede escribirse tambi\'en como 
\cite{einstlang,mottola}  

\begin{equation} e^{i S_{\rm ef}^{{\cal C}}[g,f]} = {\cal N} 
e^{i S_{\rm CGHS}^{{\cal C}}[g, f]} 
\int {\cal D} {\hat f} e^{i S_{\rm mat}^{{\cal 
C}}[g,{\hat f}]},\label{neweff}\end{equation}
donde hemos introducido el camino temporal complejo CTC 
${\cal C}= {\cal C}_+ \cup {\cal C}_-$, que va desde menos infinito a $\Sigma$
(${\cal C}_+$), y retorna, con una parte imaginaria infinitesimal decreciente
($\cal C_-$). Las integrales temporales sobre el contorno ${\cal C}$ est\'an 
definidas por $\int_{{\cal C}} 
dt =\int _{{\cal C_+}} dt -\int_{{\cal C_-}} dt$. Las fluctuaciones ${\hat f}$ 
de la ecuaci\'on (\ref{neweff}) est\'an relacionadas con aquellas presentes 
en (\ref{ctpeff}) por la relaci\'on  ${\hat f}(t, x) = 
{\hat f}_{\pm}(t,x)$ si $t \in {\cal C}_{\pm}$. Lo mismo se aplica para 
$g_{\mu\nu}$ y para la configuraci\'on cl\'asica de fondo $f$. Esta 
ecuaci\'on es \'util porque tiene la estructura de la acci\'on efectiva in-out 
o de la acci\'on Eucl\'\i dea usual. Las reglas de Feynman son las ordinarias, 
simplemente reemplazando los propagadores Eucl\'\i deos de la siguiente 
manera:
\begin{eqnarray} G(x,y) = \left\{\begin{array}{ll} 
G_{++}(x,y)=i \langle 0, in\vert T {\hat f}^+(x) {\hat f}^+(y)\vert 0, 
in\rangle,& ~t, t' ~ 
\mbox{ambos sobre} ~{\cal C}_+ \\ G_{--}(x,y)=-i \langle 0, in\vert {\tilde T}
{\hat f}^-(x) 
{\hat f}^-(y)\vert 0, in\rangle , & ~t, t' ~ \mbox{ambos sobre} 
~{\cal C}_-\\ G_{+-}
(x,y)=-
i \langle 0, in\vert {\hat f}^+(x) {\hat f}^-(y)\vert 0, in\rangle,    
&~t ~\mbox{sobre}~ 
{\cal C}_+, t'  ~\mbox{sobre} ~{\cal C}_-\\ G_{-+}(x,y)=i \langle 0, in
\vert {\hat f}^-(y) 
{\hat f}^+(x)\vert 0, in\rangle, & ~ t  ~\mbox{sobre} ~ {\cal C}_-, t'~  
\mbox{sobre}~ 
{\cal C}_+\end{array}\right. \label{prop} \end{eqnarray} 

Cada propagador se define teniendo en cuenta que las fluctuaciones de los 
campos corresponden a m\'etricas diferentes. Por ejemplo, $G_{++}$ es 
el producto ordenado temporalmente de los dos campos en la m\'etrica 
+, mientras que $G_{--}$ es el producto anti-temporalmente ordenado para los
campos en la m\'etrica $-$. Es por \'esto que la relaci\'on usual entre 
estas funciones de Green $G_{++} = G^{\star}_{--}$ ya no es v\'alida porque 
cada propagador est\'a definido sobre m\'etricas diferentes. De la misma 
forma, $G_{+-}$ se construye a partir del producto de un campo definido 
sobre la m\'etrica + con otro sobre la $-$. Los propagadores de Feynman y 
Dyson se representan por

 \begin{equation}G_{\pm\pm}(x,y)=G_{\pm}(x,y)~\theta(x^0-y^0) + G_{\mp}(x,y)
~\theta(y^0-x^0)\end{equation}
pero ahora no son v\'alidas las relaciones usuales $G_{+}=G_{+-}$ y 
$ G_{-}=G_{-+}$. 

A partir de las ecuaciones (\ref{effact}) y (\ref{prop}) la acci\'on efectiva 
de camino temporal cerrado es:
\begin{eqnarray}S_{\rm ef}^{\rm CTC} &=& S_{\rm CGHS}(g_{\mu\nu}^+, f^+) - 
S_{\rm CGHS}
(g_{\mu\nu}^-, f^-)\nonumber \\ &-& {N\over{96 \pi}}\int d^2x\sqrt{-g(x)}
\int d^2y\sqrt{-g(y)}~~R^a(x)~~ G_{ab}(x,y)~~R^b(y),\label{explctp}
\end{eqnarray}
donde los \'\i ndices $a$ y $b$ denotan cada una de las ramas CTC, + y $-$.

\subsection{Las ecuaciones de movimiento covariantes}

Aunque nuestro inter\'es fundamental radica en el estudio del proceso 
de p\'erdida de coherencia, en esta secci\'on mostramos c\'omo se 
deducen las ecuaciones de movimiento covariantes a partir de la AECTC. Estas 
ecuaciones no pueden derivarse directamente a partir de la ecuaci\'on 
(\ref{effact}) porque, como mencionamos al inicio, el formalismo CTC 
es necesario para evaluarlas correctamente a partir de una acci\'on efectiva. 
En general, las ecuaciones semicl\'asicas de movimiento han sido halladas 
por medio de \cite{18if}

 \begin{equation}
2 {\delta S_{\rm CGHS}
\over{\delta g_{\mu\nu}}} = \langle T_{\mu\nu}\rangle.
\label{scghse}
\end{equation}
Las componentes del tensor de energ\'\i a-impulso pueden determinarse 
utilizando la anomal\'\i a de traza e imponiendo la ley de conservaci\'on 
\cite{19if}. En la literatura previa, estas ecuaciones han sido escritas y 
estudiadas en la medida conforme. Pero esta manera de calcular restringe la
posibi\-lidad de hallar las componentes del tensor de energ\'\i a-impulso 
para aquellos campos de materia conformes o  no acoplados al dilat\'on, dado 
que para estos casos m\'as generales o no conocemos la anomal\'\i a de traza, 
y/o el tensor de energ\'\i a-impulso no se conserva \cite{tmunu}.

El formalismo CTC nos permite derivar las ecuaciones de movimiento covariantes
a partir de la expresi\'on:

\begin{equation}{\delta S_{\rm ef}^{\rm CTC}\over{\delta g_{\mu\nu}^+}}
\vert_{g_{\mu\nu}^+
=g_{\mu\nu}^-}
= 0.\label{fieldequation}\end{equation} En consecuencia, la principal
 dificultad est\'a en la variaci\'on funcional de las funciones de 
Green respecto a la m\'etrica. 
Despu\'es de expandir al campo en modos, podemos probar que \cite{if}

\begin{equation}\delta G_{++}= G_{\rm ret}~~ \delta \Box ~~ G_{++}+ G_{++}~~ 
\delta \Box~~ G_{\rm avan}- G_{\rm ret}~~ \delta \Box ~~ G_{\rm avan},
\label{g++}
\end{equation} 
\begin{equation}\delta G_{+-}= G_{\rm ret}~~ \delta \Box ~~ G_{+-},\label{g+-}
\end{equation}
\begin{equation}\delta G_{-+}= G_{\rm ret}~~ \delta \Box ~~ G_{-+},\label{g-+}
\end{equation}
donde $G_{\rm ret}$ y $G_{\rm avan}$ son las funciones de Green retardada y 
avanzada usuales, respectivamente. La variaci\'on funcional del 
operador dalambertiano $\delta \Box$ actuando 
sobre una funci\'on escalar arbitraria $w$ est\'a dada por \cite{if}: 

\begin{equation}\delta \Box w = - \nabla^\mu\nabla^\nu w \delta g_{\mu\nu}
-{1\over{2}} \partial^\lambda w g^{\mu\nu}\left( \delta g_{\lambda\nu;\mu}
+\delta g_{\mu\lambda;\nu} - \delta g_{\mu\nu;\lambda}\right).\label{dalvar}
\end{equation} 
En las ecuaciones que provienen de la variaci\'on funcional todos los 
propagadores 
 $G_{\rm ret}$, $G_{\rm avan}$, $G_{+-}$ y $G_{-+}$ est\'an evaluados 
en  $g_{\mu\nu}^+ = g_{\mu\nu}^-$, porque es lo que necesitamos para 
obtener las ecuaciones de movimiento. Finalmente, las ecuaciones covariantes 
est\'an dadas por:
\begin{eqnarray}{\delta S_{\rm CGHS}\over{\delta g_{\mu\nu}^+}} &\equiv &
 {1\over{2}}\langle T_{\mu\nu} \rangle = 
-{N\over{48 \pi}}\int d^2y\sqrt{-g(y)} ~R(y)~[\nabla_{\mu} 
\nabla_{\nu} - g_{\mu\nu}\Box ]_{(x)} ~G_{ret}(x,y)]\nonumber \\
&+&{N\over{192 \pi}}\int d^2x\sqrt{-g(x)} \int d^2y\sqrt{-g(y)}\left\{
2 R(x) ~ \partial_\mu (z)G_{ret}(x,z)~ \partial_\nu (z)G_{ret}(z,y)
~ R(y)\right.\nonumber \\
&-&\left. R(x) ~ g_{\mu\nu}(z) ~ \partial^{\alpha}(z) G_{ret}(x,z)
~ \partial_{\alpha}(z) G_{ret}(z,y)~R(y)\right\}  .
\label{eqmotion}\end{eqnarray}
La expresi\'on anterior es la misma que la calculada en la 
ecuaci\'on (\ref{f-v}), donde los propagadores 
Eucl\'\i deos fueron reemplazados por los retardados, por lo tanto \'estas son 
ecuaciones de movimiento no-locales, reales y causales. De 
(\ref{eqmotion}) obtenemos la anomal\'\i a conocida \cite{bd,if}

\begin{equation}\langle T_{\mu}^{\mu}\rangle = 2 g^{\mu\nu}{\delta 
S^{\rm CTC}_{\rm ef}\over{\delta g_{\mu\nu}^+}}= N ~ {R\over{24\pi}}
.\label{trace}\end{equation} En la medida conforme, 

\begin{equation}\langle T_{+-}\rangle=\langle T_{-+}\rangle = -
{N\over{12\pi}}\partial_+\partial_-\rho,\label{tmasmenos}\end{equation}
\begin{equation}\langle T_{\pm\pm}\rangle = -{N\over{12\pi}}[\partial_\pm
^2\rho - \partial_\pm \rho \partial_\pm \rho - t^\pm].\label{tmasmas}
\end{equation} Las funciones $t^{\pm}$ dependen de $x^{\pm}$ y pueden 
escribirse como 

\begin{equation}t^{\pm}= \partial_{\pm}^2 S^{\pm} - 2 \partial_{\pm}
S^{\pm}\partial_{\pm}S^{\pm},\label{tmasmenos}\end{equation} 
donde las funciones $S^{\pm}$ son:

\begin{equation}S^{+}(x^+) = \int d^2y ~\partial_{y^-} [\rho (y) 
\partial_{y^+} G_{\rm ret}(x,y)],\end{equation} 
\begin{equation}S^-(x^-) = \int d^2y ~\partial_{y^+} [\partial_{y^-} 
\rho (y) G_{\rm ret}(x,y)].\end{equation}

Las funciones $t^{\pm}$ dependen del estado cu\'antico de los campos de 
materia. En el modelo CGHS est\'an completamente determinadas por 
las condiciones de contorno en el pasado remoto que corresponden 
al estado de vac\'\i o in. Como ejemplo calculamos la expresi\'on 
expl\'\i cita de estas funciones para m\'etricas cosmol\'ogicas y de agujero 
negro. 

Para m\'etricas cosmol\'ogicas, $\rho$ es una funci\'on del tiempo conforme 
$t$, $\rho = \rho (t)$. Como el propagador retardado es

\begin{equation}G_{\rm ret}(x,y)= \theta (t_x - t_y - \vert x 
- y\vert ),\end{equation} las funciones $t^{\pm}$ son 

\begin{equation}t^{\pm}=\partial_\pm^2 \rho - \partial_\pm \rho 
\partial_\pm \rho.\end{equation} Por lo tanto, por las ecuaciones 
(\ref{tmasmenos}), (\ref{tmasmas}) y (\ref{tmasmenos}) obtenemos

\begin{equation}\langle T_{+-}\rangle=\langle T_{-+}\rangle = -
{N\over{12\pi}}\ddot{\rho},\end{equation}
\begin{equation}\langle T_{\pm\pm}\rangle = ~0.\end{equation} De donde 
es simple ver que s\'olo 
aparece la anomal\'\i a de traza \cite{22if}. 

Si consideramos la onda de choque de la secci\'on 4.1.1 

$${1\over{2}}\partial_+f \partial_+f = a \delta (x^+ - x^+_0)
,$$ podemos hallar las funciones $t^{\pm}$ en las coordenadas $\sigma^{\pm}$ 

 \begin{equation}t^{\pm}=\partial_{\sigma^\pm}^2 \rho - 
\partial_{\sigma^\pm} \rho \partial_{\sigma^\pm} \rho,\end{equation}
y las componentes del tensor de energ\'\i a-impulso son 

\begin{equation}\langle T_{\pm\pm}\rangle = -{N\over{12}}[\partial_\pm
^2\rho - \partial_\pm \rho \partial_\pm \rho - \partial_{\sigma^\pm}^2 \rho + 
\partial_{\sigma^\pm} \rho \partial_{\sigma^\pm} \rho].\label{comptmasmas}
\end{equation} Este resultado es v\'alido para todo el espacio-tiempo. En 
particular, para la regi\'on de vac\'\i o donde  $\partial_{\sigma^+}\rho$ 
se anula obtenemos \cite{cghs}

\begin{equation}t^+(\sigma^+)=0,\end{equation} 
\begin{equation}t^-(\sigma^-) = -{1\over{4}}\lambda^2 \left[ 1 - 
\left( 1+{a\over{\lambda}}e^{\lambda\sigma^-}\right)^{-2}\right].
\end{equation} Que es el resultado conocido a trav\'es de la literatura.

\subsection{Radiaci\'on de Hawking: modelo CGHS} 

Siguiendo el razonamiento empleado en la secci\'on anterior, consideramos 
la formaci\'on de un agujero negro por colapso gravitacional de una onda 
de choque situada en $x^+ = x^+_0$. Cuando $x^+ < x^+_0$, como demostramos 
anteriormente, la m\'etrica es la de Minkowski, es decir

\be ds^2_{\rm in}=-dx^-_{\rm in}dx^+_{\rm in}\ ,\ \ \ \ x^-_{\rm in}=t-x ~ ,
\ \ \ x^+_{\rm in}=t+x .\ee
Para $x^+ > x^+_0$, la geometr\'\i a est\'a representada por 

\be ds^2=-\lambda(r) dudv,\ee 
$$
u=t-r^* ~ ,\ \ \  v=t+r^* ~ ,\ \ \ \ {dr\over{dr^*}}=\lambda(r) \ ,
$$
donde $\lambda (r)$ se anula sobre el horizonte de eventos $r = r_+$. Para 
un agujero negro de Schwarzschild tenemos, por ejemplo  que 
$\lambda (r)= 1 - {2M\over{r}}$; en lo que sigue mantendremos la generalidad 
con la funci\'on $\lambda (r)$.

\begin{figure}[ht]
\centering \leavevmode
\epsfxsize=12cm
\epsfbox{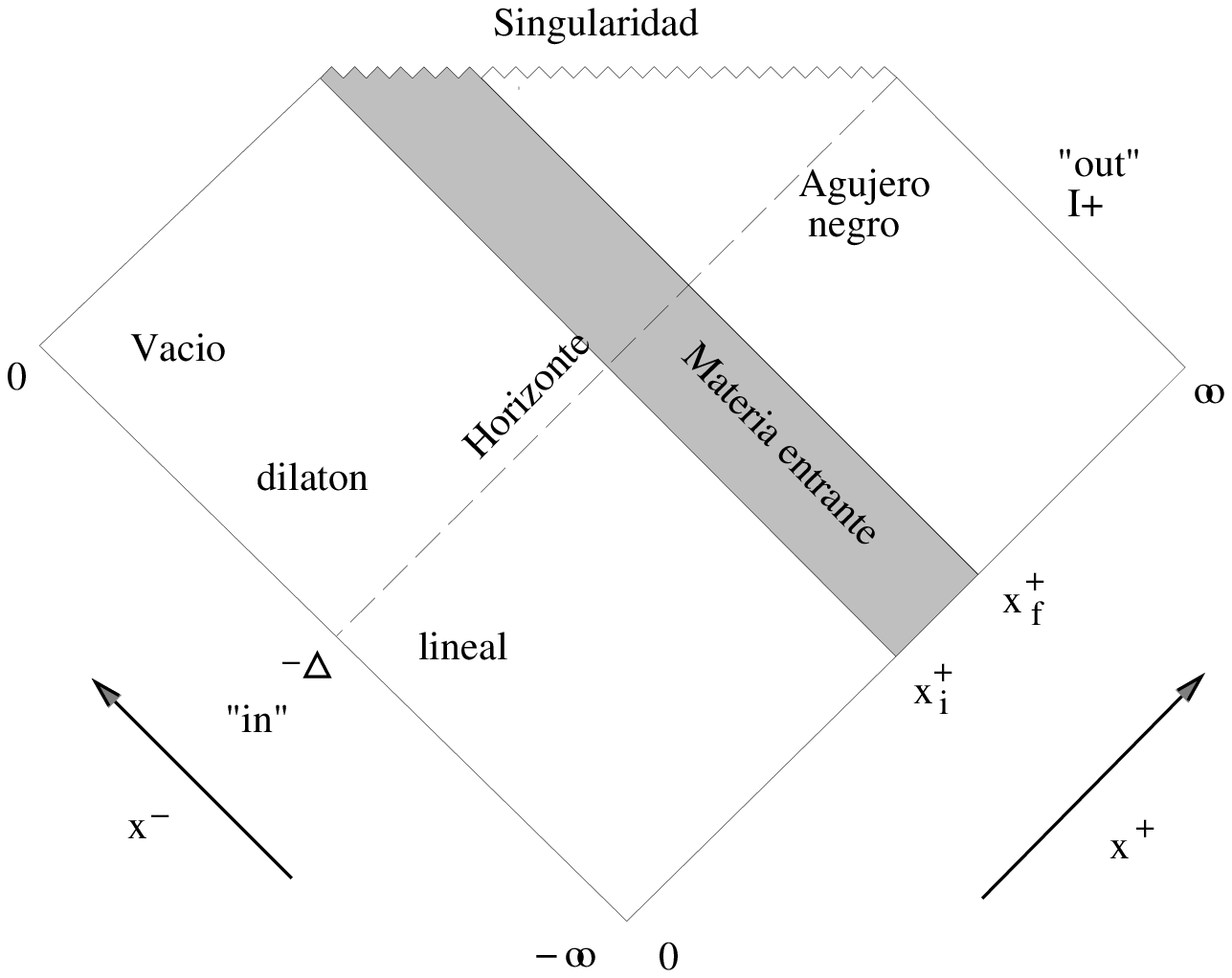}
\setlength{\captwidth}{12cm}
\capt{Diagrama de Penrose correspondiente al colapso gravitatorio de
 una onda de choque.}
\end{figure}

La relaci\'on entre las coordenadas ``in'' y ``out'' que representan uno y 
otro lado respecto de la onda de choque se obtiene de empalmar ambas 
geometr\'\i as en $x^+ = x^+_0$:

\begin{equation}
v=x^+_{\rm in}\ ,\ \ \ \ {dx^-_{\rm in}\over{du}}= \lambda\left[{1\over{2}}
(x^+_0 - x^-_{\rm in})\right]\ .
\label{inout}
\end{equation}

Como hemos visto ((\ref{eqmotion})), en la ecuaci\'on (\ref{f-v}) la 
expresi\'on formal ${1\over{\Box}} R $ 
denota al propagador retardado $G_{\rm ret}$ actuando sobre el 
escalar de Ricci. Utilizando la medida conforme, 
$ds^2=-e^{2\rho}dx^+dx^-$ tenemos que 
$-2\Box \rho = R$, por ello $\rho$ est\'a formalmente representado por 
$-2 \rho = {1\over{\Box}} R$. El propagador retardado queda definido por 
la relaci\'on $-2 \rho_{\rm in}=G_{ret} R$ donde $\rho_{\rm in}$ es el 
factor de Liouvlle en las coordenadas ``in''. La relaci\'on entre 
los factores conformes en las coordenadas ``in'' y ``out'', respectivamente 
queda determinado entonces por:

\begin{equation}
e^{2\rho_{\rm in}}=e^{2\rho_{\rm out}}{du\over dx^-_{\rm in}}{dv\over 
dx^+_{\rm in}}=e^{2\rho_{\rm out}}{du\over dx^-_{\rm in}}.
\label{rhoinout}
\end{equation}

El flujo de energ\'\i a a trav\'es de $I^+$ (ver Figura 4.1) est\'a dado por

\begin{equation}
\langle T_{uu}\rangle_{I^+}=-{1\over{12\pi}}\left[{\partial^2\rho_{\rm in}
\over\partial u^2}-\left(
{\partial\rho_{\rm in}\over\partial u}\right)^2\right]_{I^+}.
\label{flujo}
\end{equation}

Por medio de las ecuaciones (\ref{inout}) y (\ref{rhoinout}) obtenemos 

$$
2[\rho_{\rm in}]_{I^+}=\log {du\over dx^-_{\rm in}}+ \mbox{const} =
-\log [\lambda ({1\over 2}(x^+_0 - x^-_{\rm in})] + \mbox{const}
.$$
Combinando las ecuaciones anteriores podemos mostrar que \cite{13tmunu}

\begin{equation}
\langle T_{uu }\rangle_{I^+} = {1\over{192\pi}}
\lambda '^2\,\,\, ,
\label{zxxz}
\end{equation}
donde $\lambda ' = \lambda '(r_+)$. Este flujo corresponde  
a una temperatura ($\langle T_{uu}\rangle ={\pi\over{12}} T_H^2$)

\begin{equation}
T_H={1\over{4\pi}}\lambda'(r_+)
.\label{temp}\end{equation}

Este resultado es aplicable para aquellos agujeros negros 
cuya m\'etrica sea asint\'oticamente plana y de la forma 
$ds^2=-\lambda(r) dt^2+\lambda^{-1}(r) dr^2$. La temperatura de 
Hawking para un agujero negro gen\'erico tambi\'en puede calcularse trabajando 
en el espacio Eucl\'\i deo y compactificando la direcci\'on temporal. El 
resultado coincide con (\ref{temp}). Por ejemplo, para un agujero 
negro de Schwarzschild, tenemos que $T_{H}= {1\over{8\pi}} {1\over{M}}$. 
Volveremos a este punto en la 
siguiente secci\'on.

\section{Modelos m\'as realistas: acoplamiento dilat\'on-campo escalar}

Si bien el modelo de CGHS ha sido muy exitoso, es de inter\'es 
investigar otros modelos con una vinculaci\'on directa a 
modelos en cuatro dimensiones, al menos para alg\'un caso en particular. Por 
ejemplo podemos obtener modelos bi-dimensionales a partir de modelos 
en cuatro dimensiones pero restringidos a simetr\'\i a esf\'erica. En 
particular, podemos considerar la acci\'on de Einstein-Hilbert usual y campos 
m\'\i nimamente acoplados en cuatro dimensiones,

\begin{equation}
S = \int d^4x \sqrt{g^{(4)}}\left[{1\over{16\pi}} R^{(4)}
- {1\over{2}} (\partial^{(4)}f)^2\right]\ .
\label{eh4dact}
\end{equation}
Para configuraciones esf\'ericamente sim\'etricas

\begin{equation}
ds^2=g_{\mu\nu}dx^{\mu}dx^{\nu}=g_{ab}(x^a)dx^{a}dx^{b}+ e^{-2\phi (x^a)}
(d\theta^2+
\sin^2\theta \ d\varphi^2)\label{aaa}
\end{equation}
$$f=f(x^a)\ ,~a,b=0,1,$$
la acci\'on puede reducirse dimensionalmente a 

\begin{equation}
S = \int d^2x \ \sqrt{g} e^{-2\phi}\left[ {1\over{16\pi}}\left( R +
2 (\partial\phi)^2 + 2 e^{2\phi}\right) - {1\over{2}} (\partial f)^2\right].
\label{2dact}
\end{equation}
A diferencia del modelo CGHS, en esta acci\'on los campos escalares de materia 
est\'an acoplados al dilat\'on. 

De la misma manera, comenzando con campos escalares acoplados de manera 
arbitraria, cuya acci\'on est\'e dada por

\def\ha{ {\textstyle{{1\ov 2}}} }
\def\ov{\over}
\def\del{\partial }

\begin{equation}
S_{\rm materia} =-\ha \int d^4x \sqrt{g^{(4)}}\left[ (\partial^{(4)}f)^2+\xi  
R^{(4)} f^2
 \right]\ ,
\label{bbb}
\end{equation}
se obtiene, despu\'es de la reducci\'on dimensional:

\begin{equation}
S_{\rm materia} =-\ha \int d^2 x \sqrt{g} e^{-2\phi}\left[ (\partial f)^2+\xi  
f^2\big(
R^{(2)} +4\Box\phi - 6 (\del \phi )^2+2 e^{2\phi }\big)
 \right]\ ,
\label{cbb}
\end{equation}
que, en t\'erminos del campo $\psi=e^{-\phi} f$, puede re-escribirse como:

\begin{equation}
S_{\rm materia} =-\ha  \int d^2 x \sqrt{g} \left[(\partial \psi)^2+V \psi^2
\right]\ ,
\label{ccc}
\end{equation}
con la funci\'on $V$ definida por 

\begin{equation}
V=\xi R^{(2)} +(4\xi -1) \Box\phi +(1- 6\xi) (\del \phi )^2+2 \xi e^{2\phi }\ .
\label{ddd}
\end{equation}
$\xi=0$ y $\xi=1/6$ son casos especiales. Para $\xi =1/6$,
 la acci\'on es invariante conforme en cuatro dimensiones, es decir, la 
acci\'on queda invariante frente a transformaciones tales como 
$g_{\mu\nu}\to e^{2\sigma(x)} g_{\mu\nu}$ y $f \to e^{-\sigma (x)} f$. Desde 
el punto de vista del modelo bi-dimensional, la invarianza 
implica que $g_{ab}\to e^{2\sigma(x)} g_{ab}
\ ,\ \ 
\phi\to \phi-\sigma$ y $\psi\to \psi $ (o $f \to e^{-\sigma (x)} f$~). La 
acci\'on de materia  (\ref{2dact}), que corresponde al caso $\xi =0$, es 
invariante conforme en dos dimensiones, es decir, frente a las transformaciones
 $g_{ab}\to e^{2\sigma(x)} g_{ab}\ ,\ \ 
\phi\to \phi $ y $f \to f$. Para otro acoplamiento diferente de $\xi = 0$ o 
$\xi = 1/6$, no existe invarianza que implique transformaciones de Weyl en 
el modelo bi-dimensional.

Consideremos el modelo de la ecuaci\'on (\ref{2dact}). Debido a la 
simetr\'\i a conforme, la traza del tensor de energ\'\i a-impulso 
de los campos escalares se anula a nivel cl\'asico. Como ya hemos mencionado, 
aparece una 
anomal\'\i a a nivel cu\'antico. Esta anomal\'\i a 
ha sido calculada por varios autores \cite{anomal}. A partir de estos modelos, 
algunos efectos nuevos han sido discutidos, como por ejemplo la evaporaci\'on 
(y la anti-evaporaci\'on) de agujeros negros de Schwarzschild-de Sitter 
\cite{8tmunu}. Un aspecto muy importante en estos modelos es que debido al 
acoplamiento entre el dilat\'on y los campos de materia, el tensor de 
energ\'\i a-impulso bi-dimensional de los campos de materia no se 
conserva, y por ello, el conocimiento de la anomal\'\i a de traza 
no es suficiente para determinar el tensor de energ\'\i a-impulso total, 
como s\'\i \ ocurre en el modelo CGHS \cite{tmunu,9tmunu}.

\subsection{La acci\'on efectiva y el tensor de energ\'\i a-impulso}

A nivel cl\'asico, el tensor de energ\'\i a-impulso de los campos de materia 
est\'a dado por 
 
 \begin{equation}
T_{ab}= e^{-2\phi}\left[ \partial_a f \partial_b
f - {1\over{2}} g_{ab} (\partial f)^2\right]\ .
\label{tmunu2d} 
\end{equation} 
Este tensor tiene traza nula pero su divergencia no lo es; por lo tanto no 
se conserva. Efectivamente, si utilizamos la ecuaci\'on de movimiento 
cl\'asica para $f$, la divergencia est\'a dada por:

\begin{equation}
\nabla^a T_{ab}= -{1\over{2}}\partial_a (e^{-2\phi})
(\partial f)^2.
\label{divt2d}
\end{equation}
Claramente, el objeto que se conserva gracias al teorema de Noether es 
el tensor de energ\'\i a-impulso total. Como veremos ahora, La raz\'on por 
la cual $T_{ab}$ 
no se conserva es clara a partir de la conservaci\'on del tensor de 
energ\'\i a-impulso en cuatro dimensiones. 

La conexi\'on entre el modelo bi-dimensional 
y el caso general en cuatro dimensiones puede ponerse en evidencia debido a 
que gracias a la 
simetr\'\i a esf\'erica, el tensor de energ\'\i a-impulso general puede 
escribirse como:
\begin{eqnarray}
\langle  T^{(4)}_{ab}\rangle &=&{1\over 2\pi}e^{2\phi}{1\over\sqrt g}
{\delta S_{\rm ef}\over
\delta g^{ab}}= {1\over 4\pi}e^{2\phi}\langle T_{ab}\rangle \ ,\nonumber\\
\langle  T^{(4)}_{ij}\rangle &=&{1\over 8\pi}e^{2\phi}{1\over\sqrt g}
{\delta S_{\rm ef}\over
\delta \phi} g_{ij}\ ,
\label{4dt}
\end{eqnarray}
donde los \'\i ndices $i$ y $j$ indican las coordenadas angulares.

En realidad, como $
\nabla^\mu T_{\mu\nu}^{(4)}=0$, y gracias a la ecuaci\'on (\ref{aaa}), se 
obtiene que 

\begin{equation}
\nabla^a T_{ab}^{(4)}=2\del^a\phi T^{(4)}_{ab}-e^{2\phi }
(\del_b\phi T^{(4)}_{\theta\theta}
+\sin^{-2}\theta \del_b\phi T_{\varphi\varphi}^{(4)}).
\label{zaa}
\end{equation} Luego de utilizar que $T_{\mu\nu}^{(4)}=  
\partial_\mu f \partial_\nu f - {1\over{2}} g_{\mu\nu} 
(\partial f)^2 $ y $f=f(x^a)$, reproducimos  la ecuaci\'on (\ref{divt2d}). 

A nivel cu\'antico, el valor medio $\langle T_{ab}\rangle $ es una 
cantidad divergente que debe ser renormalizada. Una renormalizaci\'on 
covariante producir\'a entonces un tensor de energ\'\i a-impulso que 
no se conserva y que contiene una anomal\'\i a de traza. 

La acci\'on de materia (\ref{ccc}) puede re-escribirse para $\xi = 0$ como:

\begin{equation}
S_{\psi} = -{1\over{2}}\int d^2x ~ \sqrt{g}\left[
(\partial \psi)^2 + P~\psi^2\right],
\label{trola}\end{equation}
donde $P=(\partial\phi)^2-\Box\phi$.

Al igual que en la secci\'on anterior, la acci\'on efectiva Eucl\'\i dea puede
 calcularse usando el hecho que la anomal\'\i a de traza est\'a determinada 
por $T=2 g^{ab}{\delta S\over\delta g^{ab}}={1\over 24\pi}(R -6P)$ 
\cite{anomal}. Por lo tanto, integrando esta ecuaci\'on es posible obtener 
\cite{tmunu}
\begin{eqnarray}
S_{\rm ef} &=& -{1\over{8\pi}}
\int d^2x ~ \sqrt{g}
\int d^2y ~ \sqrt{g} \left\{{1\over{12}} R(x) {1\over{\Box}} R(y) -
 P(x) {1\over{\Box}} R(y) \right\} + S_{\rm ef}^{\rm I}\nonumber\\ 
&\equiv & S_{\rm ef}^{\rm A} + S_{\rm ef}^{\rm I}\ .
\label{niWefacc} 
\end{eqnarray}
El primer t\'ermino de (\ref{niWefacc}), $S_{\rm ef}^{\rm A}$ produce 
la anomal\'\i a de traza esperada, mientras que el segundo t\'ermino es 
invariante frente a transformaciones de Weyl. Este t\'ermino es trivial 
en el caso en que el dilat\'on sea constante y/o si los campos de materia 
no est\'an acoplados al dilat\'on. Por este motivo no lo incl\'\i mos 
al evaluar la acci\'on efectiva para el modelo CGHS (ecuaci\'on 
(\ref{effact})).

Trabajando en la medida conforme, el t\'ermino invariante de Weyl puede 
escribirse como:

\begin{equation}
e^{- S_{\rm ef}^{\rm I}}= \mbox{det} 
[-\Box_p+P_p]^{-{1\over{2}}} = {\cal N} \int {\cal D}\psi\ e^{-{1\over{2}}
\int d^2x \ \psi (-\Box_p )\psi} ~ e^{-{1\over{2}}\int d^2x\ P_p \psi^2},
\label{sefinv}
\end{equation}
donde el sub-\'\i ndice $p$ indica que la cantidad debe evaluarse sobre 
una m\'etrica plana; ${\cal N}$ es una cons\-tante de normalizaci\'on. En 
algunos trabajos esta parte invariante ha sido omitida, produciendo resultados 
contradictorios con los de los modelos en cuatro dimensiones \cite{10tmunu}. 
Una manera de calcular esta parte invariante consiste en realizar una 
expansi\'on en potencias de $P_p$ \cite{11tmunu}:

\begin{equation}
S_{\rm ef}^{\rm I} = \int d^2x P_p(x) D_1(x) + \int d^2x \int d^2y 
P_p(x) D_2(x,y) P_p(y)+... 
\label{ex}
\end{equation}
Comparando t\'erminos del mismo orden entre las ecuaciones (\ref{sefinv}) 
y (\ref{ex}) obtenemos que $D_1(x) = \ha G(x,x)$, $D_2(x,y) = {1\ov 4} 
G^2(x,y)$, donde $G$ es el propagador Eucl\'\i deo plano. Por lo tanto la 
acci\'on, a segundo orden en la expansi\'on en potencias de $P_p$ es:

\begin{equation}
S_{\rm ef}^{\rm I} = {1\over{4}}\int d^2x\int d^2y P_p(x) G^2(x,y) 
P_p(y)\ ,
\end{equation}
donde hemos omitido un t\'ermino local y divergente que puede ser 
extra\'\i do mediante la adici\'on de un contrat\'ermino. 

El propagador al cuadrado $G^2$, ha sido calculado en la Ref. \cite{12tmunu}:

\begin{equation}
 G^2(p) = {1\over{2\pi}} {1\over{p^2}}\ln {p^2\over{\mu^2}}.
\label{g2}
\end{equation} 
Finalmente, transformando Fourier la ecuaci\'on (\ref{g2}), la parte 
invariante de la acci\'on, a 
segundo orden en $P_p$ queda:
\begin{eqnarray}
S_{\rm ef}^{\rm I} &=& -{1\over{8\pi}}\int d^2x 
\int d^2y 
 P_p(x) {1\over{\Box_p}}\ln {-\Box_p\over{\mu^2}} P_p(y)
\label{avra} \\
&=& -{1\over{8\pi}}\int d^2x \sqrt{g}
\int d^2y  
\sqrt{g}\left\{ P(x){1\over{\Box}}\ln {-\Box\over{\mu^2}} P(y)+
\int d^2z\sqrt{g} P(x){1\over{\Box}}{R(y)\over{\Box}}P(z)\right\}.
\nonumber 
\end{eqnarray}
La segunda l\'\i nea de la ecuaci\'on (\ref{avra}) est\'a escrita de manera 
expl\'\i citamente covariante usando el hecho que $P_p = \sqrt{g} P$ y que 
la funci\'on de Green $1/\Box_p$ es invariante de Weyl. El par\'ametro 
$\mu$ representa una escala de corte infraroja, por lo tanto la acci\'on 
efectiva en esta aproximaci\'on depende del par\'ametro $\mu$ debido 
a que estamos evaluando perturbaciones para campos no-masivos en dos 
dimensiones. Los resultados f\'\i sicos van a depender de $\mu$. 

Si bien hemos realizado un desarrollo en potencias de $P$ para evaluar 
la parte invariante de Weyl, no hemos desarrollado en potencias de $R$, 
a diferencia de lo hecho en \cite{11tmunu}. Como veremos, \'esto ser\'a 
\'util en geometr\'\i as no-triviales con $P\ll R$. 

Una manera alternativa, basada en una  aproximaci\'on diferente, 
puede aplicarse para el c\'alculo de $S_{\rm ef}^{\rm I}$. Esta nueva 
aproximaci\'on corresponde a despreciar la retrodispersi\'on de la 
geometr\'\i a sobre la din\'amica de los campos de materia. Esta 
aproximaci\'on efectivamente supone que el t\'ermino 
de masa en la 
ecuaci\'on (\ref{sefinv}) es una constante \cite{mukha}. En este caso, la 
parte invariante de Weyl es:
\begin{eqnarray}
S_{\rm ef}^{\rm I} &=& -{1\over{8\pi}} \int d^2x  
P_p \left( 1 - \mbox{log}{P_p\over{\mu^2}}\right) 
\nonumber \\
&=& -{1\over{8\pi}} \int d^2x ~ \sqrt{g} ~ 
P(x) \left( 1 - \mbox{log}{P\over{\mu^2}}\right)-{1\over{8\pi}} 
\int d^2x \sqrt{g}\int d^2y \sqrt{g} P(x) {1\over{\Box}} R(y)\ .
\label{mukha}
\end{eqnarray} 
Nuevamente, la expresi\'on anterior fue hallada en el espacio plano y luego 
se obtuvo la expresi\'on covariante. Para ello debemos notar que en la 
medida conforme, $\log(\sqrt g)= -\Box ^{-1} R $. Est\'a claro que podemos 
calcular las correcciones que vienen del hecho de que en realidad $P$ no 
es una constante; \'estas se obtienen haciendo perturbaciones en potencias de 
derivadas de $P$ \cite{mukha}. Notar que el \'ultimo t\'ermino en 
(\ref{mukha}) se cancelar\'a con un t\'ermino 
id\'entico en la $S_{\rm ef}^{\rm A}$ (ver ecuaci\'on (\ref{niWefacc})). 

En ambas aproximaciones la acci\'on efectiva puede escribirse como 
$S_{\rm ef}=S_{\rm ef}^{\rm A}+ S_{\rm ef}^{\rm I}$. Dado que el 
tensor de energ\'\i a-impulso se obtiene a partir de la acci\'on efectiva, 
vari\'andola respecto de la m\'etrica bi-dimensional, una descomposici\'on 
similar aparece en este caso $\langle T_{ab}\rangle =\langle T_{ab}^{\rm A}
\rangle + \langle T_{ab}^{\rm I}\rangle $. La parte que corresponde a la 
acci\'on de la anomal\'\i a es independiente de la aproximaci\'on hecha para 
calcular la parte invariante:
\begin{eqnarray}
\langle T_{ab}^{\rm A}\rangle &=& {1\over{4\pi}} 
\int d^2y \sqrt{g} \left[ \nabla_a \nabla_b - g_{ab}
\Box \right]_{(x)} P(y) {1\over{\Box}} \nonumber \\
&-& {1\over{24\pi}}  \int d^2y \sqrt{g} \left[ \nabla_a \nabla_b - 
g_{ab} \Box \right]_{(x)} R(y) {1\over{\Box}}\nonumber \\
&+& {1\over{8\pi}}\int d^2y \sqrt{g} 
\left[ g_{ab} \nabla^c \phi
\nabla_c - 2 \nabla_a \phi \nabla_b + g_{ab}(\partial \phi)^2 - 
2 \nabla_a \phi \nabla_b \phi\right]_{(x)} R(y)
{1\over{\Box}} \nonumber \\
&+& {1\over{96\pi}} \int d^2x \sqrt{g} \int d^2y \sqrt{g}~\left\{
 2 \partial_a {R(x)\over{\Box}} \partial_b {R(y)\over{\Box}}
- g_{ab} \partial^c{R(x)\over{\Box}} 
\partial_c {R(y)\over{\Box}}\right\} \nonumber \\
&-& {1\over{8\pi}} \int d^2x \sqrt{g}\int d^2y \sqrt{g}~\left\{
2 \partial_a {P(x)\over{\Box}} \partial_b {R(y)\over{\Box}}
- g_{ab} \partial^c{P(x)\over{\Box}}
\partial_c{R(y)\over{\Box}} \right\}
\label{ancon}\ .
\end{eqnarray} Este tensor tiene la 
anomal\'\i a de traza correcta: $\langle T_a^a\rangle = 
{1\over{24\pi}}(R - 6 P)$. Adem\'as de la anomal\'\i a de traza, contiene una 
parte de traza nula pero que no se conserva, es decir que su 
divergencia es diferente de cero.

La parte del tensor asociada a la acci\'on efectiva invariante de Weyl, cambia 
seg\'un cada una de las aproximaciones efectuadas. En particular, en la 
aproximaci\'on en potencias de $P$ est\'a dada por, a orden lineal en $P$
\begin{eqnarray}
\langle T_{ab}^{\rm I}\rangle &=&  - {1\over{4\pi}} \int d^2y \sqrt{g} 
\left[ g_{ab}\nabla^{c} \phi
\nabla_{c} - 2 \nabla_a\phi \nabla_b +
g_{ab}(\partial \phi)^2 - 
2 \nabla_a \phi \nabla_b \phi
\right]{P(y)\over{\Box}}
\ln {-\Box\over{\mu^2}}\label{nanoool}\\
&+& {1\over{4\pi}} \int d^2y \sqrt{g}\int d^2z \sqrt{g}~\left[
g_{ab}\nabla^{c} \phi
\nabla_{c} - 2 \nabla_a\phi \nabla_b +
g_{ab}(\partial \phi)^2 - 
2 \nabla_a \phi \nabla_b \phi\right]
{R(y)\over{\Box}}{P(z)\over{\Box}} 
\nonumber \end{eqnarray}
mientras que en la aproximaci\'on en que se desprecia la retrodispersi\'on, 
la parte invariante de Weyl del tensor de energ\'\i a-impulso es:
\begin{eqnarray}
\langle T_{ab}^{\rm I}\rangle &=&  {1\over{8\pi}}\left[ g_{ab} \nabla^c \phi
\nabla_c - 2 \nabla_a \phi \nabla_b +
g_{ab}(\partial \phi)^2 - 
2 \nabla_a \phi \nabla_b \phi
\right]~\mbox{log}{P\over{\mu^2}}
\nonumber \\
&-& {1\over{8\pi}}g_{ab} 
P ~ - {1\over{4\pi}} 
\int d^2y \sqrt{g} \left[ \nabla_a \nabla_b - g_{ab}
\Box \right]_{(x)} P(y) {1\over{\Box}} \nonumber \\
&-& {1\over{8\pi}}\int d^2y \sqrt{g} \left[ g_{ab} \nabla^c \phi
\nabla_c - 2 \nabla_a \phi \nabla_b + 
g_{ab}(\partial \phi)^2 - 
2 \nabla_a \phi \nabla_b \phi
\right]_{(x)} R(y)
{1\over{\Box}}\nonumber \\
&+& {1\over{8\pi}} \int d^2x \sqrt{g}\int d^2y \sqrt{g}~\left\{
2 \partial_a {P(x)\over{\Box}} \partial_b {R(y)\over{\Box}}
- g_{ab} \partial^c{P(x)\over{\Box}}
\partial_c{R(y)\over{\Box}}\right\}\ .
\label{pepe}
\end{eqnarray}

A esta altura son necesarios algunos comentarios acerca de los 
l\'\i mites de aplicabilidad 
y alcances de cada aproximaci\'on. Debido a la presencia 
de un t\'ermino proporcional a $\mbox{log} P$, $\langle T_{ab}^{\rm I} 
\rangle $ dado por la ecuaci\'on (\ref{pepe}) tiene una singularidad 
cuando $P\rightarrow 0$. Por lo tanto, $\langle T_{ab}^{\rm I} 
\rangle $ es singular a\'un para el espacio-tiempo de Minkowski, donde 
$P\equiv 0$. Esta divergencia puede entenderse simplemente porque cuando 
estamos en el caso en que $P\cong 0$, la aproximaci\'on de 
no-retrodispersi\'on deja de ser v\'alida. El origen de esta 
singularidad est\'a asociado a la divergencia infraroja que siempre 
aparece para campos sin masa en dos dimensiones (que es lo que efectivamente 
sucede con el campo $\psi$ cuando $P=0$). O sea, la divergencia es producto 
de la aproximaci\'on donde asumimos que $P$ es una constante 
no nula. En aquellos casos donde $P\cong 0$ es m\'as adecuado utilizar 
la aproximaci\'on basada en el desarrollo en potencias de $P$, donde la 
divergencia est\'a regularizada por la aparici\'on del par\'ametro $\mu$.

Como mencionamos al comienzo de esta secci\'on, el hecho que el dilat\'on 
est\'e acoplado al campo escalar de materia, implica que el tensor de 
energ\'\i a-impulso tenga divergencia distinta de cero. En la expansi\'on 
en potencias de $P$, la divergencia est\'a dada por: 
\begin{eqnarray}\nabla^b \langle T_{ab}\rangle &=&
{1\over{8\pi}}\left[\nabla^b \phi \nabla_a\nabla_b- 
\nabla^b \nabla_a \phi \nabla_b  - 2 \nabla_a \phi \Box 
- \nabla^b\nabla_a\phi\nabla_b\phi - \nabla_a\phi\Box\phi
\right]_{(x)}\nonumber \\
&\times & \int d^2y 
\sqrt{g}\left\{ R(y) 
{1\over{\Box}}- 2 P(y) {1\over{\Box}} 
\ln {-\Box\over{\mu^2}} + 2 \int d^2z \sqrt{g}~ {R(y)\over{\Box}}
{P(z)\over{\Box}}\right\}.\label{diveravra}\end{eqnarray}
Mientras que para el caso de no-retrodispersi\'on, 
\begin{eqnarray}\nabla^b \langle T_{ab}\rangle  &=& -{1\over{8\pi}} 
\nabla_a P + {1\over{8\pi}}\left[\nabla^b \phi \nabla_a\nabla_b- 
\nabla^b \nabla_a \phi \nabla_b\right. \nonumber \\
&&\left. - 2 \nabla_a \phi \Box - \nabla^b\nabla_a\phi\nabla_b\phi
- \nabla_a\phi\Box\phi
\right]~\mbox{log}{P\over{\mu^2}}.\label{divermukha}
\end{eqnarray}
Como en el caso cl\'asico, el tensor no se conserva cuando el dilat\'on 
no es una constante. 

En las tres pr\'oximas secciones mostraremos que si ignoramos la existencia 
del t\'ermino invariante de Weyl, es decir si consideramos que 
$\langle T_{ab}\rangle = \langle T_{ab}^{\rm A} \rangle $, se obtienen 
resultados f\'\i sicos inconsistentes para los  
efectos cu\'anticos tanto en m\'etricas de agujeros negros como para 
m\'etricas cosmol\'ogicas. Lo mismo ocurre si calculamos el 
tensor de energ\'\i a-impulso $\langle T_{ab}\rangle$ utilizando la 
anomal\'\i a de traza e imponiendo  (incorrectamente) la ley de 
conservaci\'on $\nabla^b T_{ab}=0$.

\subsection{Radiaci\'on de Hawking: modelos con acoplamiento al dilat\'on}

El motivo de esta secci\'on es analizar en detalle la 
aplicaci\'on de las dos aproximaciones mencionadas en la secci\'on anterior 
al c\'alculo de la radiaci\'on de Hawking. En las secciones siguientes 
extenderemos el estudio a otros fen\'omenos f\'\i sicos de inter\'es.

En particular, para una 
m\'etrica dada por:

\begin{equation}
ds^2=-\lambda(r) dudv  ,
\end{equation}
es f\'acil demostrar que $P(r)={1\ov r}\lambda'(r)$. Por lo tanto $P$ 
es diferente de cero para todo punto fuera del horizonte de eventos 
$r>r_+$ (al menos en el caso de agujeros negros no-extremos; volveremos 
a este punto m\'as adelante). Como $P$ no se anula sobre el horizonte, 
podemos utilizar la aproximaci\'on de no-retrodispersi\'on. En consecuencia,  
sumando las ecuaciones (\ref{niWefacc}) y (\ref{mukha}) la acci\'on 
efectiva completa es:

\begin{equation}
S_{\rm ef}= -{1\over 96\pi}\int d^2x \sqrt{g} ~R{1\over\Box}R + 
\mbox{t\'erminos locales},
\label{caca}
\end{equation}  
que, a menos de los t\'erminos locales coincide con la acci\'on del 
modelo CGHS (acci\'on de Polyakov), para la cual 
hemos calculado la radiaci\'on de Hawking 
en la secci\'on anterior. 

De esta manera, el resultado general implica que la temperatura de la 
radiaci\'on de Hawking est\'a dada por la ecuaci\'on  
$T_H={1\ov 4\pi}\lambda'(r_{\rm hor})$. 

Por lo tanto, mientras $P$ sea diferente de cero, hemos demostrado que 
podemos utilizar la aproximaci\'on de no-retrodispersi\'on, y que el 
resultado para la radiaci\'on de Hawking en cuatro dimensiones es el 
calculado en la secci\'on 
anterior. La contribuci\'on a la radiaci\'on est\'a dada por el t\'ermino 
de Polyakov de la acci\'on efectiva \cite{tmunu}.

Consideremos ahora una geometr\'\i a donde $P$ se anule sobre el 
horizonte, como por ejemplo para agujeros negros de Reissner-Nordstr\"om 
($\lambda(r)=1-{2M\ov r} +{q^2\ov r^2}$) en 
el l\'\i mite extremo. A\'un en este caso, la 
aproximaci\'on de no-retrodispersi\'on contin\'ua dando el 
resultado correcto para la radiaci\'on de Hawking. Si bien $P$ se anula sobre 
el horizonte, no hay divergencia alguna en el tensor de energ\'\i a-impulso, 
dado que los t\'erminos 
que aportan a la radiaci\'on tienen derivadas no nulas de $P$.
Como $P$ es casi cero cerca del horizonte, tambi\'en podemos utilizar el 
desarrollo en potencias de 
$P$. Cerca del horizonte de eventos la contribuci\'on m\'as importante 
a la radiaci\'on de Hawking est\'a nuevamente dada por el t\'ermino de 
Polyakov que es independiente de $P$ y es exacto a todo orden en $R$. Las 
correcciones de orden superior que aparecen en este caso provienen del 
t\'ermino no-local proporcional a ${1\over{\Box}}\ln {-\Box\over{\mu^2}}$, y 
por lo tanto dependen de la escala $\mu$. 

Es importante recalcar a esta altura que la aproximaci\'on en potencias 
de $P$ no es adecuada para estimar la radiaci\'on de Hawking para 
agujeros negros de Schwarzschild. Esto se debe a que en esta 
m\'etrica $P = {2M\over{r^3}}={R\over{2}}$, por lo tanto no es 
consistente el c\'alculo de $S_{\rm ef}$ a todo orden en $R$ y restringido 
a orden cuadr\'atico en la expansi\'on en $P$.  Deber\'\i amos sumar 
un n\'umero infinito de t\'erminos no-locales para obtener el 
resultado correcto.

Finalmente, debemos notar que el hecho de no incluir el t\'ermino invariante 
de Weyl en la acci\'on efectiva produce resultados err\'oneos. El 
t\'ermino relevante para la radiaci\'on de Hawking, a partir de la variaci\'on 
funcional de la acci\'on efectiva an\'omala ser\'\i a (ver ecuaci\'on 
(\ref{ancon})):

\begin{equation}
{1\over{48\pi}} \int d^2x \sqrt{g}\int d^2y \sqrt{g}~\left\{
\partial_a {R(x)\over{\Box}} \partial_b {R(y)\over{\Box}}- 12
\partial_a {P(x)\over{\Box}} \partial_b {R(y)\over{\Box}}\right\}.
\end{equation}
Para el agujero negro de Schwarzschild $P=R/2$, el t\'ermino proporcional 
a $P$ produce un flujo ``entrante'' 
que excede por un factor 6 al flujo ``saliente''. Por lo tanto, si no se 
considera $S_{\rm ef}^{\rm I}$ se obtiene un flujo de Hawking negativo, lo 
cual es inaceptable a partir de los resultados en cuatro dimensiones. Este 
problema apareci\'o en algunos trabajos previos \cite{5y6anomal}.

\subsection{Correcciones cu\'anticas al potencial Newtoniano}

En esta secci\'on damos otro ejemplo que muestra la importancia de la 
parte invariante de Weyl en la acci\'on efectiva: calculamos las 
correcciones cu\'anticas 
al potencial Newtoniano. Aqu\'\i \ no pretendemos hacer un an\'alisis 
detallado de la manera en que estas correcciones han sido evaluadas 
anteriormente \cite{14tmunu,15tmunu}, pero si utilizarlas en el marco 
de los modelos bi-dimensionales para probar la efectividad de cada 
una de las aproximaciones efectuadas. 

Como mencionamos al inicio del presente cap\'\i tulo, las ecuaciones de 
Einstein semicl\'asicas en cuatro dimensiones est\'an dadas por 

  \begin{equation}{1\over{8\pi}}(R_{\mu\nu}-{1\over{2}}g_{\mu\nu} R) 
= {}^{\rm clas}T_{\mu\nu}^{(4)} 
+ \langle T_{\mu\nu}^{(4)}\rangle \ ,
\label{ee}
\end{equation} 
en el presente ejemplo, suponemos que la parte cl\'asica de las fuentes 
est\'a dada por una part\'\i cula puntual de masa $M$, 
${}^{\rm clas}T_{\mu\nu}^{(4)} 
= - \delta^0_\mu \delta^0_\nu M \delta^3 (\vec x)$ y que $\langle T_{\mu
\nu}^{(4)} \rangle$ es el tensor de energ\'\i a-impulso para un campo 
escalar cu\'antico sin masa. 

Estas ecuaciones pueden resolverse considerando perturbaciones alrededor del 
espacio plano $g_{\mu\nu} = \eta_{\mu\nu} + h_{\mu\nu}$. Para encontrar las 
correcciones cu\'anticas al potencial Newtoniano, es suficiente calcular 
la soluci\'on de de la ecuaci\'on para la traza de $h_{\mu\nu}$. Expandiendo 
perturbativamente, $h = h^{(0)} + h^{(1)}$, con 
$ h^{(0)}={4 M\over r}$ soluci\'on de la parte cl\'asica, la ecuaci\'on para 
$h^{(1)}$ es:

  \begin{equation}
{1\over{2\pi}}\nabla^2 h^{(1)} = 
g^{\mu\nu}\langle T_{\mu\nu}^{(4)}\rangle \ .
\end{equation}
A distancias grandes, la traza de $\langle T_{\mu\nu}^{(4)}\rangle$ es 
\cite{14tmunu}
 
\begin{equation}
\langle T^{(4)}\rangle =-{M\over 8\pi^2 r^5} \equiv {C\over r^5}\ .
\label{ffz}
\end{equation}

Para campos sin masa acoplados m\'\i nimamente en cuatro 
dimensiones, la traza del tensor de energ\'\i a-impulso depende del 
estado cu\'antico del campo. La ecuaci\'on (\ref{ffz}) corresponde a 
calcular la traza del tensor en el estado de Boulware, que es el 
estado de vac\'\i o de Minkowski a grandes distancias. 

Por lo tanto, la soluci\'on perturbativa de las ecuaciones semicl\'asicas de 
Einstein es:

\begin{equation}-{h\over 4} = - {M\over r}+ {M\over{12\pi}}
 {1\over{r^3}} + .... ,
\label{v}\end{equation}
de la cual podemos extraer las correcciones cu\'anticas al potencial 
Newtoniano. 

Para obtener el potencial Newtoniano consideramos las ecuaciones geod\'esicas  
de una part\'\i cula de prueba con coordenadas $x^\mu (\tau )$ ($d\tau^2 = 
- g_{\mu\nu} dx^\mu dx^\nu$), que se mueve en el fondo gravitatorio dado por 
$g_{\mu\nu}$. Las ecuaciones son 

\be {d^2x^\rho\over{d\tau^2}}+ \Gamma^\rho_{\mu\sigma}{dx^\mu\over{d\tau}}
{dx^\sigma\over{d\tau}},\ee donde $\Gamma^\rho_{\mu\sigma}$ es el s\'\i mbolo
 de Christoffel asociado a la m\'etrica soluci\'on de las ecuaciones 
semicl\'asicas. En el l\'\i mite de campo d\'ebil, las ecuaciones geod\'esicas
 se reducen a 

\be {d^2x\over{dt^2}}= -\nabla V = {1\over{2}}h_{00},\ee
por lo tanto, el potencial Newtoniano queda determinado por $V(r) = 
-{1\over{2}}
h_{00}$. En nuestro caso, $h_{00} = -{1\over{2}}h$ y, en consecuencia 
la ecuaci\'on (\ref{v}) determina las correcciones cu\'anticas al 
potencial Newtoniano.

Por lo tanto, queda claro que para determinar las correcciones cu\'anticas 
al potencial Newtoniano es necesario determinar la traza (en cuatro 
dimensiones) del tensor de energ\'\i a-impulso en la m\'etrica de 
 Schwarzschild. El signo de $C$ de la ecuaci\'on (\ref{ffz}) es muy 
importante. Un valor positivo de esta constante podr\'\i a implicar 
que la constante de Newton decrece con $r$, lo cual es inadmisible dado 
que implicar\'\i a efectos de apantallamiento gravitatorio. 
  
El tensor de energ\'\i a-impulso en cuatro dimensiones 
 $\langle T^{(4)} \rangle =
g^{\mu\nu}\langle T_{\mu\nu}^{(4)}\rangle =
g^{ab}\langle T_{ab}^{(4)}\rangle + g^{ij}\langle T_{ij}^{(4)}\rangle$ debe 
calcularse utilizando la aproximaci\'on en potencias de $P$, debido a que 
la aproximaci\'on que desprecia la retrodispersi\'on no es 
adecuada para regiones asint\'oticamente planas, dado que diverge cuando 
$P\rightarrow 0$. 

Por lo tanto, tenemos que evaluar las ecuaciones (\ref{ancon}) y 
(\ref{nanoool}) en la m\'etrica de colapso:

\begin{equation}ds^2 = \left(1 - {2M\over{r}}\right)\left(-dt^2 + 
dr^{\star 2}\right) + r^2 d\Omega^2,\label{schmet}\end{equation} 
donde $d\Omega^2$ es el elemento de l\'\i nea de la dos-esfera, y $r^{\star}$
est\'a dada por:

\begin{equation}r^{\star} = r + 2 M \ln{\vert {r\over{2M}} - 1\vert}
.\end{equation}

Las funciones no-locales ${R\over{\Box}}$ y ${P\over{\Box}}\ln 
{{-\Box\over{\mu^2}}}$ son evaluadas mediante tranformaciones de Fourier 
\cite{17tmunu} y resultan

$${R\over
{\Box}} = {2M\over{r}} \\\\~~~~  \mbox{and} ~~~~ \\\\  {P\over{\Box}}\ln 
{{-\Box\over{\mu^2}}} =  -{2M\over{r}}\ln {{\tilde \mu} r}.$$ 

Por lo tanto, evaluando las ecuaciones (\ref{ancon}) y (\ref{nanoool}) 
obtenemos la traza en cuatro dimensiones que, a primer orden en $M$, es

\begin{equation}
\langle T^{(4)}\rangle = - {1\over{8 \pi^2}}{M\over{r^5}} 
\ln {{\tilde \mu} r}\ = {C\over{r^5}}\ln {{\tilde \mu} r}.
\end{equation} 
Como esper\'abamos, las correcciones cu\'anticas al potencial newtoniano 
dependen de la escala $\mu$. Estas correcciones concuerdan cualitativamente 
con el resultado conocido en la literatura \cite{14tmunu}. Sin embargo, si 
la parte invariante de Weyl es ignorada, obtenemos:

\begin{equation}
\langle T^{(4)}\rangle = - {1\over{6}}{C\over{r^5}}\ ,
\end{equation}
que presenta el signo equivocado y llevar\'\i a concluir que existen 
efectos de apantallamiento gravitatorio producidos por campos de materia 
escalares.

\subsection{Creaci\'on cosmol\'ogica de part\'\i culas} 

Como otro ejemplo de la utilidad y significado f\'\i sico de la parte 
invariante de la acci\'on efectiva, en esta secci\'on calculamos la 
creaci\'on de part\'\i culas en una m\'etrica cosmol\'ogica. Consideramos la 
siguiente m\'etrica:

\begin{equation}ds^2 = a^2(t)[-dt^2 + dr^2] + a^2(t) r^2 d\Omega_2,
\end{equation}
donde $a(t) = 1 + \delta (t) $ con $\delta << 1$ y $\delta 
\rightarrow 0$ para el pasado y futuro lejanos. Llamamos $t$ al tiempo 
conforme. 

El n\'umero total de part\'\i culas creadas durante la evoluci\'on 
cosmol\'ogica est\'a determinada por la parte imaginaria de la 
acci\'on efectiva in-out. Esta acci\'on efectiva puede obtenerse a partir 
de la 
acci\'on efectiva Eucl\'\i dea reemplazando el propagador Eucl\'\i deo por 
el propagador de Feynman. Como  $P \approx \ddot \delta$, la 
aproximaci\'on en potencias de $P$ es adecuada para calcular la 
creaci\'on de part\'\i culas. Al orden m\'as bajo en $\delta$, la 
acci\'on Eucl\'\i dea est\'a dada por las ecuaciones 
(\ref{niWefacc}) y (\ref{avra}), donde los propagadores son los del 
espacio plano.

En el vac\'\i o conforme, los t\'erminos presentes en la parte an\'omala de 
la acci\'on efectiva son reales y locales (lo que traer\'a importantes 
consecuencias en relaci\'on con el l\'\i mite 
cl\'asico de los modelos cosmol\'ogicos en dos dimensiones, como se ver\'a 
en la secci\'on siguiente). S\'olo la parte invariante $S_{\rm ef}^{\rm I}$ 
es no-local y contiene una parte imaginaria que da la creaci\'on de 
part\'\i culas. 

Tomando la transformada de Fourier de la ecuaci\'on (\ref{avra}), y 
reemplazando $p^2 \rightarrow p^2 - i \epsilon$ obtenemos

\begin{equation}
S_{\rm ef}^{\rm in-out} = {1\over{16\pi^2}}\int d^2p \vert 
{\tilde P}(p)\vert^2
{1\over{p^2 - i \epsilon}} \ln{{p^2 - i \epsilon}\over{\mu^2}} + 
\mbox{t\'erminos locales}.
\end{equation} Por otra parte, utilizando el hecho que:

\begin{equation}
\ln{{p^2 - i \epsilon}\over{\mu^2}} = 
\ln \vert {{p^2}\over{\mu^2}}\vert - i \pi \theta (-p^2),
\end{equation}
el n\'umero total de part\'\i culas creadas est\'a dado por 

\begin{equation}
n_{\rm T} = {\rm Im} S_{\rm ef}^{\rm in-out} = - {1\over{16 \pi}} \int d^2p 
\vert {\tilde P}(p)\vert^2 {\theta (-p^2)\over{p^2}}.
\label{nt}
\end{equation}
Esta ecuaci\'on es general (no hemos dicho nada acerca de $P$), ahora bien, 
en el caso en que $P = P(t)$, $n_{\rm T}$ puede re-escribirse como:

\begin{equation}
n_T = {\rm Im} S_{\rm ef}^{\rm in-out} = {1\over{16 V \pi}} \int dp_0 
\vert {\tilde P}(p_0)\vert^2 {1\over{p_0^2}},
\label{zzz}
\end{equation}
donde $V$ es el volumen espacial.

Como la m\'etrica es asint\'oticamente plana para $t\rightarrow \pm \infty$, la
transformada de Fourier ${\tilde P}(p_0)$ se anula cuando $p_0 \rightarrow 0$. 
Por lo que $n_{\rm T}$ es una cantidad finita que representa el an\'alogo 
en dos dimensiones de la expresi\'on general en cuatro dimensiones (en el 
caso de $\xi =0$, $m =0$ y $C_{abcd}=0$) encontrado en \cite{18tmunu}:
\bea 
n_T^{(4D)} &=& {\pi^3\over{60}}\int ~d^4p ~\theta (p^2 - 4 m^2) \left(1 
- {4m^2\over{p^2}}\right)^{1\over{2}}\left\{\vert R(p)\vert^2
\left[60(\xi -{1\over{6}})^2
\right.\right.\nonumber \\ 
&-&\left.\left. 40 {m^2\over{p^2}}\left[\xi - {1\over{6}}+ {m^2\over{6p^2}}
\right]\right]+ \vert C_{\mu\nu\alpha\beta}(p)\vert^2 \left(1- {4m^2\over{p^2}}
\right)^2\right\}.\eea

\section{El l\'\i mite semicl\'asico}

En esta secci\'on aplicaremos el m\'etodo de la funcional de influencia 
en gravedad semicl\'asica con el objeto de iniciar el an\'alisis de la 
transici\'on cu\'antico-cl\'asica en modelos bi-dimensionales. Si\-guiendo 
la formulaci\'on desarrollada para los sistemas cu\'anticos abiertos, 
calculamos la funcional de influencia integrando los grados de libertad 
asociados a los campos de materia cu\'anticos. Este c\'alculo es exacto 
para el modelo CGHS y requiere de aproximaciones en modelos m\'as generales 
donde el dilat\'on aparece acoplado a los campos de materia. 

Como mostramos en el Cap\'\i tulo 3, la funcional de influencia 
est\'a intimamente relacionada a la acci\'on efectiva de camino 
temporal cerrado (AECTC) por lo tanto en lo sucesivo utilizaremos 
ambas denominaciones para referirnos al mismo objeto. Como la funcional de 
influencia nos provee de la evoluci\'on temporal de la matriz densidad 
reducida, es una herramienta fundamental para el estudio de la 
validez de la aproximaci\'on semicl\'asica. 

\subsection{C\'alculo exacto de la funcional de influencia}

Debido a la invariancia conforme del modelo CGHS podemos calcular de 
manera exacta la funcional de influencia y desarrollar un estudio detallado
 de la validez de la aproximaci\'on semicl\'asica. Como mostramos en la 
Secci\'on 4.1, los efectos cu\'anticos de los campos de materia producen 
la acci\'on efectiva CTC de la ecuaci\'on (\ref{explctp}) para el modelo 
CGHS \cite{cghs}. En la medida conforme esta acci\'on se reduce a:

\begin{eqnarray}S_{\rm ef}^{\rm CTC}[\rho^+,f^+,\rho^-,f^-;\Sigma ]&=& 
S_{\rm CGHS}(\rho^+,f^+) - S_{\rm CGHS}(\rho^-,f^-)\nonumber \\
&-&{N\over{6 \pi}}\int d^2x\int d^2y \partial_+\partial_-\rho^
a(x^{\pm}) ~G_{ab}(x^{\pm},y^{\pm})~\partial_+\partial_-\rho^b(y^{\pm})
.\label{cgctpefa}\end{eqnarray}

Integrando por partes, esta acci\'on efectiva puede descomponerse en la 
contribuci\'on cl\'asica, la diferencia entre los t\'erminos de Polyakov en 
cada rama, m\'as un t\'ermino de superficie:
\begin{eqnarray}S_{\rm ef}^{\rm CTC}[\rho^+,f^+,\rho^-,f^-;\Sigma ]&=& 
S_{\rm CGHS}(\rho^+,f^+) - S_{\rm CGHS}(\rho^-,f^-)\nonumber \\
&-&{N\over{12 \pi}}\int d^2x  ~~ [\rho^+\partial_+\partial_-\rho^+  
- \rho_- \partial_+\partial_-\rho^-]\nonumber \\
&-&{N\over{6 \pi}}\{\mbox{t\'erminos de superficie}\},
\label{supterms}\end{eqnarray}
donde
\begin{eqnarray}\{\mbox{term. de sup.}\}
&=&  \int_{-\infty}^{+\infty} 
dx \int_{-\infty}^{+\infty} dy \left[\partial_{x^-}\Delta (x,k(x))~ 
N_1[x,k(x);y,{\bar k}(y)]
~ \partial_{y^-}\Delta (y,{\bar k}(y))\right. \nonumber \\
&+& \left. 2 \partial_{x^-}\Xi (x,k(x))~ N_2[x,k(x);y,{\bar k}(y)] ~ 
\partial_{y^-}\Delta (y,{\bar k}(y)) 
\right. \nonumber \\
&+& \left. \partial_{x^-}\Xi (x,k(x))~ N_3[x,k(x);y,{\bar k}(y)]~ 
\partial_{y^-}\Xi (y,{\bar k}(y))\right]
\nonumber \\
&-& 2\int_{-\infty}^{+\infty} dx \int_{-\infty}^{+\infty}dy 
\left[\Xi (x,k(x)) 
~ \partial_{x^+} N_2[x,k(x);y,{\bar k}(y)] ~ \partial_{y^-} 
\Delta (y,{\bar k}(y))\right. 
\nonumber \\
&+&  \left.\Xi (x,k(x)) ~ \partial_{x^+} N_4[x,k(x);y,{\bar k}(y)] ~ 
\partial_{x^-} \Xi (y,{\bar k}(y))
\right. \nonumber \\
&-& \left. {1\over{2}}\Xi (x,k(x))~ \partial_{x^+} \partial_{y^+} 
N_4[x,k(x)
;y,{\bar k}(y)]
~ \Xi (y,{\bar k}(y))\right]\label{exbt}.\end{eqnarray}
La superficie de empalme $\Sigma$ est\'a definida por 

$t_x=k(x)$, $t_y={\bar k}(y)$; 
$\Delta ={1\over{2}}( \rho^+-\rho^-)$, $\Xi = {1\over{2}}(\rho^+ 
+ \rho^-)$, y lon n\'ucleos $N_i$ son
\begin{eqnarray}N_1&=&G_{++}+G_{+-}-G_{-+}-G_{--}\nonumber \\
N_2&=&G_{++}+G_{+-}+G_{-+}+G_{--}\nonumber \\
N_3&=&G_{++}-G_{+-}-G_{-+}+G_{--}\nonumber \\
N_4&=&G_{++}-G_{+-}+G_{-+}-G_{--}.\end{eqnarray}

La expresi\'on (\ref{supterms}) para la acci\'on efectiva es totalmente 
general, y puede aplicarse a cualquier m\'etrica en la medida 
conforme. Si la acci\'on de influencia tiene una parte imaginaria no-trivial, 
\'esta debe estar contenida en los t\'erminos de superficie. Si ambas 
m\'etricas coinciden asint\'oticamente en el futuro, y si la superficie 
de empalme pertenece a tal regi\'on, todos los t\'erminos de 
superficie se anulan debido a las relaciones usuales entre las 
funciones de Green , las cuales son v\'alidas en la regi\'on plana; $N_4$ y 
$\Delta$ son simult\'aneamente cero y s\'olo el t\'ermino de la anomal\'\i a 
sobrevive dado que todos los t\'erminos de superficie son proporcionales a 
$\Delta$ y/o a $N_4$. 

El hecho que la acci\'on de influencia pueda calcularse f\'acilmente a partir 
de la acci\'on Eucl\'\i dea es un resultado muy importante. La superficie de 
empalme $\Sigma$ tiene un rol crucial, en la medida conforme toda la 
informaci\'on relevante acerca de la transici\'on cu\'antico-cl\'asica 
est\'a contenida en los t\'erminos de superficie que son dependientes de 
$\Sigma$. 

\subsection{Funcional de influencia para historias cosmol\'ogicas: l\'\i mite
cl\'asico}

En esta secci\'on evaluamos expl\'\i citamente la funcional de influencia 
para m\'etricas cosmol\'ogicas y calculamos su parte imaginaria. Como la 
funcional de influencia depende de la superficie de empalme, mostramos 
c\'omo depende el proceso de 
p\'erdida de coherencia de la elecci\'on de tal superficie. 

Llamamos $\cal M$ y $\tilde{\cal M}$ a cada uno de los espacio-tiempos 
descriptos por  $g_{\mu\nu}^+$ y $g_{\mu\nu}^-$ respectivamente. Asumimos que
 ambos espacio-tiempos son asint\'oticamente planos en el pasado y que ellos 
coinciden sobre una hipersuperficie espacial $\Sigma$. Siempre es posible 
definir una hipersuperficie $\Sigma_{\cal M}$ en ${\cal M}$ por medio de una 
relaci\'on entre $t$ y $x$, digamos $t=k(x)$. Lo mismo para una 
superficie  $\Sigma_{\tilde{\cal M}}$ en $\tilde{\cal M}$ por medio de
 $\bar t={\bar k}(\bar x)$. Nosotros queremos identificar 
$\Sigma_{\cal M}$ y
$\Sigma_{\tilde{\cal M}}$ en una hipersuperficie com\'un $\Sigma$. Por lo 
tanto 
debemos introducir un mapa entre los puntos sobre ambas hipersuperficies  
a trav\'es de la identificaci\'on de la geometr\'\i a intr\'\i nseca local. 
En los 
modelos bi-dimensionales, una definici\'on invariante de la uno-geometr\'\i a 
est\'a provista por el valor del dilat\'on $\phi (s)$, como una funci\'on 
 de la distancia propia a lo largo de la hipersuperficie. La identificaci\'on 
de la uno-geometr\'\i a implica que para la misma distancia propia 
(medida respecto de alg\'un punto de referencia 
arbitrario) $ds^2=d\bar s^2$, el dilat\'on debe tener el mismo valor para 
cada una de las geometr\'\i as sobre $\Sigma$, es decir: $\phi^+(s)
=\phi^-(\bar s)$. Entonces,  $d\phi^+/ds=d\phi^-(\bar s)/d\bar s$. Dados 
dos espacio-tiempos y la funci\'on $k$ que define a la superficie 
$\Sigma_{\cal M}$ en  $\cal M$, las condiciones impuestas por la 
identificaci\'on permiten determinar $\bar k$, y por ende 
$\Sigma_{\tilde{\cal M}}$ en $\tilde {\cal M}$. 

Consideremos dos m\'etricas cosmol\'ogicas caracterizadas por las funciones 
$\rho^+(t)$ y $\rho^-(t)$. Para calcular la funcional de influencia es 
necesario conocer expl\'\i citamente las funciones de Green $G_{ab}$. Como 
ambas m\'etricas son planas en el pasado remoto y son conformes al 
espacio de Minkowski, los propagadores en el estado de vac\'\i o in tienen 
la misma estructura funcional que para el espacio plano, por ejemplo, 
el propagador de Feynman est\'a dado por 
\begin{eqnarray}G_{++}(x,y)&=&i\langle 0, in\vert T{\hat f}^+(x)
{\hat f}^+(y)\vert 0,
 in\rangle\nonumber \\  
&=&{1\over{{2 \pi}^2}}\int d^2p {e^{ip(x-y)}\over{p^2 + i \epsilon}}
=- {{2\pi i}\over{{2 \pi}^2}}\int_0^\infty {dp\over{p}} e^{-i p 
(x - y)} e^{-i p\vert t_x - t_y\vert}\nonumber \\
&=& {\pi\over{2}}Sgn[\vert t_x - t_y\vert + x - y] - i Log\vert 
t_x - t_y + x - y\vert + C,\end{eqnarray}
donde $C$ es una constante indeterminada (que proviene de la divergencia 
infraroja cuando $p\rightarrow 0$). Expresiones similares se pueden 
escribir para las dem\'as funciones de Green. Es importante notar 
que en $G_{+-}(x,y)$ y $G_{-+}(x,y)$ las coordenadas $x$ e $y$ corresponden 
a diferentes espacio-tiempos.

\subsection{Hipersuperficies de tiempo constante}

Consideramos una superficie $\Sigma_{\cal M}$ en ${\cal M}$, definida por 
$t=T$. Para realizar el ``empalme'' entre las hipersuperficies, debemos 
imponer que  $\phi^+=\phi^-$ sobre $\Sigma$. Como $\phi^-$ es una 
constante sobre $\Sigma_{\tilde{\cal M}}$, esta hipersuperficie debe 
ser tambi\'en de tiempo constante ${\bar t} = {\bar T}$. Cambiando la escala 
temporal siempre podemos tomar ${\bar T} = T$. 

La AECTC para m\'etricas cosmol\'ogicas puede escribirse como
\begin{eqnarray}S_{\rm ef}^{\rm CTC}&=& S_{\rm CGHS}(\rho^+,f^+) - S_{\rm CGHS}
(\rho^-,f^-)\nonumber \\
&-&{N\over{6 \pi}}\int_{-\infty}^T dt_x\int_{-\infty}^T dt_y 
~~\ddot{\rho}^a(t_x)~~\ddot{\rho}^b(t_y)~~\int_{-\infty}^{+\infty}
 dx\int_{-\infty}^{+\infty} dy ~~G_{ab}(x,y)
,\label{cosmoctpefa}\end{eqnarray} donde los \'\i ndices $a$ y $b$ denotan 
nuevamente las diferentes ramas del CTC. Utilizando regularizaci\'on 
dimensional, podemos evaluar 

\begin{equation}\int_{-\infty}^{+\infty} dx\int_{-\infty}^{+\infty} 
 dy ~~G_{++}= {\Omega\over{2}}~
\vert t_x - t_y\vert,\label{intprop}\end{equation}
donde $\Omega$ es un factor de volumen global. Similares expresiones 
pueden obtenerse para $G_{--}$ y $G_{+-}$. Reemplazando esta expresi\'on 
en (\ref{cosmoctpefa}) podemos demostrar que:
\begin{eqnarray}S_{\rm ef}^{\rm CTC}&=& S_{\rm CGHS}(\rho^+,f^+) - 
S_{\rm CGHS}
(\rho^-,f^-)\nonumber \\
&-&{N\over{12 \pi}}\int d^2x  ~~ [\rho^+\partial_+\partial_-
\rho^+  - \rho_- \partial_+\partial_-\rho^-].\end{eqnarray}

Como se ve de esta expresi\'on no existe parte imaginaria y/o no-local en 
esta acci\'on efectiva. La \'unica correcci\'on al t\'ermino cl\'asico 
proviene de la anomal\'\i a de traza. La consecuencia de este hecho 
es que la funcional de p\'erdida de coherencia es identicamente uno. Para que 
la aproximaci\'on semicl\'asica sea v\'alida, la funcional de p\'erdida de 
coherencia debe ser diagonal para geometr\'\i as del espacio-tiempo 
macrosc\'opicamente diferentes, a\'un si ellas coinciden sobre una 
hipersuperficie espacial. Por lo tanto podemos concluir que 
debido a la invarianza conforme, los modelos cosmol\'ogicos en 
dos dimensiones no tienen un l\'\i mite cl\'asico bien definido. 

\subsection{Hipersuperficies m\'as generales}

Para mostrar expl\'\i citamente la dependencia de los resultados 
con la hipersuperficie, calculamos la funcional de influencia para 
hipersuperficies m\'as generales. 

Debemos calcular expresiones del tipo 

\begin{equation}\int dx\int dy \int_{-\infty}^{\Sigma} dt_x\int_{-\infty}^{
\Sigma} dt_y \ddot{\rho}^a(t_x)\ddot{\rho}^b(t_y)G_{ab}(x,y),
\label{4t}\end{equation}
donde hemos definido las hipersuperficies en cada rama como:
   
\begin{equation}k(x) = T + \Delta k^+(x),\nonumber\end{equation}
y
\begin{equation}{\bar k}(x) = T + \Delta k^-(x).\nonumber\end{equation}
 Consideremos que $\Delta k^+(x)$ y $\Delta k^-(x)$ son peque\~nas 
fluctuaciones alrededor de la superficie $t=T$ y calculamos la funcional de
influencia a segundo orden en una expansi\'on en potencias de dichas 
fluctuaciones. Obviamente, el orden cero s\'olo da la anomal\'\i a de traza. 
A primer orden la funcional de influencia s\'olo tiene t\'erminos reales 
\cite{if} (t\'erminos de superficie que no contriuyen a las 
ecuaciones de movimiento). El segundo orden est\'a dado por

 \begin{equation}\ddot{\rho}^a(T) \ddot{\rho}^b(T)\int_{-\infty}
^{+\infty} dx 
\int_{-\infty}^{+\infty} dy ~~G_{ab}(x,T;y,T)\Delta k^a(x) 
\Delta k^b(y).
\end{equation} Introduciendo los propagadores y haciendo las integrales 
espaciales, la parte imaginaria resulta ser proporcional a:
\begin{eqnarray}&&4 \pi i \int_0^{+\infty}{dp\over{p}}\left[
 \ddot{\rho}^{2+}(T) \Delta k^{+2}(p)-  \ddot{\rho}^+(T)
\ddot{\rho}^-(T) \Delta k^{+\star}(p) \Delta k^-(p)\right.
\nonumber \\
&&-\left. \ddot{\rho}^-(T)
\ddot{\rho}^+(T)\Delta k^+(p) \Delta k^{-\star}(p)+ 
\ddot{\rho}^{-2}(T) \Delta k^{-2}(p)\right],
\label{im}\end{eqnarray}
donde $\Delta k^+(p)$ y $\Delta k^-(p)$ son las transformadas de Fourier 
de las funciones de perturbaci\'on. Para describir el empalme entre las 
hipersuperficies debemos resolver las ecuaciones, 

\begin{equation}\phi^+[k(x)]= \phi^+[T + \Delta k^+(x)] = \phi^-
[{\bar k}(y)]= 
\phi^-[T + \Delta k^-(y)].\label{embedding}\end{equation} Esta 
identificaci\'on puede describirse por medio de la funci\'on $y(x)$ 
entre las coordenadas sobre $\Sigma$ en cada uno de los espacio-tiempos. 
Adem\'as debemos imponer que los intervalos sean iguales en cada 
espacio-tiempo sobre $\Sigma$, por lo tanto:

 \begin{equation}\left[{dx\over{dy}}\right]^2={1 - \left({d{\bar k}
\over{dy}}
\right)^2\over{1 - \left({dk\over{dx}}\right)^2}}
{e^{{\rho}^-[{\bar k}(y)]}\over{e^{\rho^+[k(x)]}}}.\end{equation}

Expandiendo la ecuaci\'on (\ref{embedding}) para $\Delta k^+(p)$ y 
$\Delta k^-(p)$ peque\~nos, y teniendo en cuenta que 
$y=x+{\cal O}(\Delta k^2)$, encontramos que 

\begin{equation}\Delta k^+(x) \cong \Delta k^-(y) {\dot{\phi}^-(T)
\over{\dot{\phi}^+(T)}}\cong \Delta k^-(x) {\dot{\phi}^-(T)
\over{\dot{\phi}^+(T)}}.\label{impordess}\end{equation}
Por lo tanto la parte imaginaria puede re-escribirse como

\begin{equation}4 \pi i \left[\ddot{\rho}^+(T)-{\dot{\phi}^+(T)
\over{\dot{\phi}^-(T)}}\ddot{\rho}^-(T)\right]^2\int_0^\infty 
{dp\over{p}}
\vert \Delta k^+(p)\vert^2.\label{imif}\end{equation} En consecuencia 
existe una parte 
imaginaria de la funcional de influencia distinta de cero para peque\~nas 
fluctuaciones alrededor 
de la superficie de 
tiempo constante. El valor absoluto de la funcional de p\'erdida de 
coherencia est\'a dado por 

\begin{equation}\left\vert {\cal D}[\rho^+,\rho^-;\Sigma]\right\vert
 \approx e^
{- 4\pi\left[\ddot{\rho}^+(T)
-{\dot{\phi}^+(T)
\over{\dot{\phi}^-(T)}}\ddot{\rho}^-(T)\right]^2\int_0^\infty {dp\over{p}}
\vert \Delta k^+(p)\vert^2   }.\end{equation}

En este punto debemos aclarar que un criterio razonable para asegurar 
que la transici\'on cu\'antico-cl\'asica es posible es exigir que 
$\vert {\cal D}\vert \ll 1$ para 
toda hipersuperficie de empalme $\Sigma$. Est\'a claro que esta 
condici\'on no se satisface para la superficie m\'as simple en el 
caso cosmol\'ogico. El objetivo de los c\'alculos realizados en esta secci\'on 
es mostrar la dependencia con $\Sigma$ de la acci\'on de influencia 
(ecuaci\'on (\ref{imif})). El ejemplo de la ecuaci\'on (\ref{imif}) muestra 
la dependencia con $\Sigma$ de la acci\'on de influencia en un ejemplo 
sencillo. La importancia de tal ejemplo se pone de manifiesto tambi\'en 
cuando se estudia la validez de la aproximaci\'on 
semicl\'asica en las cercan\'\i as de un agujero negro \cite{ortiz,ortiz2}. Si
 la 
aproximaci\'on semicl\'asica es correcta, la funcional de onda de los 
campos cu\'anticos no puede depender fuertemente de la masa del agujero 
negro. En particular, si consideramos dos espacio-tiempos diferentes, uno 
descripto por el colapso de un agujero negro de masa $M$ y el otro 
el de un agujero negro con masa $M + \Delta M$, funcionales de 
onda similares al inicio del colapso no pueden diferir mucho cuando 
el agujero negro ya se ha formado (si $\Delta M$ es peque\~no). Para 
comparar ambas funcionales debemos hacer coincidir la superficie $\Sigma$ 
en ambos espacio-tiempos y calcular el producto interno de las funcionales 
de onda sobre $\Sigma$. En \cite{ortiz} fue demostrado que para ciertas 
superficies, este producto es arbitrariamente chico para geometr\'\i as 
de colapso cl\'asicas; mientras que en \cite{ortiz2} el producto es 
de orden uno si se considera la retro-reacci\'on cu\'antica. 

El producto interno referido en el p\'arrafo anterior (calculado en 
\cite{ortiz} y \cite{ortiz2}) es exactamente la funcional de influencia 
evaluada 
a trav\'es de $S_{\rm ef}^{\rm CTC}$, para geometr\'\i as de colapso de 
agujeros negros con masas $M$ y $M + \Delta M$ y superficie de empalme 
$\Sigma$. Por lo tanto, toda la informaci\'on acerca de la aproximaci\'on 
semicl\'asica en las cercan\'\i as de un agujero negro debe estar contenida
en los t\'erminos de superficie que aparecen en la funcional de influencia 
de la ecuaci\'on (\ref{supterms}). En realidad, el resultado de la ecuaci\'on 
(\ref{supterms})
 es completamente general y vale para todo par de m\'etricas y para toda 
hipersuperficie de empalme $\Sigma$.       
 
Volviendo a la situaci\'on cosmol\'ogica, la interpretaci\'on f\'\i sica de 
los resultados obtenidos en esta secci\'on 
es la siguiente: nosotros hemos elegido el vac\'\i o in como estado 
cu\'antico de los campos de materia. 
Para hipersuperficies $t=T$, siempre podemos elegir una base out tal que 
sea el vac\'\i o conforme en ambos espacio-tiempos. Por lo
 tanto, los coeficientes de Bogoliubov entre las bases in y out son triviales 
en ambas geometr\'\i as. La funcional de influencia es real 
y no hay p\'erdida de coherencia. Para hipersuperficies m\'as generales, 
podemos elegir como base out el vac\'\i o conforme en uno de los 
espacio-tiempos, pero esta base, en general, no corresponde al 
vac\'\i o conforme en el otro espacio-tiempo. Por lo tanto las bases 
in y out son diferentes en ambos espacio-tiempos, existe creaci\'on de 
part\'\i culas y por ende aparece una parte imaginaria. 

Hemos mostrado que la funcional de influencia tiene una parte 
imaginaria para algunas hipersuperficies, y que esta parte imaginaria se 
anula para la m\'as com\'un de todas las posibles elecciones, las de tiempo 
constante. El valor absoluto de la funcional de p\'erdida de 
coherencia tambi\'en depende de la hipersuperficie $\Sigma$. 

\section{C\'alculos aproximados de la funcional de influencia}
  
En la secci\'on anterior utilizamos la acci\'on efectiva para el modelo 
bi-dimensional de gravedad dilat\'onica, donde s\'olo los campos de 
materia fueron cuantizados, para evaluar de manera exacta la funcional de 
influencia y discutir el mecanismo de p\'erdida de coherencia cu\'antica. Si 
queremos tener una descripci\'on completa de 
los efectos cu\'anticos en el modelo CGHS, debemos considerar las 
fluctuaciones cu\'anticas del dilat\'on y de la m\'etrica. En la literatura, 
estas fluctuaciones fueron consideradas para el c\'alculo de la acci\'on 
efectiva in-out \cite{24if}. La cuantizaci\'on de estos campos implica 
la necesidad de desarrollar c\'alculos perturbativos en lazos y expansiones 
en potencias de la curvatura. Por ejemplo, para evaluar las 
correcciones cu\'anticas que provienen del campo dilat\'on, podemos 
separarlo en don contribuciones: un fondo cl\'asico m\'as una 
fluctuaci\'on cu\'antica peque\~na

\begin{equation}\phi (x) =\phi_0 (x) + \hat{\phi}(x),\end{equation}   
por lo tanto, introduciendo esta divisi\'on en la acci\'on cl\'asica, y 
a orden cuadr\'atico en las fluctuaciones cu\'anticas podemos escribir la 
contribuci\'on dilat\'onica a la acci\'on cl\'asica como 
\begin{eqnarray}S_{\phi} &=& {1\over{2 \pi}} \int d^2x \sqrt{-g(x)} 
\left\{e^{-2\phi_0}[R + 4 (\partial \phi_0)^2 + 4 \lambda^2]\right. 
\nonumber \\
&+&\left. 4 (\partial \psi)^2+ 2 \left[R(x) + 2 (\partial \phi_0)^2
+ 4 \partial^2\phi_0\right]\psi^2+8 \lambda^2 \psi^2\right\},\label{psidil}
\end{eqnarray}
donde re-definimos $\psi = e^{-\phi_0}\hat{\phi}$. 

En consecuencia, la acci\'on para las fluctuaciones del dilat\'on $\psi$ 
corresponde a la de un campo escalar masivo, acoplado de manera 
arbitraria a la curvatura y al dilat\'on cl\'asico de fondo \cite{if}. En la 
Secci\'on 4.2.1, obtuvimos un modelo bi-dimensional de 
juguete a partir de modelos en cuatro dimensiones con simetr\'\i a esf\'erica, 
lo que representa otro ejemplo en la misma direcci\'on.
 En ese caso la acci\'on para los campos de materia es Ec. 
(\ref{trola}) 

\be
S_{\psi} = -{1\over{2}}\int d^2x ~ \sqrt{g}\left[
(\partial \psi)^2 + P~\psi^2\right],\label{psidil2} 
\ee
que corresponde a un campo escalar no-masivo acoplado a la geometr\'\i a.

Por lo tanto, para obtener las contribuciones cu\'anticas en casos m\'as 
generales, hay que tener en cuenta acciones del tipo de (\ref{psidil}) o 
(\ref{psidil2}). Dado que en la Secci\'on 4.2 calculamos la acci\'on 
efectiva Eucl\'\i dea 
correspondiente a la ecuaci\'on (\ref{psidil2}), en esta secci\'on, a modo 
de ejemplo de los casos m\'as generales,  
vamos a calcular la correspondiente funcional de influencia y mostraremos 
c\'omo aparece una parte imaginaria 
que produce los efectos de p\'erdida de coherencia a trav\'es de la funcional 
de Hartle y Gell-Mann (para el caso de la acci\'on (\ref{psidil}) ver la Ref. 
\cite{if}).

Partimos de la acci\'on efectiva Eucl\'\i dea, a orden 
cuadr\'atico en $P$ y $R$ de (\ref{niWefacc}) y 
(\ref{avra}):
\begin{eqnarray}
S_{\rm ef} &=& -{1\over{8\pi}}
\int d^2x ~ \sqrt{g}
\int d^2y ~ \sqrt{g} \left\{{1\over{12}} R(x) {1\over{\Box}} R(y) -
 P(x) {1\over{\Box}} R(y) \right\} \nonumber\\ 
&-& {1\over{8\pi}}\int d^2x ~ \sqrt{g}
\int d^2y ~ 
\sqrt{g} ~ P(x) {1\over{\Box}}\ln {-\Box\over{\mu^2}} P(y),
\label{1}\end{eqnarray}
que corresponde a la expansi\'on en potencias de $P$ de la secci\'on anterior.
 Para obtener la acci\'on efectiva de camino temporal cerrado, seguimos el 
mismo procedimiento que en el c\'alculo exacto de la funcional de influencia 
explicado anteriormente. Debemos reemplazar los propagadores Eucl\'\i deos por
los correspondientes al CTC. Antes de hacer este reemplazo es necesario 
tener en cuenta que en el \'ultimo t\'ermino de (\ref{1}) aparece un 
n\'ucleo no-local que no es el propagador Eucl\'\i deo estrictamente. Debemos 
notar que este n\'ucleo puede ser escrito de la 
siguiente manera en t\'erminos de propagadores masivos en el espacio 
Eucl\'\i deo:

\be
{1\over{\Box}}\ln {-\Box\over{\mu^2}} = \lim_{\delta \rightarrow 0}\left[ 
-\int_0^\infty dz {G_{z\mu^2}\over{(z + \delta)}} + G_{-\delta \mu^2}
\ln\delta \right],\label{2}\ee
donde 

$$G_{z\mu^2}= {1\over{p^2 + z \mu^2}}$$ $$G_{-\delta\mu^2}=
{1\over{p^2-\delta\mu^2}}.$$

Ahora si, reemplazando los propagadores Eucl\'\i deos por los del CTC, la 
acci\'on de influencia para este caso es:

\be
\Gamma^{\rm CTC}= -{1\over{8\pi}}\int d^2x ~\int d^2y ~\left\{{1\over{12}}
R_a ~ G_{ab}~ R_b - ~P_a ~ G_{ab}~ P_b + ~P_a ~ F_{ab} ~ 
P_b\right\},
\ee
donde $F_{ab} = \lim_{\delta \rightarrow 0}[ 
-\int_0^\infty dz {G^{z\mu^2}_{ab}\over{(z + \delta)}} + G^{-\delta \mu^2}_{ab}
\ln\delta ]$. Los \'\i ndices $a$ y $b$ denotan las ramas temporales 
del CTC. En la medida conforme, los dos primeros t\'erminos se localizan 
aportando t\'erminos de superficie. Por lo tanto la acci\'on de influencia 
en este caso es:
\bea 
\Gamma^{\rm CTC} &=& -{1\over{12\pi}}\int d^2x~\left[ \rho^+~\partial_+
\partial_-
\rho^+ - ~\rho^- ~\partial_+\partial_-\rho^-\right]\nonumber \\
&-& {1\over{8\pi}}\int d^2x~\left[\rho^+ ~ \Delta P - \rho^- ~ \Delta P\right]
\nonumber \\
&-& {1\over{\pi}}\int d^2x~\int d^2y~ \Delta P(x) ~\Sigma P(y)~ 
\theta (x^0 -y^0)
~\int {d^2p\over{(2\pi )^2}}~ e^{ip(x - y)}~ {1\over{p^2}}~
\ln \vert {p^2\over{\mu^2}}\vert \nonumber \\
&+& {i\over{2\pi}}\int d^2x~\int d^2y~ \Delta P(x)~ \Delta P(y) 
~\int {d^2p\over{(2\pi )^2}}~ e^{ip(x - y)}~ {\theta (-p^2)
\over{p^2}}\nonumber \\
&+& \{\mbox{t\'erminos de superficie}\},\label{3}\eea 
donde $\Delta P = P^+ - P^-$. 

Finalmente, el valor absoluto de la funcional de p\'erdida de coherencia 
tiene dos contribuciones: la parte imaginaria 
de los t\'erminos de superficie m\'as el \'ultimo t\'ermino de (\ref{3})

\begin{equation}\left\vert {\cal D}[\rho^+,\rho^-]\right\vert 
\approx e^{-{1\over{16\pi}}
\int d^2p~\vert {\Delta\tilde P}(p)\vert^2 {\theta (-p^2)
\over{p^2}}}~~e^{-\mbox{Im}~
\left[\mbox{t\'erminos de superficie}\right]}
.\label{compltDF}
\end{equation}
Es importante ver que el primer factor del lado derecho de esta ecuaci\'on 
tiene la misma forma que el t\'ermino de creaci\'on de part\'\i culas para 
m\'etricas cosmol\'ogicas (ecuaci\'on (\ref{nt})). Aqu\'\i \ $P$ ha 
sido reemplazada por la 
funci\'on $\Delta P = P^+ - P^-$ debido a que 
estamos haciendo el c\'alculo a partir de la acci\'on efectiva de 
camino temporal cerrado. La ecuaci\'on (\ref{nt}) fue obtenida a partir de la 
acci\'on efectiva in-out. 
 
En particular, para aquellas situaciones donde los t\'erminos de 
superficie se anulen, los efectos de p\'erdida de coherencia 
persisten debido al t\'ermino invariante proporcional a $P^2$. Un modelo 
m\'as realista deber\'\i a incluir los efectos de los gravitones. En ese 
caso m\'as realista donde las fluctuaciones cu\'anticas de la m\'etrica 
son tenidas en cuenta, aparecer\'a una parte imaginaria extra, pero la 
conclusi\'on acerca de c\'omo el efecto de p\'erdida de 
coherencia aparece en estos modelos seguir\'a siendo v\'alida.  

\section{Discusi\'on}

En este cap\'\i tulo hemos aplicado el m\'etodo de la funcional de influencia 
a modelos de gravedad escalar-tensorial en dos dimensiones. Estos modelos
 presentan simplificaciones importantes res\-pecto a los modelos generales en 
cuatro dimensiones. Para el modelo CGHS hemos mostrado que la 
acci\'on de influencia puede calcularse exactamente y que es un 
objeto fuertemente dependiente de la hipersuperficie donde se 
empalman las geometr\'\i as. En particular, en la medida conforme 
la AECTC puede escribirse como la diferencia entre las acciones de Polyakov 
en cada rama temporal m\'as una integral sobre $\Sigma$. 

Utilizamos esta acci\'on de influencia para derivar las ecuaciones de 
movimiento semicl\'asicas. Estas ecuaciones son reales, causales y 
no-locales, y se tornan locales en la medida conforme. La derivaci\'on de 
estas ecuaciones no es trivial debido a la dependencia con la 
m\'etrica de las funciones de Green. Este es un hecho muy importante porque 
permite extender el c\'alculo a casos donde no se puede determinar el 
$\langle T_{\mu\nu}\rangle$ utilizando la ley de conservaci\'on y la 
anomal\'\i a de traza \'unicamente. En referencia a este \'ultimo caso 
mostramos c\'omo obtener el tensor de energ\'\i a-impulso para modelos 
donde el dilat\'on aparece acoplado al campo escalar de materia. En estos 
modelos el $\langle T_{\mu\nu}\rangle$ no se conserva y la anomal\'\i a de 
traza no lo determina completamente. Seguidamente extendimos la discusi\'on al 
c\'alculo correcto de la radiaci\'on de Hawking y otros observables 
f\'\i sicos en este modelo. 

Hemos estudiado la transici\'on cu\'antico-cl\'asica en modelos 
cosmol\'ogicos. La funcional de influencia no tiene una parte imaginaria para 
algunas superficies de empalme. Por lo tanto la aproximaci\'on semicl\'asica 
no es v\'alida en estos modelos.  

La aproximaci\'on semicl\'asica puede obtenerse de manera consistente 
cuando inclu\'\i mos en la cuantizaci\'on al dilat\'on  
para el modelo CGHS \cite{if}, o cuando estudiamos la funcional de 
influencia para 
campos escalares masivos y/o acoplados al dilat\'on o a la geometr\'\i a de 
manera no conforme. 

Para finalizar, es necesario mencionar que la geometr\'\i a en los 
modelos bi-dimensionales est\'a determinada por la m\'etrica y el 
dilat\'on. Por ejemplo, cuando restringimos los modelos en cuatro dimensiones 
con simetr\'\i a esf\'erica (como el \'ultimo ejemplo del cap\'\i tulo), el 
dilat\'on es parte de la geometr\'\i a dado que $e^{-2\phi}$ es el radio de 
la dos-esfera. En general, en el modelo CGHS, los lazos asociados a las 
fluctuaciones cu\'anticas
 del dilat\'on y de la m\'etrica eran ignorados argumentando el l\'\i mite 
de N grande (N es el n\'umero de campos escalares de materia presentes en el 
modelo). Como pudimos demostrar en este cap\'\i tulo no importa cu\'an 
grande es N, no hay p\'erdida de coherencia si uno no considera el efecto de 
las fluctuaciones cu\'anticas del dilat\'on y/o de la geometr\'\i a para el 
modelo CGHS. Para modelos m\'as generales, donde existe un acoplamiento entre 
el dilat\'on y el campo de materia, la misma interacci\'on asegura la 
aparici\'on de una parte imaginaria en la acci\'on de influencia y el 
efecto de p\'erdida de coherencia est\'a asegurado.

%% file: tesis7
\newpage

\thispagestyle{empty}
~
\newpage

\chapter{Las ecuaciones de Einstein - Langevin}
  
\thispagestyle{empty}

Como dijimos anteriormente, las ecuaciones semicl\'asicas para la gravedad  
son la generalizaci\'on de las ecuaciones de Einstein cuando la materia, 
fuente de estas ecuaciones, es considerada cu\'anticamente. La fuente de 
estas ecuaciones es el valor de expectaci\'on de vac\'\i o del tensor 
de energ\'\i a-impulso de los campos de materia. Debido a la presencia de 
divergencias en la teor\'\i a (dado que la relatividad general no es 
una teor\'\i a renormalizable), la ecuaci\'on semicl\'asica (\ref{see}) del 
cap\'\i tulo anterior usualmente se escribe como

\begin{equation}
{1\over{8\pi G}}\left[R_{\mu\nu}-\frac{1}{2}Rg_{\mu\nu}\right]- 
\alpha H^{(1)}_{\mu\nu} - \beta H^{(2)}_{\mu\nu}= T_{\mu\nu}^{\rm clas}+ 
\langle T_{\mu\nu}\rangle,
\label{see2}
\end{equation}
donde los t\'erminos proporcionales a 

\begin{equation} 
H^{(1)}_{\mu\nu} = \left[4 R_{;\mu\nu}-4g_{\mu\nu}\Box R\right]+ 
O(R^2),\end{equation} 
\begin{equation} H^{(2)}_{\mu\nu}=\left[4R_{\mu}{}^\alpha{}_{;\nu\alpha}-2\Box 
R_{\mu\nu}- g_{\mu\nu}\Box R\right]+ O(R^2),\end{equation}
provienen de t\'erminos cuadr\'aticos en la curvatura presentes en la 
acci\'on gravitatoria, los cuales son necesarios para renormalizar la 
teor\'\i a. 

Este teor\'\i a, que asume 
a la gravedad como a un campo cl\'asico, es v\'alida para escalas mayores que
la de Planck y cuando las fluctuaciones cu\'anticas del tensor de 
energ\'\i a-impulso son peque\~nas respecto al valor medio, es decir que 
cuando vale \cite{ford}

\be \langle T_{\mu\nu}(x)T_{\rho\sigma}(y)\rangle \approx \langle 
T_{\mu\nu}(x)\rangle \langle T_{\rho\sigma}(y)\rangle,\ee
por lo tanto la descripci\'on semicl\'asica no ser\'\i a apropiada para 
aquellos estados de campo en los cuales el tensor de energ\'\i a-impulso 
presenta fluctuaciones grandes. Estas fluctuaciones, sin embargo, pueden 
tenerse en cuenta incluyendo un t\'ermino estoc\'astico adicional del lado 
derecho de (\ref{see2}) \cite{einstlang,verda1,verda2}. Estos t\'erminos de 
ruido adicionales provienen de la parte imaginaria de la acci\'on efectiva 
de camino temporal cerrado; por lo tanto las ecuaciones de Einstein 
semicl\'asicas pasan a ser ecuaciones de {\it Einstein-Langevin}, que incluyen
 efectos tanto disipativos como difusivos de los campos de materia sobre 
la geometr\'\i a del espacio-tiempo.

Estas ecuaciones de Einstein-Langevin (EEL) fueron derivadas 
en la literatura, para perturbaciones peque\~nas de la m\'etrica 
con acoplamiento conforme a campos escalares sin masa sobre un 
fondo plano \cite{verda2}, y en escenarios cosmol\'ogicos, para 
campos masivos en universos de Robertson-Walker planos \cite{7el}, o 
en modelos de Bianchi-I \cite{8el}. En este cap\'\i tulo presentamos 
una derivaci\'on de las EEL para un campo cu\'antico, en una expansi\'on 
covariante en potencias de la curvatura. En particular, mostramos que 
cuando el campo escalar no tiene masa, las EEL est\'an completamente 
determinadas por la dependencia de las constantes de la teor\'\i a 
con la escala de energ\'\i a, a trav\'es de las ecuaciones del grupo de 
renormalizaci\'on (que llamaremos ``escaleo'' de las constantes). Para 
campos masivos, el escaleo no es suficiente para obtener la EEL y es necesario
 utilizar m\'etodos m\'as complejos.  

\section{La acci\'on efectiva Eucl\'\i dea}

Consideramos un campo escalar masivo sobre un fondo curvo Eucl\'\i deo 
cl\'asico en cuatro dimensiones. La acci\'on cl\'asica est\'a dada por 

 \begin{equation}
S=S_{\rm grav} + S_{\rm materia},
\label{sclas}
\end{equation}
donde 

\begin{equation}
S_{\rm grav}=-\int d^4x\sqrt g\left [{1\over 16\pi G_0}(R-2\Lambda_0)
+\alpha_0 R^2 +\beta_0 R_{\mu\nu}R^{\mu\nu}\right ],\label{bare}
\end{equation}
y
\begin{equation}
S_{\rm mat}={1\over 2}\int d^4x\sqrt g [\partial_{\mu}\phi\partial^{\mu}\phi
+m^2\phi^2+\xi R\phi^2].
\end{equation}
Como antes, $\xi$ es la constante de acoplamiento a la curvatura del 
espacio-tiempo. $G_0$, $\Lambda_0$ y las constantes adimensionales $\alpha_0$ 
y $\beta_0$ son las constantes desnudas. 

La acci\'on efectiva para esta teor\'\i a es un objeto no-local complicado. 
Est\'a definida integrando funcionalmente sobre el campo escalar de materia, 
y como vimos en el cap\'\i tulo anterior, formalmente podemos definirla como:

\begin{equation} e^{-S_{\rm ef}} = N \int {\cal D}\phi ~e^{-S[g_{\mu\nu}, 
\phi]},\end{equation} donde $N$ es una constante de normalizaci\'on. 

Este c\'alculo es obviamente imposible de realizar de manera exacta, y en 
consecuencia uno de los problemas fundamentales consiste en hallar esquemas 
aproximados para las ecuaciones efectivas en teor\'\i a de campos. Dos 
de estos esquemas est\'an representados por los desarrollos perturbativos 
de campo d\'ebil y por la t\'ecnica de Schwinger-de Witt \cite{bd,dewitt65}. 
Cada una de ellas tiene sus ventajas y desventajas. En particular, la 
teor\'\i a de perturbaciones no es covariante, y la t\'ecnica de 
Schwinger-de Witt no puede producir t\'erminos no-locales en la acci\'on 
efectiva. A partir de estas observaciones, Barvinsky y Vilkovisky propusieron 
un m\'etodo aproximado para evaluar la acci\'on efectiva, basado en una 
expansi\'on covariante en potencias de la curvatura \cite{barvil}. 

En general, la acci\'on efectiva a orden cuadr\'atico en la expansi\'on 
en potencias de la curvatura tendr\'a la siguiente estructura \cite{11tmunu}
\begin{eqnarray} S_{\rm ef} &=& -\int d^4x\sqrt g\left 
[{1\over 16\pi G}~ R +\alpha ~R^2 +\beta ~R_{\mu\nu}~R^{\mu\nu}\right]
\nonumber\\ 
&+&\frac{1}{32\pi^2} \int d^4x\sqrt g~\left[F_0~ R +  R~F_1(\Box)~R+R_{\mu\nu} 
~F_2(\Box)~R^{\mu\nu}+...\right], \label{seffel} \end{eqnarray}
donde los puntos suspensivos representan t\'erminos c\'ubicos en la 
curvatura. Por simplicidad hemos omitido el t\'ermino relacionado con la 
constante cosmol\'ogica. Adem\'as en la ecuaci\'on (\ref{seffel}) las 
constantes desnudas de (\ref{bare}) han sido reemplazadas por constantes 
``vestidas''. 

Toda la informaci\'on acerca de los efectos de los campos cu\'anticos 
est\'a contenida en la constante $F_0$ y en los llamados factores 
de forma $F_1$ y $F_2$. Los factores de forma, en general, son funciones 
no-locales constru\'\i das con el operador $\Box$ y los par\'ametros $\xi$ y 
$m^2$. Adem\'as, $F_1$ y $F_2$ dependen tambi\'en de la escala de 
energ\'\i a $\mu$, usualmente introducida en el proceso de 
regularizaci\'on.

\subsection{Caso no-masivo}

Por medio de las ecuaciones del grupo de renormalizaci\'on, las constantes de 
acoplamiento vestidas dependen de la escala de energ\'\i a $\mu$ de la 
siguiente forma \cite{bd}:
\begin{eqnarray}
\mu{dG\over d\mu}&=&
 \frac{G^2 m^2}{\pi} \left(\xi -  \frac{1}{6}  \right),\label{rge1} \\
\mu{d\alpha\over d\mu}&=&
 -\frac{1}{32 \pi^2}
 \left[
 \left( \frac{1}{6} - \xi \right)^2 -\frac{1}{90}
 \right], \label{rge2}\\
\mu{d\beta\over d\mu}&=&
 -\frac{1}{960 \pi^2}. \label{rge3}
\end{eqnarray}

La dependencia de $F_0$, $F_1$ y $F_2$ con $\mu$ debe ser tal que la 
ecuaci\'on completa no dependa de $\mu$. Por ejemplo, a partir de las 
ecuaciones (\ref{seffel}) y (\ref{rge1}) es inmediato demostrar que 
  $F_0=m^2\ln\left({m^2\over\mu^2}\right)(\xi-{1\over 6})+ {\rm constante}$. 

Cuando el campo escalar no tiene masa, la informaci\'on provista por las 
ecuaciones del grupo de renormalizaci\'on es suficiente para determinar 
completamente los factores de forma. En particular, como las funciones 
 $F_i,\,\, i=1,2$ son funciones de dos puntos adimensionales, mediante 
an\'alisis dimensional es simple ver que $F_i(\Box,\mu^2,\xi)=
F_i({\Box\over\mu^2},\xi)$. Usando este hecho en la ecuaci\'on 
(\ref{seffel}), las ecuaciones (\ref{rge2}) y (\ref{rge3}) y el hecho 
que $S_{\rm ef}$ debe ser independiente de $\mu$, es posible obtener

$$-\mu {d\alpha\over{d\mu}}+\mu {dF_1(-\Box/\mu^2)\over{d\mu}}=0,$$
$$-\mu {d\beta\over{d\mu}}+\mu {dF_2(-\Box/\mu^2)\over{d\mu}}=0,$$
por lo tanto, es simple demostrar que:
\begin{eqnarray}
F_1(\Box)&=&{1\over 2}\left[(\xi-{1\over 6})^2-{1\over 90}\right ]
\ln \left [{-\Box
\over\mu^2}\right] + {\rm constante},\nonumber\\                   
F_2(\Box)&=&{1\over 60} 
\ln \left [{-\Box
\over\mu^2}\right] + {\rm constante}.
\label{m=0}
\end{eqnarray}

En consecuencia, la acci\'on efectiva en este caso adquiere una clara
 interpretaci\'on: es la acci\'on cl\'asica donde las constantes de 
acoplamiento $\alpha$ y $\beta$ se reemplazan por funciones de 
dos puntos (no-locales) que tienen en cuenta el ``escaleo'' en el 
espacio de configuraciones \cite{einstlang}.

\subsection{Caso masivo}

La situaci\'on se vuelve m\'as compleja cuando el campo escalar es masivo. Los
 factores de forma dependen tambi\'en del par\'ametro $m^2/\mu^2$, y por lo 
tanto no alcanza con imponer que la acci\'on efectiva sea independiente de 
$\mu$ para conocer los factores de forma completamente. Estos factores de 
forma han sido calculados expl\'\i citamente por medio de 
 desarrollos covariantes en potencias de la curvatura \cite{barvil}, o 
mediante una resumaci\'on de la expansi\'on de Schwinger-de Witt 
\cite{11tmunu}. La expresi\'on de dichas funciones es: 

 \begin{equation}
F_i(\Box)=\int_0^1 d\gamma ~\chi_i(\xi, \gamma)~ \ln \left 
[{m^2-{1\over 4}(1-\gamma^2)\Box
\over\mu^2}\right]\label{fgrav},\end{equation}
donde 
\begin{eqnarray}
\chi_1(\xi, \gamma)&=&{1\over 2} \left [ \xi^2-{1\over 2}\xi(1-\gamma^2)
+{1\over 48}(3-6 \gamma^2-\gamma^4)\right ],\nonumber\\
\chi_2(\xi, \gamma)&=&{1\over 12}\gamma^4.
\label{chi}
\end{eqnarray}
Obviamente estos factores de forma coinciden con los de la ecuaci\'on 
(\ref{m=0}) en el caso no-masivo.

A esta altura es muy \'util introducir una representaci\'on integral 
para las funciones de dos puntos de las ecuaciones (\ref{m=0}) y (\ref{fgrav})
 para poder interpretarlas es t\'erminos del propagador Eucl\'\i deo 
(de la misma manera que obtuvimos la acci\'on efectiva CTC al final del 
cap\'\i tulo anterior). B\'asicamente, el logaritmo del operador $\Box$ puede 
escribirse en t\'erminos del propagador Eucl\'\i deo masivo 

\begin{equation}
\ln\left[{m^2-{1\over 4}(1-\gamma^2)\Box\over\mu^2}\right]
=\left[\ln
{(1-\gamma^2)\over 4}+\int_0^{\infty}dz\left({1\over z+\mu^2}- 
G_{\rm E}^{(z)}\right)\right],\label{rep}
\end{equation}
con
\be G_{\rm E}^{(z)} = {1\over{z+{4 m^2\over (1-\gamma^2)}-\Box }}.\ee
A partir de esta representaci\'on, la construcci\'on de la acci\'on efectiva 
de camino temporal cerrado sigue el mismo camino que el que empleamos 
anteriormente.

\section{La acci\'on efectiva de camino temporal cerrado}

Reemplazando el propagador Eucl\'\i deo por el de Feynman en la 
representaci\'on integral (\ref{rep}) obtenemos la acci\'on efectiva 
in-out usual. Sin embargo, nosotros queremos obtener la acci\'on efectiva CTC 
 que provee de ecuaciones de movimiento reales y causales, a la vez de 
hacer expl\'\i cita la presencia de fluctuaciones estoc\'asticas. Como 
ya mencionamos, esta acci\'on est\'a definida por la ecuaci\'on 
(\ref{neweff}), reemplazando los propagadores Eucl\'\i deos por
los del CTC, como realizamos en (\ref{prop}).

Las condiciones de contorno, como antes, implican que $\phi^+$ y $\phi^-$ 
deben tener modos de frecuencia negativa y positiva respectivamente, en 
el pasado remoto y deben coincidir sobre una hipersuperficie espacial 
en el futuro; por lo tanto la dependencia con esta superficie $\Sigma$ parece 
ser nuevamente crucial para el c\'alculo de la AECTC. Sin embargo, debemos 
notar que en este caso estamos calculando la acci\'on efectiva utilizando 
un desarrollo covariante en potencias de la curvatura hasta el orden 
cuadr\'atico. Entonces, como los factores de forma est\'an presentes en los
 t\'erminos de orden $R^2$ de la acci\'on, los propagadores que intervienen 
en el c\'alculo de la AECTC, son simplemente los correspondientes al 
espacio-tiempo plano; la discusi\'on del cap\'\i tulo anterior 
(donde estabamos calculando la AECTC de manera exacta  para modelos 
bi-dimensionales) acerca de la dependencia con $\Sigma$ deja de tener 
importancia en este caso. 

En consecuencia, podemos introducir coordenadas normales de Riemann y 
escribir los propagadores como los planos (con signatura ($-$,+,+,+))

\begin{equation} G_{\rm F}(x,y)= \int {d^4p\over{(2 \pi)^4}} {e^{ip(x-
y)}\over{p^2+m^2-i\epsilon}}= G_{\rm D}^{\star}(x,y),\end{equation} 

\begin{equation} 
G_{\pm}(x,y)=\mp \int {d^4p\over{(2 \pi)^4}} e^{ip(x-y)}2 \pi i 
\delta(p^2-m^2)\theta(\pm p^0),
\label{gpm}
\end{equation} reemplazando la masa $m^2$ por ${4m^2\over{1-\gamma^2}}
+z$. 

Realizando la integraci\'on en el par\'ametro $z$ de la representaci\'on 
integral, podemos escribir 
\begin{eqnarray}
\ln\left[{{4 m^2\over{(1-\gamma^2)}}- \Box\over\mu^2}\right]_{CTC}
= \left\{\begin{array}{ll}
 \int {d^4p\over{(2 \pi)^4}}e^{ip(x-y)}  \ln\left[{(1-\gamma^2)
(p^2-i\epsilon)+4m^2\over{\mu^2}}\right] & t, t'~ \mbox{en} 
~{\cal C}_+
\\
\int {d^4p\over{(2 \pi)^4}}e^{ip(x-y)}  \ln\left[{(1-\gamma^2)
(p^2+i\epsilon)+4m^2\over{\mu^2}}\right] & t, t' ~ 
\mbox{en} ~{\cal C}_-\\
\int {d^4p\over{(2 \pi)^4}}e^{ip(x-y)}2 \pi i \theta (p^0)\theta 
\left(-p^2 -{4m^2\over{1-\gamma^2}}\right)    & t ~\mbox{en}~ 
{\cal C}_-, t'  
~\mbox{en} ~{\cal C}_+\\
-\int {d^4p\over{(2 \pi)^4}}e^{ip(x-y)}2 \pi i \theta (-p^0)\theta 
\left(-p^2 -{4m^2\over{1-\gamma^2}}\right) & t  ~\mbox{en} 
~ {\cal C}_+, t'~  
\mbox{en}~ {\cal C}_-\end{array}\right.\nonumber\end{eqnarray} 

Gracias a esta representaci\'on podemos escribir la AECTC como
\begin{eqnarray}
S_{\rm ef}^{\rm CTC}[g^+,g^-] &=& S^{\rm r}_{\rm grav}[g^+] - 
S^{\rm r}_{\rm grav}[g^-]
 \nonumber \\
&+&{i\over{8 \pi^2}}\int d^4x \int d^4y ~\Delta(x)~ \Delta(y)~N_1(x,y)
\nonumber \\
&-& {1\over{8 \pi^2}}\int d^4x \int d^4y ~\Delta(x)~\Sigma(y)~D_1(x,y)
\nonumber\\
&+&{i\over{8 \pi^2}}\int d^4x\int d^4y~\Delta_{\mu\nu}(x)
~\Delta^{\mu\nu}(y)~N_2(x,y)\nonumber \\
&-&{1\over{8 \pi^2}}\int d^4x\int d^4y ~\Delta_{\mu\nu}(x)~\Sigma^{\mu\nu}(y)
~D_2(x,y),
\label{seffctp}
\end{eqnarray}
donde $\Delta = {{R^+ - R^-}\over{2}}$, $\Sigma = {{R^+ + R^-}\over{2}}$, 
$\Delta_{\mu\nu} = {{R^+_{\mu\nu} - R^-_{\mu\nu}}\over{2}}$ y 
$\Sigma_{\mu\nu} = {{R^+_{\mu\nu}
 + R^-_{\mu\nu}}\over{2}}$. La acci\'on gravitatoria cl\'asica $S^{\rm r}
_{\rm grav}$ contiene las constantes de acoplamiento vestidas, las cuales 
dependen de $\mu$, y $F_0$ se absorve en la constante $G$. 

La parte real e imaginaria de la acci\'on efectiva $S_{\rm ef}^{\rm CTC}$ 
pueden asociarse con la disipaci\'on y el ruido 
respectivamente, como vimos en el Cap\'\i tulo 3. Los n\'ucleos de disipaci\'on
 $D_i$ y de ruido $N_i$ son:

\begin{equation} D_i(x,y) = \int_0^1 d\gamma \chi_i(\xi, \gamma)\int 
{d^4p\over{(2 \pi)^4}} \cos[p (x-y)]\ln\left|{{(1-\gamma^2)p^2 + 
4 m^2}\over{\mu^2}}\right|,\end{equation}

\begin{equation} N_i(x,y) = \int_0^1 d\gamma \chi_i(\xi, \gamma)\int 
{d^4p\over{(2 \pi)^4}} \cos[p(x-y)] \theta\left(- p^2 - {4 m^2\over{1-
\gamma^2}}\right). \end{equation} 

\section{Disipaci\'on y ruido: las ecuaciones de Einstein-Langevin}

Dada su definici\'on, la parte imaginaria de la AECTC es definida 
positiva. Esto puede ponerse de manifiesto escribiendo la parte 
imaginaria en t\'erminos del tensor de Weyl $C_{\mu\nu\alpha\beta}$ y el 
escalar de curvatura $R$ por medio de la relaci\'on 
$C_{\mu\nu\alpha\beta}C^{\mu\nu\alpha\beta}=2R_{\mu\nu}R^{\mu\nu} - 2/3 R^2$. 
 Por lo tanto es sencillo ver que la parte imaginaria de la acci\'on 
efectiva es definida positiva:

\be \int d^4p ~\vert \Delta (p)\vert^2 ~ A ~ (\xi - {A^2\over{12}} - 
{1\over{4}})^2  
+\int d^4p ~\vert \Delta_{\mu\nu\alpha\beta} (p)\vert^2 ~ B,\ee 
donde $A = \int_0^1 ~d\gamma ~\theta (p^2 - {{4m^2}\over{
{(1-\gamma^2)}}}) > 0$, y $B = \int_0^1 d\gamma ~{\gamma^4\over{12}} 
~\theta (p^2 - {{4m^2}\over{{(1-\gamma^2)}}})>0$.

La parte imaginaria de la AECTC puede asociarse con la presencia de fuentes 
estoc\'asticas cl\'asicas de ruido  $\eta(x)$ y
$\eta^{\mu\nu\alpha\beta}(x)$, donde este \'ultimo tensor tiene las 
simetr\'\i as del de Weyl. Como es usual, esta parte imaginaria puede 
re-escribirse como:
\begin{eqnarray} &&\int {\cal D}\eta(x)\int {\cal 
D}\eta^{\mu\nu\alpha\beta}(x) P[\eta, 
\eta^{\mu\nu\alpha\beta}]\exp\left(i\left\{ \Delta(x)~ \eta(x) + 
\Delta_{\mu\nu\alpha\beta} ~\eta^{\mu\nu\alpha\beta}\right\}\right) \nonumber 
\\ &=& \exp\left\{-\int d^4x\int d^4y \left[\Delta(x) {\tilde N}(x-y) 
\Delta(y) + \Delta_{\mu\nu\alpha\beta}(x) N_2(x-y) 
\Delta^{\mu\nu\alpha\beta}(y)\right]\right\}, \end{eqnarray} 
donde ${\tilde 
N}(x,y) = N_1(x,y)+ 1/3 N_2(x,y)$ y $\Delta_{\mu\nu\alpha\beta} = 
{1\over{2}}(C^+_{\mu\nu\alpha\beta} - C^-_{\mu\nu\alpha\beta})$. Adem\'as 
$P[\eta,\eta^{\mu\nu\alpha\beta}]$ es una distribuci\'on de probabilidad 
Gaussiana determinada por
\begin{eqnarray} P[\eta, \eta^{\mu\nu\alpha\beta}] &=& A\,\, 
\exp\left\{-{1\over{2}} \int d^4x\int d^4y \eta(x) \left[ {\tilde N}(x, 
y)\right]^{-1} \eta(y)\right\} \nonumber \\ &&\times\exp\left\{-
{1\over{2}}\int d^4x\int d^4y \eta_{\mu\nu\alpha\beta}(x) \left[ N_2(x, 
y)\right]^{-1} \eta^{\mu\nu\alpha\beta}(y)\right\},\label{noise}\end{eqnarray}
donde $A$ es un factor de normalizaci\'on.

Por lo tanto, podemos definir la acci\'on efectiva $A_{\rm ef}$ como

\begin{equation} \exp\{iS_{\rm ef}\}= \int {\cal D}\eta {\cal 
D}\eta_{\mu\nu\alpha\beta} P[\eta ,\eta_{\mu\nu\alpha\beta}] \exp\left\{i 
A_{\rm ef}[\Delta, \Delta_{\mu\nu\alpha\beta}, \Sigma, \Sigma_{\mu\nu}, \eta, 
\eta_{\mu\nu\alpha\beta}]\right\},\end{equation} donde 
\begin{equation} A_{\rm ef} = {\mbox Re} 
S_{\rm ef}^{\rm CTC} + \int d^4x [\Delta(x) \eta(x) 
+ \Delta_{\mu\nu\alpha\beta}(x) 
\eta^{\mu\nu\alpha\beta}].\end{equation} 

Las ecuaciones de movimiento $\left.{\delta A_{\rm ef}
\over{\delta g^+_{\mu\nu}}} 
\right|_{g^+_{\mu\nu} = g^-_{\mu\nu}} = 0$, que son las ecuaciones de 
{\it Einstein-Langevin} quedan 
\begin{eqnarray}
&&{1\over 8 \pi G}\left( R_{\mu\nu}- 
{1\over{2}} g_{\mu\nu} R \right)
- \tilde{\alpha}~H^{(1)}_{\mu\nu}
- \tilde{\beta}~H^{(2)}_{\mu\nu}\nonumber \\
&&= -{1\over{32 \pi^2}}\int d^4y  ~ D_1(x,y)~ H_{\mu\nu}^{(1)}(y)
-{1\over{32 \pi^2}}\int d^4y ~ D_2(x,y) ~ H_{\mu\nu}^{(2)}(y)\nonumber \\ 
&&+ g_{\mu\nu}\Box \eta - \eta_{;\mu\nu}
+2 \eta_\mu{}^\alpha{}_\nu{}^\beta{}{}{}{}_{;\alpha\beta}~~,
\label{ele}
\end{eqnarray}
donde $\tilde{\alpha}$ y $\tilde{\beta}$ difieren de 
$\alpha$ y $\beta$ en una constante finita que depende de $\xi$. Como 
fuentes de estas ecuaciones tenemos al valor medio del tensor de 
energ\'\i a-impulso del campo escalar m\'as una correcci\'on 
estoc\'astica caracterizada por las funciones de correlaci\'on 
\begin{eqnarray}
\langle \eta (x)\eta (y)\rangle &=& {\tilde N}(x,y)\nonumber\\
\langle \eta_{\mu\nu\alpha\beta} (x)\eta_{\rho\sigma\lambda\theta} (y)\rangle 
&=& T_{\mu\nu\alpha\beta\rho\sigma\lambda\theta} N_2(x,y) ,\label{corrfunct} 
\end{eqnarray} donde el tensor $T_{\mu\nu\alpha\beta\rho\sigma\lambda\theta}$ 
es una conbinaci\'on lineal de productos de las m\'etricas que conserva 
las simetr\'\i as de Weyl \cite{verda2}:
 \begin{eqnarray}
  T_{\mu\nu\alpha\beta\rho\sigma\lambda\theta}
    &\equiv & \frac{1}{24}
          \left\{ 8\left[ \eta_{\rho[\mu}\eta_{\nu]\sigma}
                          \eta_{\lambda[\alpha}\eta_{\beta]\theta}
                         +\eta_{\rho[\alpha}\eta_{\beta]\sigma}
                          \eta_{\lambda[\mu}\eta_{\nu]\theta}
                         +\eta_{\alpha[\mu}\eta_{\nu]\beta}
                          \eta_{\lambda[\rho}\eta_{\sigma]\theta}
                   \right]\right.
          \nonumber \\
&+&4\left[ \eta_{\rho[\mu}\eta_{\beta]\sigma}
                          \eta_{\lambda[\alpha}\eta_{\nu]\theta}
                         +\eta_{\rho[\mu}\eta_{\alpha]\sigma}
                          \eta_{\lambda[\nu}\eta_{\beta]\theta}
                         +\eta_{\rho[\nu}\eta_{\alpha]\sigma}
                          \eta_{\lambda[\beta}\eta_{\mu]\theta}
                         +\eta_{\rho[\beta}\eta_{\nu]\sigma}
                          \eta_{\lambda[\alpha}\eta_{\mu]\theta}
                   \right]
          \nonumber \\
&-&3\left[\eta_{\mu\alpha}
                      \left( \eta_{\rho\lambda}\eta_{\sigma(\nu}
                             \eta_{\beta)\theta}
                            +\eta_{\sigma\theta}\eta_{\rho(\nu}
                             \eta_{\beta)\lambda}
                            -\eta_{\sigma\lambda}\eta_{\rho(\nu}
                             \eta_{\beta)\theta}
                            -\eta_{\rho\theta}\eta_{\sigma(\nu}
                             \eta_{\beta)\lambda}
                      \right)
                   \right.
          \nonumber \\
&+&\eta_{\nu\beta}
                      \left( \eta_{\rho\lambda}\eta_{\sigma(\mu}
                             \eta_{\alpha)\theta}
                            +\eta_{\sigma\theta}\eta_{\rho(\mu}
                             \eta_{\alpha)\lambda}
                            -\eta_{\sigma\lambda}\eta_{\rho(\mu}
                             \eta_{\alpha)\theta}
                            -\eta_{\rho\theta}\eta_{\sigma(\mu}
                             \eta_{\alpha)\lambda}
                      \right)
          \nonumber \\
&-&\eta_{\nu\alpha}
                      \left( \eta_{\rho\lambda}\eta_{\sigma(\mu}
                             \eta_{\beta)\theta}
                            +\eta_{\sigma\theta}\eta_{\rho(\mu}
                             \eta_{\beta)\lambda}
                            -\eta_{\sigma\lambda}\eta_{\rho(\mu}
                             \eta_{\beta)\theta}
                            -\eta_{\rho\theta}\eta_{\sigma(\mu}
                             \eta_{\beta)\lambda}
                      \right)
          \nonumber \\
     & & \left.
                   \left.-\eta_{\mu\beta}
                      \left( \eta_{\rho\lambda}\eta_{\sigma(\nu}
                             \eta_{\alpha)\theta}
                            +\eta_{\sigma\theta}\eta_{\rho(\nu}
                             \eta_{\alpha)\lambda}
                            -\eta_{\sigma\lambda}\eta_{\rho(\nu}
                             \eta_{\alpha)\theta}
                            -\eta_{\rho\theta}\eta_{\sigma(\nu}
                             \eta_{\alpha)\lambda}
                      \right)
                   \right]
          \right\}.
\end{eqnarray}

A partir de la ecuaci\'on (\ref{ele}) podemos definir el tensor de 
energ\'\i a-impulso efectivo 

\begin{equation} 
T_{\mu\nu}^{\rm ef}=\langle T_{\mu\nu}\rangle + T_{\mu\nu}^{\rm est}= 
\langle T_{\mu\nu}\rangle + 
g_{\mu\nu}\Box \eta - \eta_{;\mu\nu}
+2 \eta_\mu{}^\alpha{}_\nu{}^\beta{}{}{}{}_{;\alpha\beta}~~,\end{equation}
donde $\langle T_{\mu\nu}\rangle$ es el valor de espectaci\'on cu\'antico 
para el campo de materia y $T^{\rm est}_{\mu\nu}$ es la 
contribuci\'on de la fuerza estoc\'astica, la que aparece debido a 
la presencia de los n\'ucleos de ruido escalar ${\tilde N}$ y tensorial $N_2$.
  En el 
caso no-masivo y de acoplamiento conforme, el n\'ucleo de ruido escalar 

\begin{equation} {\tilde N}(x,y)={1\over{2}}\int_0^1 d\gamma \left[\left(\xi-
{(1-\gamma^2)\over{4}}\right)^2-{\gamma^4\over{36}}\right]\int {d^4p\over{(2 
\pi)^4}} cos[p(x-y)]\theta\left(-p^2-{4m^2\over{1-
\gamma^2}}\right),\end{equation}
se anula y por lo tanto $(T_\mu{}^\mu)^{\rm est}= 0$, debido a que la fuente 
de ruido $\eta^{\mu\nu\alpha\beta}$ tiene traza nula. Esto significa que 
no existe correcci\'on estoc\'astica a la anomal\'\i a de traza 
\cite{einstlang}. 

\section{Discusi\'on}

En el presente cap\'\i tulo hemos mostrado c\'omo los efectos de 
ruido aparecen en los modelos semicl\'asicos en cuatro dimensiones a trav\'es 
del c\'alculo de la acci\'on efectiva de camino temporal cerrado. Un nuevo 
t\'ermino estoc\'astico no trivial se agrega a las 
fuentes de las ecuaciones de Einstein semicl\'asicas. 

Como hemos mostrado, la parte imaginaria de la acci\'on efectiva (o 
alternativamente de la funcional de influencia) juega un rol muy 
importante en el estudio de la transici\'on cu\'antico-cl\'asica, dado que 
el proceso de p\'erdida de coherencia queda puesto en evidencia a 
partir de la funcional de Hartle y Gell-Mann. En el presente 
caso, la parte imaginaria ha sido escrita en t\'erminos de dos n\'ucleos de 
ruido. El an\'alisis de dichos n\'ucleos permite un mayor entendimiento del
 proceso de p\'erdida de coherencia y por lo tanto de la validez de la 
aproximaci\'on semicl\'asica.

En particular, estudiamos algunos casos de inter\'es. En el caso 
no-masivo, $ {\tilde N}$ es proporcional a $(\xi - 1/6)^2$ y se anula 
para acoplamiento conforme. Por lo tanto este t\'ermino est\'a presente 
s\'olo cuando el campo es masivo y/o cuando el acoplamiento 
no es conforme. Esto es de esperar, dado que la parte imaginaria de la 
AECTC est\'a directamente relacionada con la creaci\'on de part\'\i culas. 
Para campos sin masa y acoplados de manera conforme, la creaci\'on de 
part\'\i culas s\'olo tiene lugar si el espacio-tiempo no es 
conformemente plano. En consecuencia, en ese caso la \'unica 
contribuci\'on a la parte imaginaria de la AECTC es la proporcional al 
cuadrado del tensor de Weyl. Cuando los campos son masivos y/o no 
est\'an acoplados conformemente, la creaci\'on de part\'\i culas existe 
a\'un cuando el tensor de Weyl es nulo. Esto explica la aparici\'on de 
un t\'ermino proporcional a $R^2$ en la parte imaginaria de la AECTC.

%% file: tesis8
\newpage

\thispagestyle{empty}
~
\newpage

\chapter{Conclusiones}

\thispagestyle{empty}

La motivaci\'on general de la presente Tesis fue avanzar en la 
comprensi\'on del origen y de los mecanismos por los cuales la transici\'on 
cu\'antico-cl\'asica tiene lugar en teor\'\i a de campos. En particular, 
utilizando una extensi\'on del formalismo de la funcional de influencia de 
Feynman y Vernon para teor\'\i a de campos,  
estudiamos este proceso para campos escalares en el espacio de Minkowski y 
para campos 
escalares acoplados a geometr\'\i as arbitrarias 
con el objeto de entender la transici\'on ``a lo cl\'asico'' de modelos 
de gravedad cu\'antica. Estudiamos el mecanismo de p\'erdida de coherencia 
como primer paso hacia un entendimiento global 
del proceso de transici\'on cu\'antico-cl\'asica en teor\'\i a de campos. 

A partir de los resultados conocidos para la part\'\i cula Browniana 
cu\'antica, hemos
obtenido la acci\'on efectiva de granulado grueso para los modos de 
frecuencia baja integrando los modos de 
alta frecuencia de un campo escalar con autointeracci\'on. En consecuencia, 
gracias a la funcional de influencia de Feynman y Vernon obtuvimos los 
coeficientes de difusi\'on de la 
ecuaci\'on maestra. Adem\'as, evaluamos las ecuaciones de movimiento para 
el campo-sistema, las cuales incluyen efectos tanto disipativos como 
 difusivos gracias al acoplamiento con el entorno. 

Como sistema y entorno son dos sectores de un mismo campo 
escalar, los resultados dependen fuertemente del ``tama\~no'' de estos 
sectores, el cual est\'a determinado por la longitud de onda cr\'\i tica 
$\Lambda^{-1}$. Los 
resultados obtenidos para el espacio de Minkowski 
pueden generalizarse inmediatamente para el espacio-tiempo de de Sitter 
cuando el campo escalar es no-masivo y est\'a acoplado de manera 
conforme a la geometr\'\i a. En ese caso mostramos  que, cuando la longitud 
de onda cr\'\i tica es igual al radio de Hubble, todos los modos del 
sistema sufren 
p\'erdida de coherencia y por lo tanto la transici\'on a lo cl\'asico 
se torna posible. 

Posteriormente, como primer paso hacia la comprensi\'on del proceso de 
transici\'on cu\'antico-cl\'asica en modelos de gravedad cu\'antica, 
aplicamos el m\'etodo de la funcional de influencia 
a modelos de gravedad escalar-tensorial en dos dimensiones, considerando 
el r\'egimen semicl\'asico, donde la geometr\'\i a es un objeto 
cl\'asico, mientras que los campos de materia se consideran de 
natu\-raleza cu\'antica. Estos modelos bi-dimensionales 
 presentan simplificaciones importantes respecto a los modelos generales en 
cuatro dimensiones. 

En primer lugar consideramos el modelo CGHS. En ese caso hemos mostrado que 
la acci\'on de influencia puede calcularse exactamente y que es un 
objeto fuertemente dependiente de la hipersuperficie  $\Sigma$ donde se 
empalman las geometr\'\i as. En particular, en la medida conforme 
la acci\'on efectiva de camino temporal cerrado puede escribirse como la 
diferencia entre las acciones de Polyakov 
en cada rama temporal m\'as una integral sobre $\Sigma$. 

Utilizamos esta acci\'on de influencia para derivar las ecuaciones de 
movimiento semicl\'asicas. Estas ecuaciones son reales, causales y 
no-locales, y se tornan locales en la medida conforme. La derivaci\'on de 
estas ecuaciones no es trivial debido a la dependencia con la 
m\'etrica de las funciones de Green. Este es un hecho muy importante porque 
permite extender el c\'alculo a casos donde no se puede determinar el 
$\langle T_{\mu\nu}\rangle$ utilizando la ley de conservaci\'on y la 
anomal\'\i a de traza \'unicamente. En referencia a este \'ultimo caso 
mostramos
 c\'omo obtener el tensor de energ\'\i a-impulso para modelos donde el 
dilat\'on aparece acoplado al campo escalar de materia. En estos modelos 
el $\langle T_{\mu\nu}\rangle$ no se conserva y la anomal\'\i a de traza 
no lo determina completamente. Seguidamente extendimos la discusi\'on al 
c\'alculo correcto de la radiaci\'on de Hawking y otros observables 
f\'\i sicos en este modelo. 

Hemos estudiado la transici\'on cu\'antico-cl\'asica en modelos 
cosmol\'ogicos. La funcional de influencia no tiene una parte imaginaria para 
algunas superficies de empalme, y en consecuencia la aproximaci\'on 
semicl\'asica no es v\'alida en estos modelos.  

La aproximaci\'on semicl\'asica puede obtenerse de manera consistente 
cuando inclu\'\i mos en la cuantizaci\'on al dilat\'on  
para el modelo CGHS \cite{if}, o cuando estudiamos la funcional de 
influencia para 
campos escalares masivos y/o acoplados al dilat\'on o no conformemente a 
la geometr\'\i a. 

Por otro lado, es necesario mencionar que la geometr\'\i a en los 
modelos bi-dimensionales est\'a determinada por la m\'etrica y el 
dilat\'on. Por ejemlo, cuando restringimos los modelos en cuatro dimensiones 
con simetr\'\i a esf\'erica (como el \'ultimo ejemplo del cap\'\i tulo), el 
dilat\'on es parte de la geometr\'\i a dado que $e^{-2\phi}$ es el radio de 
la dos-esfera. En general, en el modelo CGHS, los lazos asociados a las 
fluctuaciones cu\'anticas
 del dilat\'on y de la m\'etrica eran ignorados argumentando el l\'\i mite 
de N grande (N es el n\'umero de campos escalares de materia presentes en el 
modelo). Como pudimos demostrar en  Cap\'\i tulo 4 no importa cu\'an 
grande es N, no hay p\'erdida de coherencia si uno no considera el efecto de 
las fluctuaciones cu\'anticas del dilat\'on y/o de la geometr\'\i a para el 
modelo CGHS. Para modelos m\'as generales, donde existe un acoplamiento entre 
el dilat\'on y el campo de materia, la misma interacci\'on asegura la 
aparici\'on de una parte imaginaria en la acci\'on de influencia y el 
efecto de p\'erdida de coherencia est\'a asegurado.  

Finalmente, en el Cap\'\i tulo 5 hemos estudiamos modelos semicl\'asicos 
en cuatro dimensiones, donde la m\'etrica cl\'asica est\'a 
acoplada arbitrariamente a campos escalares masivos. Como 
mencionamos a lo largo de la presente Tesis, las ecuaciones de Einstein 
semicl\'asicas han sido postuladas para entender algunos de los efectos 
que se producen al acoplar campos cu\'anticos con geometr\'\i as de naturaleza
 cl\'asica. Estas ecuaciones no proveen una descripci\'on completa del 
problema, dado que podemos considerar estados cu\'anticos de los 
campos de materia con fluctuaciones tales que dicha descripci\'on se 
vuelva totalmente inadecuada. Una manera de tener en cuenta la presencia 
de dichas fluctuaciones implica evaluar las correcciones estoc\'asticas 
que se inducen en las ecuaciones de movimiento de las variables cl\'asicas. 
De la misma manera que la ecuaci\'on asociada de Langevin aparece en el 
movimiento Browniano cu\'antico (Cap\'\i tulo 2), o las ecuaciones del tipo 
Langevin que dedujimos para el modelo de teor\'\i a campos en el espacio plano 
(Cap\'\i tulo 3), estas fluctuaciones deben tenerse en 
cuenta como nuevas fuentes en las ecuaciones de Einstein semicl\'asicas. 

Las fuentes estoc\'asticas provienen de la parte imaginaria de 
la acci\'on efectiva de camino temporal cerrado que, en este modelo 
semicl\'asico es b\'asicamente la acci\'on de influencia. Esta parte 
imaginaria de la acci\'on efectiva juega tambi\'en un rol muy 
importante en el estudio de la transici\'on cu\'antico-cl\'asica, dado que 
el proceso de p\'erdida de coherencia queda puesto en evidencia a 
partir de los n\'ucleos de ruido de la acci\'on de influencia y se 
ve claramente debido a la relaci\'on de \'esta con la funcional de 
p\'erdida de coherencia de Hartle y Gell-Mann. 

El an\'alisis de los n\'ucleos de ruido permite una comprensi\'on mayor 
del proceso de p\'erdida de coherencia y por lo tanto de la validez de la 
aproximaci\'on semicl\'asica.
 
Para campos sin masa y acoplados de manera conforme, la creaci\'on de 
part\'\i culas s\'olo tiene lugar si el espacio-tiempo no es 
conformemente plano. En consecuencia, en ese caso la \'unica 
contribuci\'on a la parte imaginaria de la AECTC es la proporcional al 
cuadrado del tensor de Weyl. Cuando los campos son masivos y/o no 
est\'an acoplados conformemente, la creaci\'on de part\'\i culas existe 
a\'un cuando el tensor de Weyl es nulo. Esto explica la aparici\'on de 
un t\'ermino proporcional a $R^2$ en la parte imaginaria de la AECTC.

Por \'ultimo quisi\'eramos se\~nalar que el estudio de las transiciones de 
fase y la formaci\'on de defectos topol\'ogicos en teor\'\i a de campos, 
ha cobrado gran inter\'es en los \'ultimos tiempos, debido quizas, a 
que los resultados esperados de estos modelos te\'oricos tienen relevancia, 
tanto para f\'\i sica de altas energ\'\i as como para sistemas de 
materia condensada. La funcional de influencia 
extensamente estudiada en los cap\'\i tulos previos, es una herramienta 
fundamental para el an\'alisis te\'orico de tales fen\'omenos. Es necesaria
 para la comprensi\'on de por qu\'e el par\'ametro de orden de la 
transici\'on se vuelve cl\'asico y para estudiar la din\'amica de los 
defectos topol\'ogicos que se generan, donde el objeto de importancia 
es el valor del campo (cl\'asico) promediado 
en una regi\'on finita del espacio. La funcional de influencia, por lo 
visto en esta Tesis, puede aportar importantes resultados al estudio de las
transiciones de fase en teor\'\i a de campos. Actualmente estamos investigando
 este tema \cite{vorti}.

\clearpage

%% file: tesis11
\renewcommand{\baselinestretch}{1} 

\small
\fancyhead{}
\fancyhead[LE]{\bf \thepage}
\fancyhead[RE]{\sl Bibliograf\'{\i}a}
\fancyhead[RO]{\bf \thepage}

\renewcommand{\baselinestretch}{1.5}

%% file: tesis2
\normalsize

\renewcommand{\baselinestretch}{1.2}

\thispagestyle{empty}

~
\newpage
\thispagestyle{empty}
~
\newpage
\thispagestyle{empty}

\vspace{1cm}

{\bf AGRADECIMIENTOS}

\vspace{1cm}

En primer lugar deseo expresar mi gratitud a Diego Mazzitelli 
por haberme permitido realizar el doctorado bajo su direcci\'on; por 
su permanente apoyo; su incre\'\i ble capacidad docente y por volcar 
en nuestro trabajo todo su entusiasmo e idoneidad. Por otro lado deseo 
destacar las cualidades humanas de Diego, gracias a las cuales trabajar 
con \'el ha sido un placer. 

Agradezco a Juan Pablo Paz su permanente atenci\'on, inter\'es  y
toda la ayuda prestada. De la misma manera deseo agradecer la constante 
predisposici\'on para atender mis consultas de Esteban Calzetta y los 
gratos momentos de trabajo y ``chimentos'' con Mario Castagnino.

Gracias a Mart\'\i n Ruiz de Az\'ua aprend\'\i \ mucho acerca de c\'omo 
desempe\~narme en la docencia, lo cual ha sido de gran importancia en mi 
formaci\'on. 

Todo el esfuerzo puesto en mi trabajo durante este tiempo hubiera sido 
en vano de no contar con el amor incondicional de Silvia, a la que 
debo muchas horas de atenci\'on y con la que disfruto cada instante. S\'olo 
una persona como Pilvi puede escuchar acerca de f\'\i sica, monta\~nas, 
Contingente Frontera, pol\'\i tica, futbol,
 basket, etc, y al mismo tiempo ser tan dulce. Este 
trabajo es, en gran parte de ella; y tambi\'en de mi mam\'a Telly y de 
Antonio, mi pap\'a, quienes nos han dado, a mi hermano Daniel y a m\'\i \, 
todas las herramientas necesarias para afrontar las realidades que 
aparecen a cada momento y por haber motivado, permanentemente, 
nuestra libertad de pensamiento. Gracias a ellos hemos podido afrontar los 
problemas y disfrutar 
los buenos momentos. Gracias Dany por los boletos de tren, libros y 
fotocopias cuando las cosas eran dif\'\i ciles. Los momentos que en cada
 monta\~na hemos pasado juntos han sido un pilar fundamental para 
crecer. 

Finalmente los amigos, Rom\'an, Mat\'\i as, Diego, Chicho, Lucho, Fecho, Tano,
 Rolo, Miguel, C\'esar, La Lacranet, Carlitos, etc, han colaborado de 
diferentes maneras y han aportado en general, uno de los ingredientes m\'as 
importantes para el trabajo: el humor.